\documentclass{iagtese_en}	
\usepackage{color}
\definecolor{mygreen}{rgb}{1, 0, 0}
\newcommand{\grifa}[1] {#1}
\newcommand{\grifab}[1] {#1}
\newcommand{\grifac}[1] {#1}

\begin{document}			
\institution{University of São Paulo \\ Institute of Astronomy, Geophysics and Atmospheric Sciences \\ Department of Astronomy}

\title{\grifa{Galaxy} Evolution in Clusters}

\translator{{Thesis presented to the Department of Astronomy of the Institute of Astronomy, Geophysics and Atmospheric Sciences of the University of São Paulo, as a partial requirement for the title of Doctor in Sciences.\\ \\
Field of Study: Astronomy\\
Advisor: Prof. Dr. Gastão B. Lima Neto}}

\author{Rafael Ruggiero}

\date{São Paulo \\ 2018}
						%

\pagestyle{empty}			
						%
\maketitle					%
						%
\Agradecimentos			%
I thank the São Paulo Research Foundation, FAPESP, for financing this project with grants 15/13141-7 and 16/19586-3, and also for partially financing the Laboratory of Astroinformatics at IAG/USP (grant 2009/54006-4), which was widely used in this project -- in particular, the Alphacrucis cluster.

I also thank my advisor Prof. Dr. Gastão Lima Neto, whose expertise was essential for the development and quality of this project; my advisor at the University of Zurich, Prof. Dr. Romain Teyssier, whose teachings allowed the project to reach a far greater degree of technical sophistication; Prof. Dr. Rubens Machado, who very patiently introduced me to the field of numerical simulations when I was still an undergraduate student; my teachers at IAG for the shared knowledge; and my colleagues at IAG for sharing both the struggle of the graduate courses and qualification exam, and also the fun at La Silla (where I'd like to register here for eternity that I drove the official ESO pickup truck) and at the Observatório Pico dos Dias. 

Most of all, I thank my family and friends for everything.

\vfill

\begin{flushleft}
\rule{6cm}{0.5pt}\\
{\footnotesize{This thesis has been written in \LaTeX{} using the \textsc{IAGTESE} class, for thesis and dissertations of the IAG.}}
\end{flushleft}	
						%
\Epigrafe					%
\vfill
\begin{flushright}

``\textit{Understanding is, after all, what science is all about -- and science is a great deal more than mindless computation.}''\\

\vspace{0.4cm}

Roger Penrose

\end{flushright}

\vspace{2cm}
		%
						%
						%
\Abstract					%
\vspace{-2.2cm}

Ruggiero, Rafael. Galaxy Evolution in Clusters. 2018. 103 p. Tese
(Doutorado em Ciências) – Instituto de Astronomia, Geofísica e Ciências Atmosféricas,
Universidade de São Paulo, São Paulo, 2018. Versão original.

In this thesis, we aim to further elucidate the phenomenon of galaxy evolution in the environment of galaxy clusters using the methodology of numerical simulations. For that, we have developed hydrodynamic models in which idealized gas-rich galaxies move within the ICM of idealized galaxy clusters, allowing us to probe in a detailed and controlled manner their evolution in this extreme environment. The \grifac{main} code used in our simulations is \textsc{ramses}, and our results concern the changes in gas composition, star formation rate, luminosity and color of infalling galaxies. Additionally to processes taking place inside the galaxies themselves, we have also described the dynamics of the gas that is stripped from those galaxies with unprecedented resolution for simulations of this nature (122 pc in a box including an entire $10^{14}$ M$_\odot$ cluster), finding that clumps of molecular gas are formed within the tails of ram pressure stripped galaxies, which proceed to live in isolation within the ICM of a galaxy cluster for up to 300 Myr. Those molecular clumps possibly represent a new class of objects; similar objects have been observed in both galaxy clusters and groups, but no comprehensive description of them has been given until now. We additionally create a hydrodynamic model for the A901/2 multi-cluster system, and correlate the gas conditions in this model to the locations of a sample of candidate jellyfish galaxies in the system; this has allowed us to infer a possible mechanism for the generation of jellyfish morphologies in galaxy cluster \grifab{collisions} in general.

\vspace{0.35cm}

\Resumo					%
\vspace{-2.2cm}
Ruggiero, Rafael. Evolução de Galáxias em Aglomerados. 2018. 103 p. Tese
(Doutorado em Ciências) – Instituto de Astronomia, Geofísica e Ciências Atmosféricas,
Universidade de São Paulo, São Paulo, 2018. Versão original.

Nesta tese, nós visamos a contribuir para o entendimento do fenômeno da evolução de galáxias no ambiente de aglomerados de galáxias usando a metodologia de simulações numéricas. Para isso, desenvolvemos modelos hidrodinâmicos nos quais galáxias idealizadas ricas em gás movem-se em meio ao gás difuso de aglomerados de galáxias idealizados, permitindo um estudo detalhado e controlado da evolução destas galáxias neste ambiente extremo. O \grifac{principal} código usado em nossas simulações é o \textsc{ramses}, e nossos resultados tratam das mudanças em composição do gás, taxa de formação estelar, luminosidade e cor de galáxias caindo em aglomerados. Adicionalmente a processos acontecendo dentro das próprias galáxias, nós também descrevemos a dinâmica do gás que é varrido dessas galáxias com resolução sem precedentes para simulações dessa natureza (122 pc em uma caixa incluindo um aglomerado de $10^{14}$ M$_\odot$ inteiro), encontrando que aglomerados de gás molecular são formados nas caudas de galáxias que passaram por varrimento de gás por pressão de arraste, aglomerados estes que procedem a viver em isolamento em meio ao gás difuso de um aglomerado de galáxias por até 300 Myr. Esses aglomerados moleculares possivelmente representam uma nova classe de objetos; objetos similares foram previamente observados tanto em aglomerados quanto em grupos de galáxias, mas um tratamento compreensivo deles não foi apresentado até agora. Nós adicionalmente criamos um modelo hidrodinâmico para o sistema multi-aglomerado A901/2, e correlacionamos as condições do gás nesse modelo com a localização de uma amostra de galáxias jellyfish nesse sistema; isso nos permitiu inferir um possível mecanismo para a geração de morfologias jellyfish em colisões de aglomerados de galáxias em geral.

\vspace{0.35cm}

Palavras-chave: aglomerados de galáxias, evolução de galáxias, interações de galáxias,
simulações numéricas.

		%

						%
\listoffigures 				
\listoftables 				
\tableofcontents 			
\cleardoublepage			%
\pagestyle{fancy}			
\chapter{Introduction} \label{c:introduction}

In the 20th century, with the discovery of celestial objects outside the Milky Way, a branch of Astronomy called \emph{Extragalactic Astronomy} \grifa{began}. The interest of astronomers in this field has been to explain the formation and evolution of objects in the Universe at large, such as galaxies and galaxy clusters. This is the context of this thesis, and in this chapter we will define what those objects are, as well as introduce the methodology we have employed in our research, which is that of numerical simulations. This will set up the context for the remainder of this thesis.

\section{Galaxies} \label{sec:galaxies}

\begin{figure}%
    \centering
    \subfloat[]{{\includegraphics[width=0.30\textwidth]{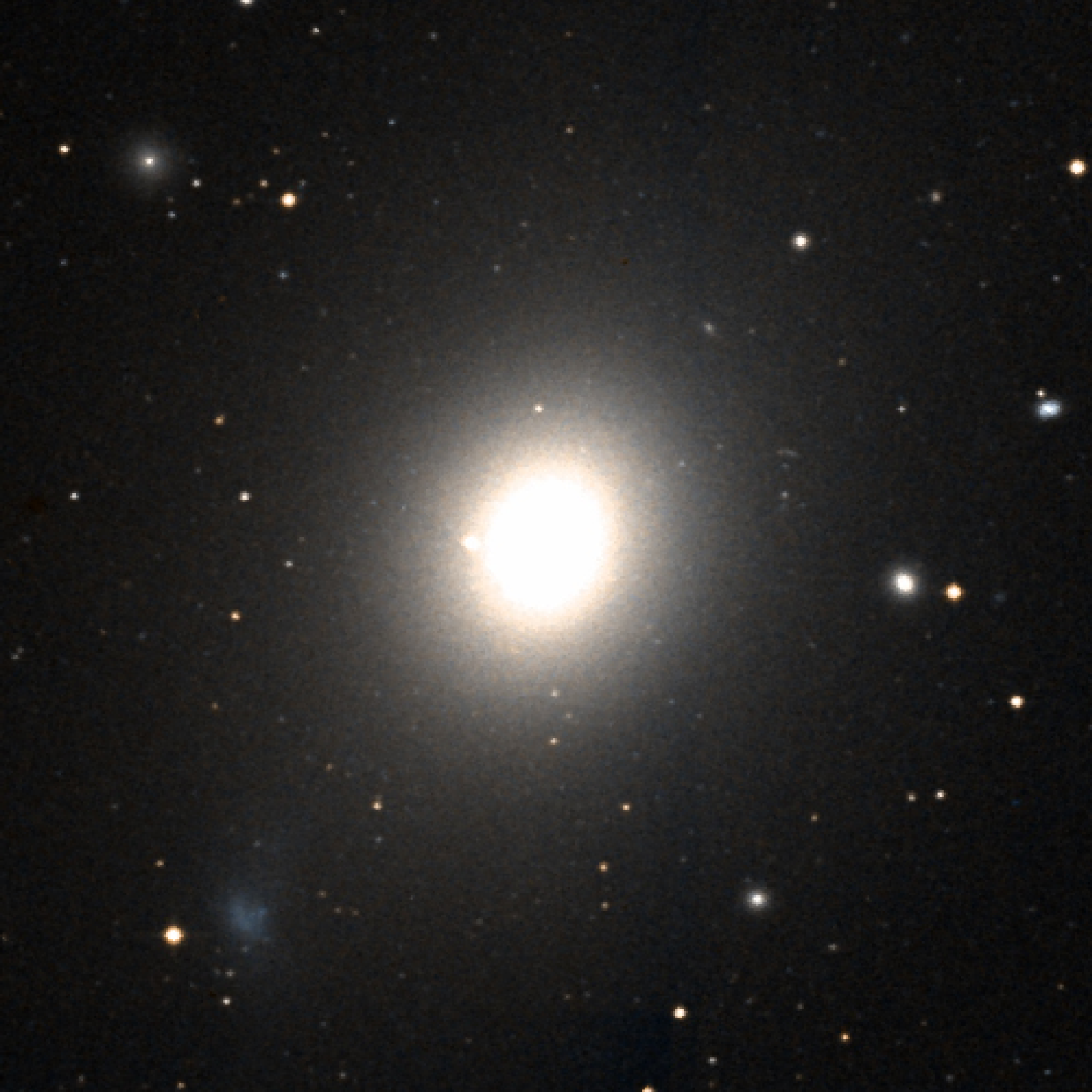} \label{fig:M49} }}%
    \hspace{10pt}
    \subfloat[]{{\includegraphics[width=0.30\textwidth]{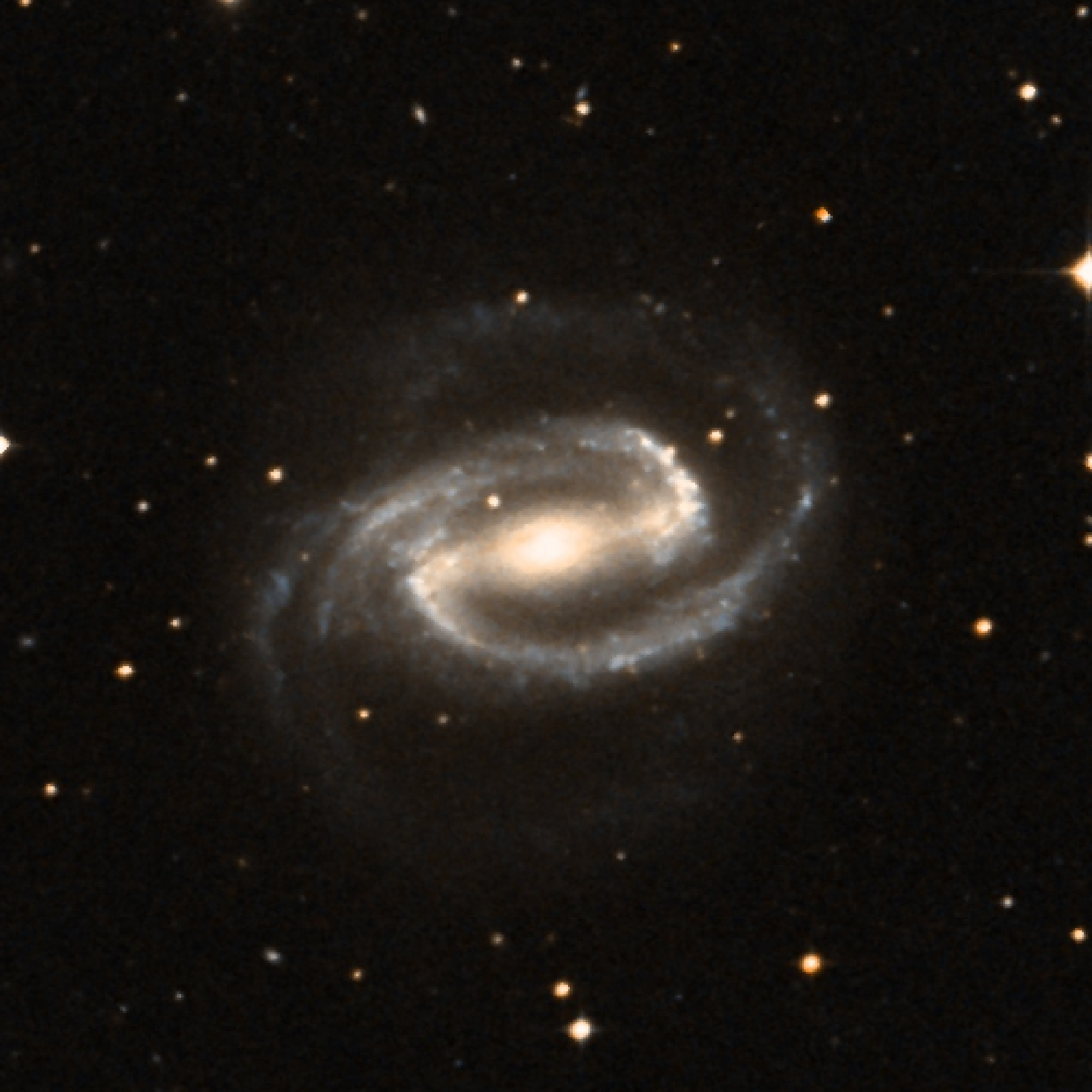} \label{fig:NGC1300} }}%
    \hspace{10pt}
    \subfloat[]{{\includegraphics[width=0.30\textwidth]{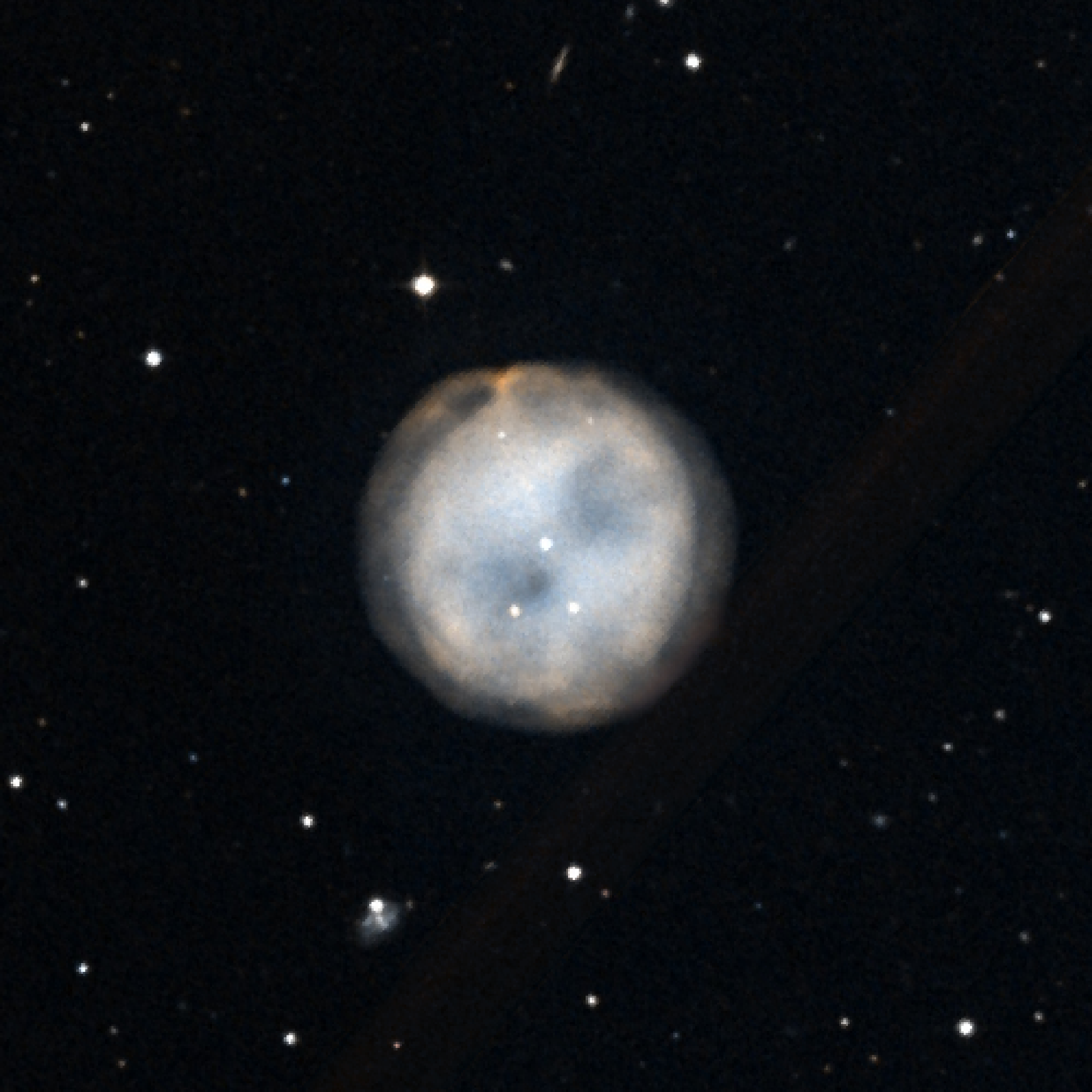} \label{fig:M97} }}%
    \caption{(a) M49, an example of elliptical galaxy. (b) NGC 1300, an example of spiral galaxy. (c) M97, a nebula inside the Milky Way. Until the beginning of the 20th century, it was unclear whether galactic nebulae resided inside the Milky Way, or if they were distant objects with sizes comparable to our own galaxy. \grifa{The images are in arbitrary scales.} Credit: DSS.}%
    \label{fig:nebulae}%
\end{figure}

Galaxies are objects observed in the sky as faint nebulae, much like nebulae within our own Milky Way, but with their own peculiar morphologies. Whereas nebulae are seen as clouds of gas with somewhat random shapes, \grifab{bright} galaxies fall into two major morphological groups: ellipticals and spirals. A comparison between those three kinds of objects can be seen in Figure \ref{fig:nebulae}. Until the beginning of the 20th century, the distances to galaxies and their sizes were unknown, and the Astronomy community was divided into those who believed galaxies resided inside the Milky Way and those who thought they were actually distant objects with sizes similar to our own galaxy. This division gave rise to the iconic ``Great Debate'' between representatives of the two groups (namely Harlow Shapley and Heber Curtis), at the National Academy of Sciences in Washington, in 1920. 

This debate was settled \grifa{a few years later} with the use of a special type of star called \emph{Cepheid stars}. Those are variable stars (i.e. stars which luminosity varies over time) with a well defined relation between the period of their variability and their luminosity \mbox{\grifa{\citep{1912HarCi.173....1L}}}, which allows the determination of the distance to any such star from its apparent magnitude and its period of oscillation. \grifa{By observing these stars individually} inside nearby galaxies in the sky, it was possible to infer that galaxies actually reside at distances much larger than the size of the Milky Way itself, making them \emph{extragalactic} objects. This work was originally carried out by Edwin Hubble, and published in a series of papers between 1925 and 1926, which are summarized in \citet{1926ApJ....64..321H}. Hubble is also famous for having \grifa{contributed to the discovery of} the expansion of the Universe using the same methodology \citep{1929PNAS...15..168H}, \grifa{although \citet{1927ASSB...47...49L} was the first to obtain this result two years before Hubble.}

The study of galaxies has historically been based on the electromagnetic radiation they emit. The most obvious source of such radiation are stars, since part of their emission is in the visible range of the electromagnetic spectrum. But sources of radiation other than stars also exist within a galaxy. For instance, cold interstellar gas emits radio waves in the 21 cm band from a spin transition of neutral hydrogen atoms, and hot coronal gas emits X-rays through thermal bremsstrahlung. Recently, non-electromagnetic information has started to be acquired from galaxies with the development of gravitational wave detectors, which have already detected emission from black hole mergers \citep[e.g.][]{2016PhRvL.116f1102A} and from a neutron star merger \citep{2017PhRvL.119p1101A}, but this new window of astronomical research is still in its infancy.

\grifa{Historically, galaxies have been classified primarily by their morphologies, although more recently color classifications have become equally popular since photometric data for large numbers of galaxies has become available, starting with the Sloan Digital Sky Survey \citep[SDSS, ][]{2000AJ....120.1579Y}}. The two major morphological groups, as we have mentioned earlier, are ellipticals and spirals -- \grifa{this comes from the classic Hubble classification scheme \citep{1936rene.book.....H}, which has the caveat of being valid primarily for bright galaxies and in the visible range of the electromagnetic spectrum}. Ellipticals are characterized by being composed primarily of old and red stars; featuring a cloud-like and relatively featureless morphology; and being very poor in neutral gas. Spirals on the other hand are typically actively forming stars, due to a high presence of dense and cold gas, and for this reason tend to have a bluer color. Their morphologies are much more intricate: their disks may sustain spiral arms, a central bar, star forming molecular clouds, etc. Other than ellipticals and spirals, galaxies halfway between the two groups exist \grifa{in this classification scheme}, namely \emph{lenticulars}, and also galaxies without a clear morphology, called \emph{irregulars}. 

\begin{figure} 
 \centering
 \includegraphics[width=0.5\textwidth]{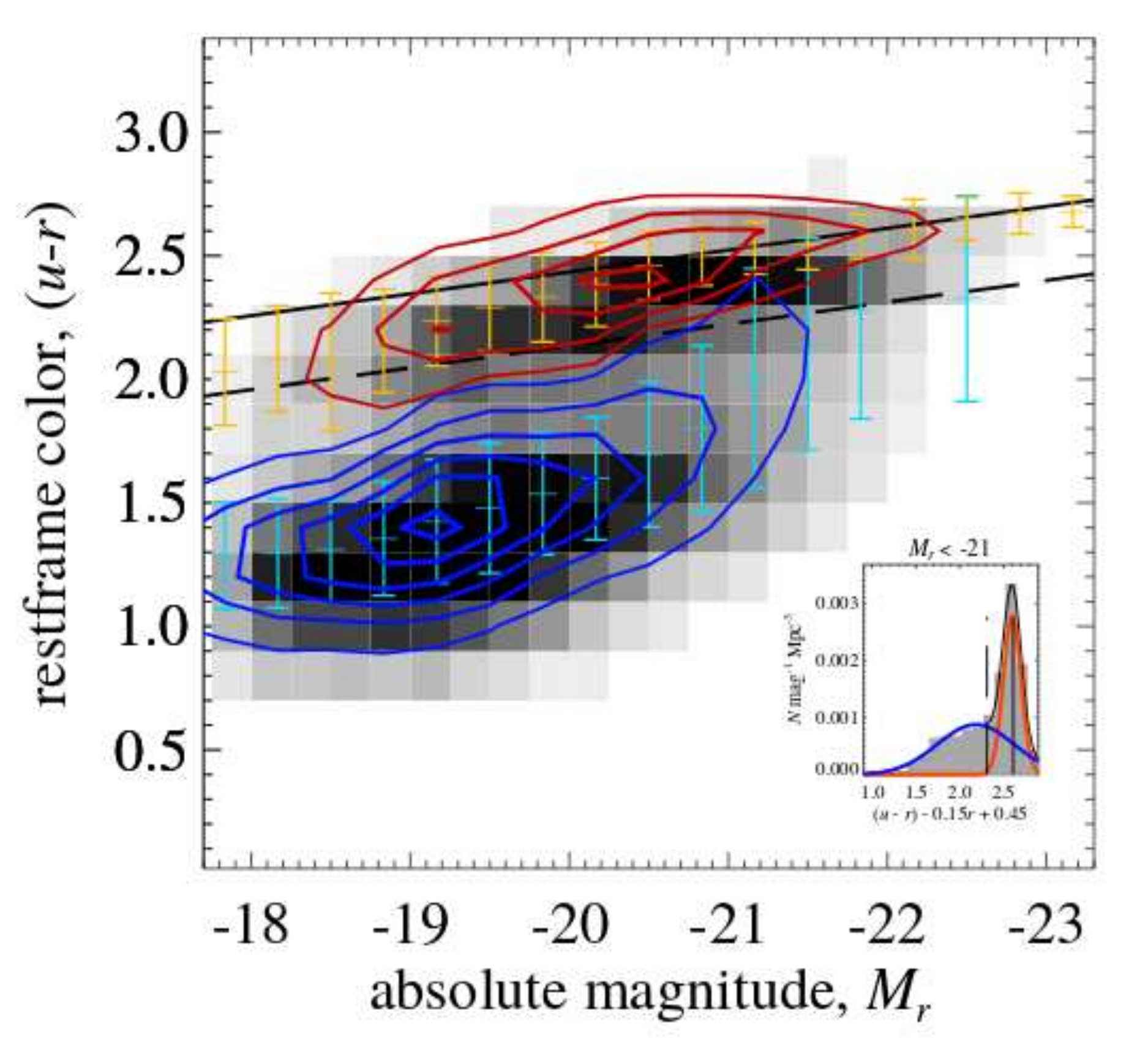}
 \caption{An example of color-magnitude diagram. The upper locus is the red sequence, and the lower locus is the blue cloud. Reproduced from \citep{2009ApJ...694.1171T}.}
 \label{fig:colormagnitude}
\end{figure}

\grifa{Regarding their colors, galaxies are commonly classified in a so called \emph{color-mangitude diagram}, such as the one shown in Figure \ref{fig:colormagnitude}. In this kind of diagram, galaxies in the Universe at large fall into two major regions: the \emph{blue cloud}, which consists of galaxies featuring blue color due to active star formation, and the \emph{red sequence}, consisting of red galaxies which have little gas left to form stars. In between the two regions, a more sparsely populated transition region exists, called the \emph{green valley}. Spiral galaxies are mostly found in the blue cloud, and ellipticals in the red sequence; since ellipticals tend to be more massive, they display on average larger luminosities in this diagram. A collaborative work for classifying large numbers of galaxies in the SDSS survey called ``Galaxy Zoo'' has demonstrated that blue ellipticals and red spirals are not uncommon \citep{2014MNRAS.440..889S}, adding depth to the oversimplified classification proposed in the beginning of the 20th century by Hubble.}

\grifa{The process of a galaxy ceasing to form stars and moving from the blue cloud to the red sequence is known as \emph{star formation quenching}. Several physical processes can cause the quenching of a galaxy. For instance, it has been suggested that galaxy mergers may trigger feedback processes which lead to the quenching of the merged system \citep{2005Natur.433..604D}, and that the shock heating of cold gas infalling into a galaxy may prevent it from replenishing its cold gas supply, also contributing to its quenching \citep{2005MNRAS.363....2K,2006MNRAS.368....2D}. More important for this thesis is the possibility of \emph{environmental quenching}, in which the density in the surroundings of the galaxy affects its star formation. The results we will present in Chapter \ref{c:interstellar} relate directly to this.}

An important piece of observational data on spiral galaxies are their rotation curves, which are a measure of the rotation speed of their disks as a function of cylindrical radius $R$. Using radio observations, it is possible to measure the rotation speed even outside the optical disk of a galaxy; such observations show that the rotation curves of spirals tend to be flat for large radii, instead of sharply falling after the optical radius, as would be expected if most of the mass in the galaxy was in its stars. The fact that this does not happen implies that some unseen mass is present in galaxies, which is now assumed to be in the form of a \emph{dark matter} component. This same conclusion also holds for elliptical galaxies, which display larger velocity dispersions than one would expect based on their stellar masses.

\section{Galaxy clusters}

\begin{figure} 
 \centering
 \includegraphics[width=0.8\textwidth]{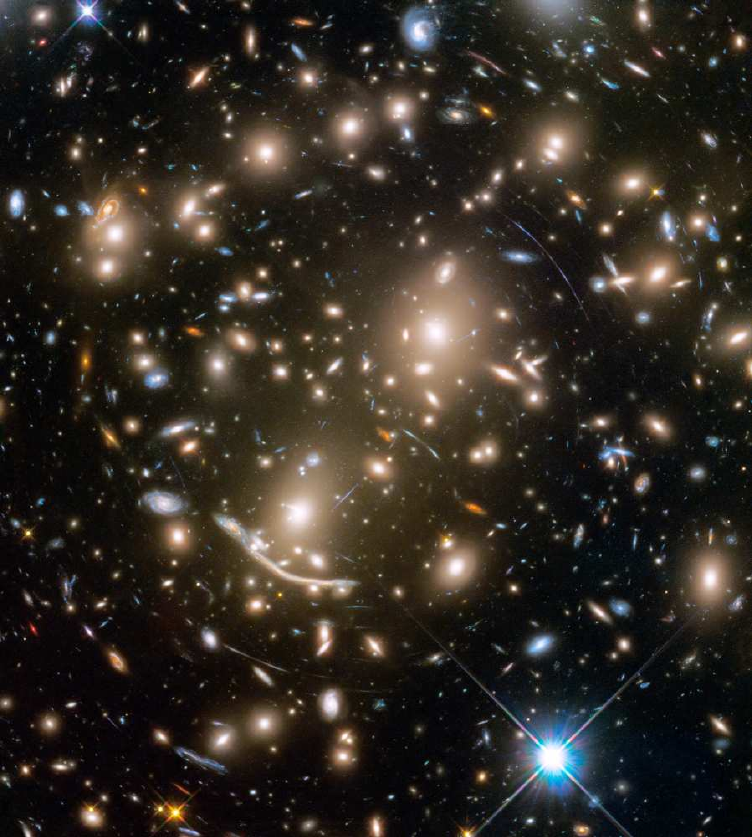}
 \caption{Abell 370, an example of galaxy cluster. This image is a very long exposure taken with the Hubble Space Telescope. Inside the cluster, galaxies are visibly more red and more often elliptical than in the field. Credit: NASA, ESA, Jennifer Lotz and the HFF Team (STScI). }
 \label{fig:galaxycluster}
\end{figure}

Upon surveying the sky for galaxies, one notices that certain regions are much denser than others. This has been noticed as far back as 1784, when Charles Messier reported an exceptional concentration of nebulae in the direction of what we now call the Virgo Cluster. A famous catalog of such \emph{galaxy clusters} is the Abell catalog, which was published in 1958 for the northern hemisphere and in 1989 for the southern, totaling 4.073 different objects. An example of galaxy cluster is shown in Figure \ref{fig:galaxycluster}.

A simple visual survey of galaxies in the sky leads to false positives, where many unrelated galaxies simply lie in the same line of sight, appearing as a cluster upon projection; but it turns out that many such observed clusters are actual gravitationally bound objects. When that is the case, the population of galaxies inside those objects is typically observed to be more red and to contain a greater population of ellipticals than the population of \emph{field} galaxies (i.e. isolated galaxies).

Despite their landmark being the high concentration of galaxies, galaxies actually represent a small fraction of the mass within galaxy clusters. An early technique which has been used to verify that is to apply the virial theorem to observed clusters. This theorem states that, for a self-gravitating system in equilibrium, the relation between its total gravitational energy $U$ and its total kinetic energy $K$ is
\begin{equation}
2K + U = 0.
\end{equation}
\grifa{For a galaxy cluster with mass $M$, radius $R$ and velocity dispersion $\sigma$ (which can be estimated from the redshifts of the member galaxies), the terms in the equation above can be roughly approximated by
\begin{equation}
2 \left(\frac{M \sigma^2}{2}\right) - \frac{1}{2} \frac{G M^2}{R} = 0,
\end{equation}
or
\begin{equation}
M \sim \frac{2 R \sigma^2}{G}.
\end{equation}
}Upon applying this relation a real cluster, one finds that most ($\gtrsim 98$\%) of its mass is not in the stars present in it. Even though this is a very simple and approximated way to estimate the mass of a cluster, the conclusion is sustained with independent estimates of the cluster mass, such as by gravitational lensing.

Part of this missing mass has been revealed by X-ray observations, which show the presence of a hot and diffuse gaseous medium permeating clusters, now known as the \emph{intracluster medium}, or ICM for short\footnote{\grifa{The ICM also contains a population of free-floating stars, responsible for the so called \emph{diffuse light} observed in galaxy clusters, but it is habitual to use the term ``ICM'' to refer to the diffuse gas exclusively.}}. But even the sum of the ICM mass inferred from its X-ray emission plus the total stellar mass of a cluster still represents a small fraction of its total mass ($\sim$10\%). Nowadays, the most popular solution for this discrepancy is the same as for the missing mass in galaxies we discussed in Section \ref{sec:galaxies}, namely that most of the mass in a cluster is attributed to the unknown type of matter called dark matter.

\begin{figure} 
 \centering
 \includegraphics[width=0.5\textwidth]{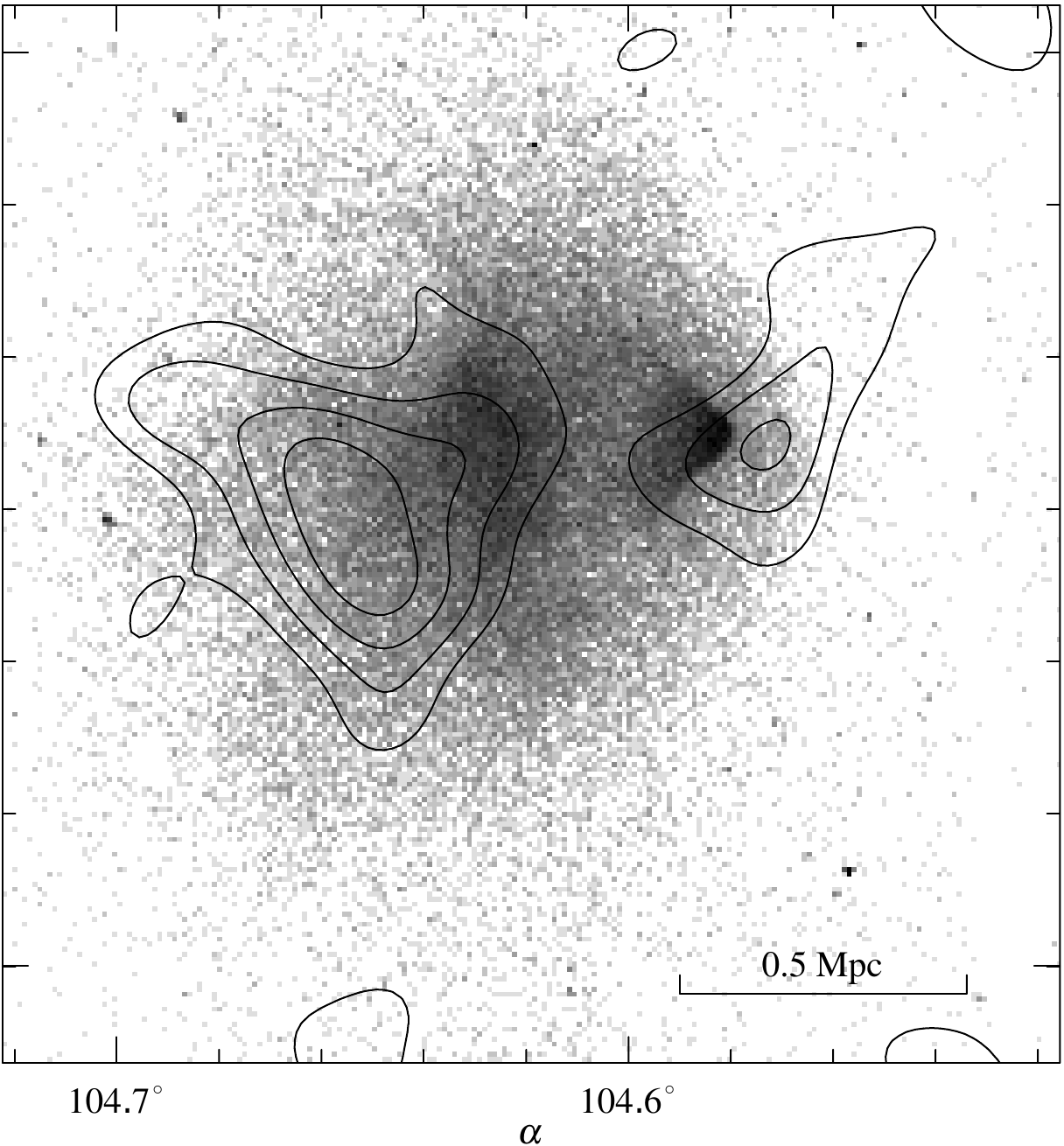}
 \caption{The Bullet Cluster \grifab{merging} system in X-rays, with mass contours obtained from gravitational lensing overlaid. This shows the ICM gas lagging behind an additional matter component present in the system, which is attributed to dark matter. Reproduced from \citet{2004ApJ...606..819M}.}
 \label{fig:bulletcluster}
\end{figure}

Although there are alternative explanations to the missing mass in both galaxy clusters and galaxies, such as modified gravity schemes, some pieces of evidence strongly favor the existence of dark matter. One of those is the so called \emph{Bullet Cluster}, which is a galaxy cluster \grifab{collision} system \citep[see e.g.][]{2002ApJ...567L..27M}. Its mass distribution inferred from gravitational lensing \citep{2004ApJ...606..819M} suggests that a collisionless (or nearly so) mass component was originally present in each of the two original clusters, and that it dissociated itself from the intracluster gas after the \grifab{collision} due to the latter being strongly self-interacting, leading it to decelerate and fall behind this collisionless component. In this scenario, which is shown in Figure \ref{fig:bulletcluster}, it is natural to conclude that a dark matter component must be present within the system.

The ICM of observed clusters is known to show a variety of features. An important one, which will be relevant later in this thesis and which is present in about 50\% of observed galaxy clusters \citep[see e.g.][]{2017ApJ...843...76A}, are so called \emph{cool-cores}, which consist of a concentration of dense and cold gas at the very center of certain galaxy clusters. \grifa{The inner region of galaxy clusters is known to have a cooling time smaller than the age of the cluster itself, which led to the theoretical prediction that a radial cooling-flow of gas must take place in clusters \citep{1977MNRAS.180..479F,1984Natur.310..733F}. But observations have demonstrated that the amount of cold gas present in observed clusters does not match the cooling-flow model prediction \citep{2001ApJ...557..546D,2003ApJ...590..207P}; those clusters which nonetheless have cold gas in their centers have been named \emph{cool-core clusters} \citep{2001ApJ...560..194M}. A peculiar characteristic of cool-core clusters is that the higher gas density in their central regions makes their X-ray emission particularly intense there, due to the high dependence of bremsstrahlung (thermal) emission on density -- the emission is proportional to $\rho^2 \,T^{1/2}$, where $\rho$ is the density and $T$ is the temperature.}

\section{\grifa{Interaction between galaxies and galaxy clusters}}

\grifa{Galaxies which reside within galaxy clusters have different properties relative to field galaxies. A relationship between local density and morphology exists, with early-type morphologies becoming more common as the density of the environment increases \citep{1980ApJ...236..351D,1984ApJ...285..426B}. The star formation of a galaxy is also related to the local density, and tends to become lower as the density increases \citep{2004MNRAS.353..713K}. Related to this is the fact that cluster galaxies tend to be more often red than field galaxies \citep{2006MNRAS.373..469B}.}

\grifa{Several mechanisms have been proposed for the evolution of galaxies in the environment of galaxy clusters. Galaxy-galaxy interactions play a role through \emph{harassment} \citep{1996Natur.379..613M}, which is a process in which successive high speed encounters between galaxies disturb their disks, making those disks thicker and consequently changing the morphology of the galaxy; and also through galaxy mergers \citep{1972ApJ...178..623T}, although these are not very common in clusters because of the large relative velocity between galaxies there. The interaction of a galaxy with the ICM can be associated with gas loss by ram pressure stripping \citep{1972ApJ...176....1G}, which can both directly remove gas from the disk of the galaxy and also remove gas from its halo, leading to a cessation in the replenishment of new gas within the disk in a process called \emph{starvation} \citep{1980ApJ...237..692L}.}

\grifa{Other than global relationships between galaxy properties and local density, two further pieces of evidence can be pointed out for the case of evolution of galaxies in the environment of galaxy clusters. The first is the \emph{Butcher-Oemler Effect} \citep{1978ApJ...219...18B}, which consists of a trend of cluster galaxies being on average bluer as redshift grows; this suggests that, as time passes, the environment of clusters tends to reduce the star formation rate of member galaxies. The second piece of evidence are the so called \emph{backsplash galaxies}, which consist of galaxies which have crossed the cluster for the first time and are observed near their apocentric passage \citep[see e.g.][]{2006MNRAS.366..645P,2011MNRAS.411.2637P}. Those galaxies are systematically redder than the infalling population \citep{2014A&A...564A..85M}, suggesting that a single crossing of of a cluster is enough to significantly affect the evolution of a galaxy.}

\section{Cosmology} \label{sec:cosmology}

Current understanding of the structure and evolution of the universe as a whole is based on the theory of General Relativity. At the core of the theory are the so called \emph{Einstein equations}, which can be compactly written as
\begin{equation} \label{eq:einstein}
G_{\mu\nu} + \Lambda g_{\mu \nu} = 4 \pi G\  T_{\mu \nu}.
\end{equation}
The terms on the left hand side of the equation are all function of the \emph{metric} $g_{\mu \nu}$, which is a tensor that completely specifies the space-time geometry of the universe, and consequently all gravitational effects taking place in it. The right hand side of the equation is proportional to the \emph{stress-energy tensor} $T_{\mu \nu}$, which on the other hand is a function of the matter density and pressure within the universe.

To see how these equations are used in cosmology, let's begin with the basic assumptions, which are that the universe in large scales is both homogeneous and isotropic. Those assumptions are commonly referred to as the \emph{Cosmological Principle}. If the Cosmological Principle is valid, then the metric takes the following simple form, for the case of a plane universe and using Cartesian spacial coordinates $(t, x, y, z)$:
\begin{equation}
g_{\mu \nu} = 
\left( \begin{array}{cccc}
-1 & 0    & 0 &    0    \\
0  & a(t)^2 & 0 &    0    \\
0  & 0    & a(t)^2 & 0    \\
0  & 0    & 0    & a(t)^2 \\
\end{array} \right).
\end{equation}
Here $a(t)$ is the \emph{scale factor}, which is the ruler that defines physical lengths in the universe, and which is allowed to vary over time. For an universe with arbitrary curvature, this metric is generalized into the following one, more conveniently written in spherical spacial coordinates $(t, r, \theta, \phi)$ \grifa{\citep{2006asci.book.....F}}:
\begin{equation}
g_{\mu \nu} = 
\left( \begin{array}{cccc}
-1 & 0    & 0 &    0    \\
0  & (1-kr^2)^{-1}a(t)^2 & 0 &    0    \\
0  & 0    & r^2 a(t)^2 & 0    \\
0  & 0    & 0    & r^2 \sin^2(\theta) a(t)^2 \\
\end{array} \right).
\end{equation}
The curvature of the universe is parametrized by $k$, which can be either $-1$ (negative curvature, hyperboloid universe), $0$ (no curvature, plane universe), or $1$ (positive curvature, hypersphere surface universe). By replacing this generalized metric into Equation \ref{eq:einstein}, one gets \grifa{(after some tensor algebra)} to the \emph{Friedmann Equation}, which gives the evolution of the scale factor $a$ of the universe over time. It is usually written as:
\begin{equation} \label{eq:friedmann}
\frac{H^2}{H_0^2} = \Omega_{0,R} a^{-4} + \Omega_{0,M} a^{-3} + \Omega_{0,k} a^{-2} + \Omega_{0,\Lambda}.
\end{equation}
Here $H \equiv \dot{a}/a$ is the \emph{Hubble parameter}, and the parameters $\Omega_i$ are defined as $\Omega_i \equiv \rho_i/\rho_c$, where $i$ denotes an energy component ($R$ for radiation, $M$ for mass, $k$ for curvature and $\Lambda$ for dark energy). The quantity
\begin{equation}
\rho_c \equiv \frac{3 H^2}{8 \pi G}
\end{equation}
is called the \emph{critical density} of the universe, and the subscripts $0$ denote the current values of the parameters.

It is currently believed, based on several pieces of observational evidence, \grifa{e.g. cosmological parameters obtained from the observation of distant supernovae \citep{1999ApJ...517..565P} and of the Cosmic Microwave Background \citep{2016A&A...594A..13P,2011ApJS..192...18K}}, that the Universe is expanding at an accelerating rate. This is consistent with either a non-zero\footnote{\grifa{In fact, $\Lambda$ must positive for the universe to expand with an acceleration.}} value of $\Lambda$ on Equation \ref{eq:einstein}, which would mean a modification of gravity, or alternatively with \grifa{an additional} energy density $\Omega_\Lambda$ incorporated into the right side of the equation, which would imply in the presence of a \emph{dark energy} density (as in Equation \ref{eq:friedmann}) of unknown nature in the Universe. 

As we have mentioned in previous sections, the presence of a dark matter component in the Universe also seems to be necessary, in order to explain the structures present in it. Due to these two factors, the current cosmological paradigm is referred to as $\Lambda$CDM, where the CDM refers to Cold Dark Matter (meaning a form of dark matter which does not exert thermal pressure, which seems to be the most viable candidate of dark matter). 

Cosmological simulations of a $\Lambda$CDM universe show that it is consistent with the dark matter in the universe clumping over time into halos, which progressively merge with each other to form larger halos \citep{1978MNRAS.183..341W,1985ApJ...292..371D}. In this context, \grifa{collapsed objects observed in the Universe, from galaxies to galaxy clusters,} are then nothing but concentrations of baryonic matter trapped inside the potential well of these dark matter halos. In this sense, both galaxies and galaxy clusters are the same kind of object, but manifested on a different scale -- \grifa{although of course the baryon properties change vastly with scale, primarily because in smaller halos the amount of gravitational energy gained by those baryons as the halo is formed is smaller, allowing the gas to cool by radiative emission more easily later on.}

The $\Lambda$CDM model also implies that, looking back in time, the universe used to be progressively smaller than it is today, until a point about 13.8 Gyr ago when it was found in the state of a singularity indescribable with current Physics. In its initial moments, all energy in the universe was in the form of high energy photons, quarks and electrons; as the expansion went on, it got progressively cooler, allowing protons and neutrons to be formed, which along with the electrons later gave rise to the first atoms. At this point, the matter field became transparent to the remaining primordial photons, and they can still be seen today in the form of a \emph{Cosmic Microwave Background} \citep{2003ApJS..148..175S,2016A&A...594A..11P}. Later, with the birth of stars, the then cooled primordial gas was progressively reionized, giving rise to the plasma that is currently found permeating the Universe on large scales. The $\Lambda$CDM theory must be additionally supplemented with an initial period of exponential expansion called \emph{inflation} in order to explain the homogeneity \grifa{and flatness} of the universe on large scales\grifa{, as well as the initial density perturbations necessary for the formation of the observed structures}; the cause of the inflation process is also not clear. 

\section{Numerical simulations} \label{sec:intronumerical}

Overall, the formation and evolution of structures in the universe at large are ruled by a simple set of equations, which have as their stage the expanding space-time described in Section \ref{sec:cosmology}. The first equation is the Poisson equation for gravity:
\begin{equation}
\nabla^2 \phi = 4 \pi G \rho,
\end{equation}
where $\phi$ is the gravitational potential and $\rho$ is the matter density. We note that this is an equation of Newtonian mechanics, which is known to be valid only at weak field regimes, where relativistic corrections to gravity can be neglected. It turns out that in the vast majority of astrophysical scenarios this is the case, with the exception of compact objects like black holes and their surroundings. The remaining equations are the fluid equations in the presence of gravity:

\begin{equation} \label{eq:fluidconservation}
\frac{\partial\rho}{\partial t} + \vec{\nabla} \cdot \left(\rho\vec{v}\right) = 0;
\end{equation}

\begin{equation}  \label{eq:fluidmotion}
\frac{\partial\vec{v}}{\partial t} + (\vec{v} \cdot \vec{\nabla})\vec{v} + \frac{\vec{\nabla} P}{\rho} + \vec{\nabla} \phi = 0; \ \mathrm{and}
\end{equation}

\begin{equation} \label{eq:fluidenergy}
\frac{\partial}{\partial t}  \left[\rho\left(\frac{1}{2} v^2 + \phi + \frac{3 k T}{2 \bar{m}} \right)\right] + \vec{\nabla}\cdot\left[\rho \vec{v} \left(\frac{v^2}{2} + \phi + \frac{5 k T}{2 \bar{m}}\right)\right] = 0, \\[12pt]
\end{equation}
where we have assumed an ideal fluid (i.e. a fluid without viscosity), and neglected magnetic fields as a first approximation. \grifa{Here $\rho$ is the gas density, $\vec{v}$ is the gas velocity, $P$ is the gas pressure, $T$ is the gas temperature and $\bar{m}$ is the average mass of the particles which make up the fluid}. Equation \ref{eq:fluidconservation} is the \emph{mass conservation equation}, which simply expresses the conservation of total fluid mass over time; Equation \ref{eq:fluidmotion} is the equation of motion for the fluid, which is analogous to Newton's second law; and Equation \ref{eq:fluidenergy} is the energy equation, used to calculate the temperature evolution over time.

\begin{figure} 
 \centering
 \includegraphics[width=0.6\textwidth]{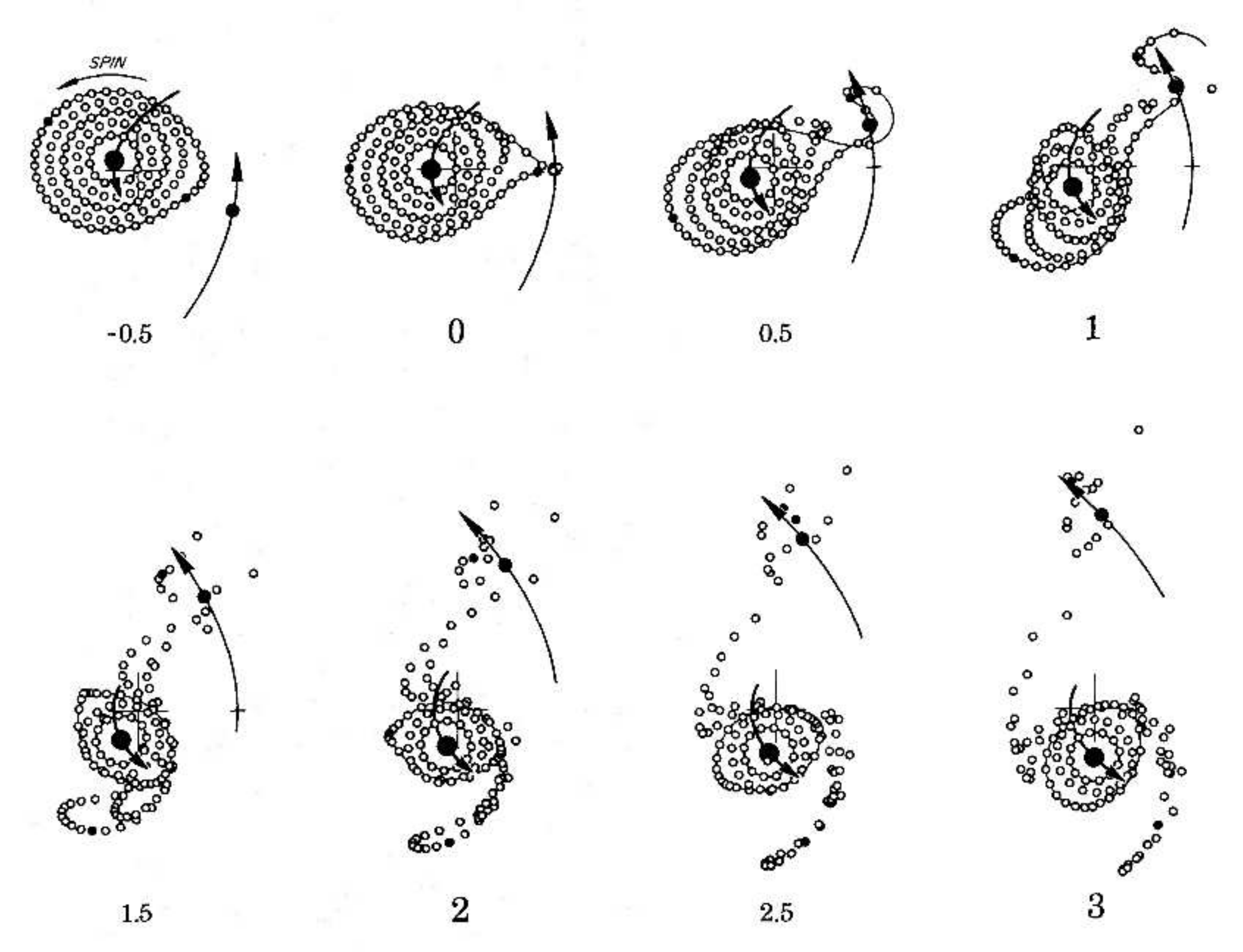}
 \caption{Numerical simulation by \citet{1972ApJ...178..623T}, showing the formation of a tidal arm as a result between a galaxy-galaxy interaction.}
 \label{fig:toomre}
\end{figure}

In astrophysical scenarios, those equations are not solvable\footnote{At least with current Mathematics. Nowadays it is not even known if Equations \ref{eq:fluidconservation} -- \ref{eq:fluidenergy} always have a solution, or if the solution is unique -- that is in essence one of the famous \emph{Millenium problems}, \grifa{formulated in detail in the following official address:} \url{https://www.claymath.org/millennium-problems/navier\%E2\%80\%93stokes-equation}.} using any simple analytical method, given the high level of non-linearity involved, with multiple-body orbits \grifa{present in the Poisson equation}, \grifa{turbulence present in the fluid equations}, etc. For this reason, after computers became available, the methodology of numerical simulations was soon adopted in the study of astronomical objects. \grifa{As soon as 1963, \citet{1963MNRAS.126..223A} modeled the dynamics of a set of idealized galaxy clusters containing from 25 to 100 particles using N-body simulations. Another} classic early simulation was that of \citet{1972ApJ...178..623T}, in which the authors studied the collision of galaxies using just $\sim$10$^2$ particles and a simplified gravity scheme -- the disk particles were not affected by each other's gravity, only by the point masses at the center of each disk. Figure \ref{fig:toomre} shows one of those simulations.

Nowadays, simulations have become much more sophisticated, and have been used to characterize in detail systems across a very wide range of scales. Some examples in ascending size are:

\begin{itemize}
\item Planet collisions ($10^{-9}$ pc). 
\item The evolution of a single star ($10^{-8}$ pc). 
\item Star formation inside a molecular cloud ($10$ pc). 
\item Galaxy dynamics ($10^{5}$ pc). 
\item Galaxy cluster collisions ($10^{6}$ pc). 
\item The evolution of the entire observable universe ($10^{9}$ pc). 
\end{itemize}

The typical dynamic range (i.e. the size of the simulated system divided by the size of the smallest resolution element) of a simulation nowadays is $\sim$10$^{5}$, much smaller than the $10^{18}$ \grifa{required for all the examples we have just given to be considered simultaneously}. Given that the evolution of large scale structures (galaxies, galaxy clusters, etc) is not independent of phenomena taking place on small scales (e.g. star formation and feedback), which cannot be resolved in simulations of those structures, the introduction of sub-grid recipes becomes necessary, in order for the simulations to be realistic. We will come back to this in Section \ref{sec:subgrid}.

\section{Open problems}

\grifac{So far, we have focused on introducing concepts and methods in this section. But the question remains: what are the open problems which motivate the study of galaxy evolution in the environment of galaxy clusters?}

\grifac{As we have mentioned in Section \ref{sec:intronumerical}, the formation and evolution of structures in the universe is a highly nonlinear process which cannot be fully accounted for by any available analytical method. In the context of galaxy evolution in clusters, a consequence of this fact is that simple questions cannot be answered in a straightforward way, such as:}

\grifac{
\begin{itemize}
\item What will a gas rich, disk galaxy look like after crossing a galaxy cluster once?
\item How much gas will it still have?
\item How much of the original gas mass will have been converted into stars?
\item What happens to the gas removed by ram pressure stripping? Is it immediately mixed with the ICM?
\item What is the effect of a galaxy cluster interaction on the ram pressure stripping of member galaxies?
\end{itemize}}

\grifac{All of those questions require the interplay of complex hydrodynamics and often small scale processes (such as star formation and feedback) to be answered, while also being dependent on a multitude of additional variables -- the details of the orbits of the galaxies, the masses and density distributions of the clusters, etc.}

\grifac{While several models for such processes were already present in the literature, we felt that simulations in which an entire galaxy cluster, with radial density and temperature profiles, were lacking, since most numerical models consisted of wind tunnel setups with simplifying assumptions about the ICM conditions -- e.g. that its density, temperature and velocity relative to the galaxy are constant. In this thesis, we aim to compliment these past results with ram pressure simulations including an entire galaxy cluster, allowing the simulation to naturally account for a more complex and realistic ram pressure variation over time. Additionally, we have also created a model to explain the formation of jellyfish galaxies in galaxy cluster collisions -- many observations of such galaxies have been reported in recent years, but a theoretical treatment of ram pressure stripping in the environment of cluster collisions had never been presented.}

\section{Overview of this thesis}

This thesis is based on the three papers that we have \grifa{published} during this PhD. On Chapter \ref{c:numerical}, we will first describe the overall numerical methodology used throughout those papers. The first paper is \citet[][hereafter \hyperlink{Paper I}{Paper I}]{2017MNRAS.468.4107R}; in it, we have constrained the possible final states of a Milky Way-like galaxy after crossing a galaxy cluster, as well as characterized its star formation rate and gas mass evolution across a range of infall scenarios. Its results will be described in Chapter \ref{c:interstellar}. The second paper is \citet[][hereafter \hyperlink{Paper II}{Paper II}]{PlaceholderPaper2}, and in it we have described the population of clumps of molecular gas left behind free-floating within the ICM of a galaxy cluster after infalling galaxies undergo a ram pressure stripping event, and also described the contamination with ICM gas of the gaseous disks of those galaxies as they cross the cluster. The results regarding the galaxies themselves will also be described in Chapter \ref{c:interstellar}, along with some additional analysis of the color and luminosity evolution of those infalling galaxies not included in the paper; the gaseous clumps will be separately described in Chapter \ref{c:clumps}. The third paper is \citet[][hereafter \hyperlink{Paper III}{Paper III}]{ruggieroa901}, in which we have proposed a mechanism for the generation of jellyfish galaxies in galaxy cluster \grifab{collisions}, based on the comparison between the gas conditions in a galaxy cluster \grifab{collision} simulation tailored to mimic the A901/2 system and the location of a sample of candidate jellyfish galaxies in that system. The thesis will be summarized in Chapter \ref{c:summary}, where some possible extensions of the work we present here will also be discussed.

\chapter{Numerical simulations} \label{c:numerical}

In this chapter, we will use the term \emph{numerical simulation} or \emph{astrophysical simulation} to refer to a 2D or 3D simulation of the time evolution of a given system, considering hydrodynamics and/or gravitation. In such kind of simulation, the system under consideration must somehow be discretized so that it can be represented in the limited memory of a computer. Regarding hydrodynamic simulations, there are two major approaches for that. The first is a Lagrangian/pseudo-Lagrangian discretization, in which the fluid is represented by mass elements which move in space following the flow. The most famous representative of this class of discretization is the \emph{Smoothed Particle Hydrodynamics} (SPH) method \citep{1977MNRAS.181..375G,1977AJ.....82.1013L}. The second option is an Eulerian discretization, in which the flow is represented into a mesh composed of unmovable cells instead, and the hydrodynamic evolution of the system is calculated considering the fluxes at the faces of the cells at each timestep. The methodology used in this thesis is the latter -- we have employed the code \textsc{ramses} \citep{2002A&A...385..337T} in our simulations, which in particular uses the \emph{Adaptive Mesh Refinement} method. This method will be described below, and after that we will introduce other key aspects of the numerical methodology employed in our work.

\section{The AMR framework}

\grifa{For the purpose of this thesis, we have chosen to use the simulation code \textsc{ramses}. Our criteria for choosing a code were that it had to be open source, well documented, widely used in the literature and, for convenience, that it had to have a set of sub-grid recipes already implemented (we will define what this means in Section \ref{sec:subgrid}). The two codes which we found to be most aligned with those criteria were \textsc{ramses} and \textsc{enzo} \citep{2014ApJS..211...19B}, and we believe this is still the case; \textsc{ramses} was chosen for no particular reason other than personal preference.}

The code \textsc{ramses} uses a special type of mesh discretization called \emph{Adaptive Mesh Refinement}, or AMR. Before getting into the details of AMR, let's first consider the simplest case of mesh simulation: one using an uniform mesh. Suppose we want to run a typical hydrodynamic cosmological simulation \grifa{(we will discuss cosmological simulations in Section \ref{sec:cosmologicalsimulations})} using this setup, where a resolution of 1 kpc is intended in a box with side length of 100 Mpc. Given that the resolution is uniform, the total number of cells in the simulation is simply
\begin{equation}
\frac{\left(100\,\mathrm{Mpc}\right)^3}{\left(1\,\mathrm{kpc}\right)^3} = 10^{15}.
\end{equation}
For comparison, the largest hydrodynamic cosmological simulations run to this date have used $\sim$10$^{10}$ resolution elements \citep[e.g.][]{2018MNRAS.475..624N}, a factor of 100.000 less. Clearly, uniform resolution simulations are impractical.

\begin{figure} 
 \centering
 \includegraphics[width=\textwidth]{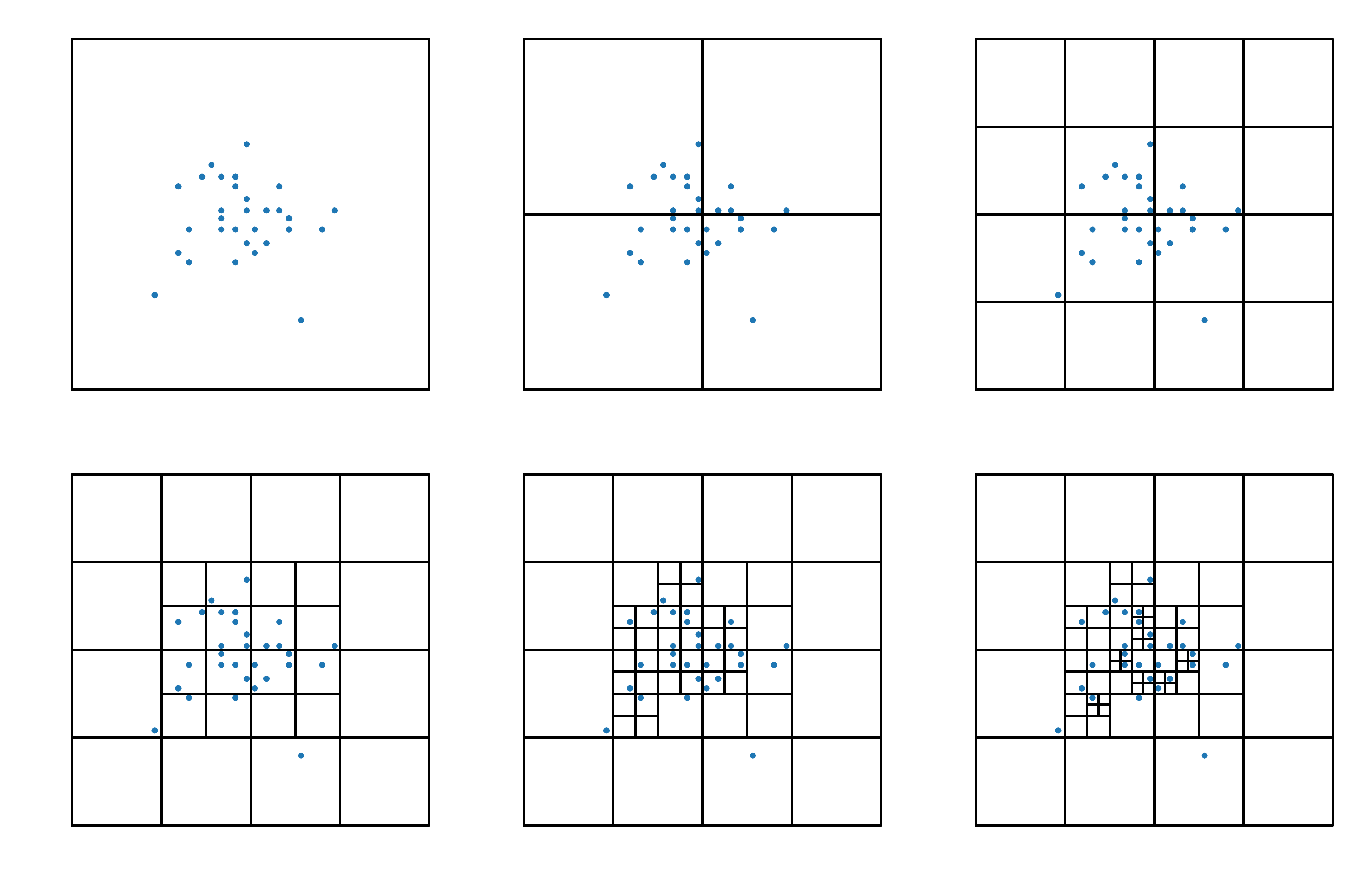}
 \caption{Illustration of how an AMR grid is built. In this toy case, the refinement criterion is that a cell is refined when it contains more than one particle; the grid is recursively refined until the criterion is no longer satisfied for any cell, in this case reaching 5 levels of refinement.}
 \label{fig:amr}
\end{figure}

It turns out that in large scale scenarios \grifac{(e.g. at cosmological scales)}, density generally does not distribute homogeneously over space, so that some regions can be represented with less resolution elements than others with minor loss of accuracy in the time evolution of the system. The AMR method seeks to exploit this fact by representing the system into a grid in which the cells are recursively refined according to some refinement criterion. An illustration of how the refinement process works is shown in Figure \ref{fig:amr}. The most obvious and widely used refinement criterion is a mass-based one, in which a cell is refined if its mass exceeds a given threshold. This way, regions with more mass will naturally have more resolution elements. Another refinement criterion that is often useful is based on the local Jeans length in each cell, defined by
\begin{equation}
\lambda_J \equiv \sqrt{\frac{15kT}{4\pi G \mu \rho}}.
\end{equation}
The cell is refined if its size exceeds this length divided by a certain factor $n$. The Jeans length is associated with star formation; the physical meaning of this criterion is that a star formation region will be resolved by $n$ cells. It is also useful for aggressively refining regions of dense and cold gas, while leaving hot regions unrefined.

Other than the refinement criterion, the other essential parameters of an AMR simulation are:

\begin{itemize}
\item The box size. Contrary to a particle based simulation, an AMR simulation necessarily must be executed into a box with predefined size.
\item The maximum level of refinement, which must be set because usually the gas in a hydrodynamic simulation will be refined until no more memory is available, so that some resolution limit is necessary.
\end{itemize}

The following parameters are not necessary, but are commonly used:

\begin{itemize}
\item The number of cells to be used for padding between two levels of refinement. For instance, if this number is set to 5 for level of refinement $n$, then in the surroundings of a level $n+1$ cell, a minimum of 5 level $n+1$ cells which would otherwise not have been refined will be refined from level $n$. This is useful for reducing numerical errors in the gravity calculation due to abrupt changes in refinement levels.
\item The minimum level of refinement, which is useful for resolving unrefined regions at a minimum desired resolution.
\end{itemize}

\grifa{\grifac{The gravity calculation in an AMR simulation requires special techniques, since cells with inhomogeneous internal resolutions are present. In \textsc{ramses}, this is done using} a so called ``one-way interface'' scheme \citep{1994JCoPh.115..339J,1997ApJS..111...73K}, in which the Poisson equation is solved at a parent level, and then the boundary conditions at the lower levels of refinement are interpolated from the parent solution, without those lower levels ever directly affecting the upper level calculation. At the minimum level of refinement of the simulation (called the \emph{coarse level}), the Poisson equation is solved using the Fast Fourier Transform technique, and at lower levels a relaxation technique based on the Gauss-Seidel method is employed. This iterative technique has a free parameter $\epsilon$, which sets the relative error below which the iteration is stopped; a typical value is $\epsilon = 10^{-4}$. In case the AMR grid also contains collisionless particles, the mass of each of those particles is deposited into the grid cells using a \emph{Cell-In-Cloud} \citep[CIC][]{1981csup.book.....H} interpolation scheme before the calculation is carried out.}

\grifa{The gas fluxes at the cell interfaces in grid simulations in general, including \textsc{ramses} and most AMR codes, are found at each timestep using a \emph{Riemann solver}. A Riemann solver is any algorithm for solving the \emph{Riemann problem}, which is defined as a problem in which conservation equations must be solved in a piecewise constant discretization \grifac{with discontinuities at the boundaries of the elements of the discretization}. This is the case for a fluid simulation because mass must be conserved, \grifac{while the gas velocity is in principle different at each side of each cell face.} A common solver, and also the one which we have used in all simulations presented in this thesis, is the Harten-Lax-van Leer-Contact (HLLC) solver \citep{1994ShWav...4...25T}.}

\section{Idealized simulations} \label{sec:idealizedsimulations}

A common class of numerical simulations are \emph{idealized simulations}, \grifa{which we define as simulations} in which objects of interest (galaxies, galaxy clusters, stars, etc) are represented by \grifa{simplified density profiles}. For instance, instead of the halo of a galaxy being represented by a cosmological model with substructures, asymmetries, etc, in an idealized setup, it might simply be represented by a spherically symmetric model. Those simulations have the advantage of allowing a higher level of control over the initial conditions, which translates into a more comprehensive interpretation of the results of the simulation; but they can also leave out relevant details that a more realistic setup would naturally include, which can lead to biased results.

The first step in this kind of simulation is to generate representative initial conditions. This is done by choosing a density profile for each mass component that will be present in the simulation. Let's consider some commonly used profiles. For dark matter halos, a Navarro-Frenk-White (NFW) profile \citep{1997ApJ...490..493N} can be used, which was originally proposed as a good fit for the density profile of dark matter halos formed in $\Lambda$CDM cosmological simulations:
\begin{equation}
\rho(r) = \frac{\rho_0}{\frac{r}{r_0}\left(1 + \frac{r}{r_0}\right)^2},
\end{equation}
where $\rho_0$ is \grifa{a density which is used} to specify the desired mass of the profile within a given radius, and $r_0$ is a scale factor. An alternative to the NFW is the Hernquist density profile \citep{1990ApJ...356..359H}:
\begin{equation}
\rho(r) = \frac{M}{2\pi} \frac{a}{r} \frac{1}{(r+a)^3},
\end{equation}
where $M$ is the total mass and $a$ is a scale factor. The parameter $a$ of the Hernquist profile can be adjusted to very closely match the shape of a given NFW profile. \grifa{This is done in the following way. First we define the $R_{200c}$ of a halo as the radius within which the average density is equal to 200 times the critical density of the universe (see Section \ref{sec:cosmology}), and the $M_{200c}$ as the mass within that radius. It turns out that, for $\Lambda$CDM halos, a relationship exists between the \emph{concentration} of a NFW profile, defined by $c \equiv R_{200c}/r_0$, and its $M_{200c}$; one example can be found in \citet{2008MNRAS.390L..64D}. Based on this relationship and on the desire $M_{200c}$ of a halo, one first finds the associated concentration of the NFW profile, and then optimizes the parameter $a$ of the Hernquist profile to minimize the distance to the resulting profile according to some metric.} This conversion is often useful in numerical simulations because the Hernquist profile, contrary to the NFW, has a finite mass, meaning it does not require any ad-hoc truncation. The Hernquist profile was originally proposed to fit the density of galaxy bulges, and it is a special case of the more general Dehnen density profile \citep{1993MNRAS.265..250D}:
\begin{equation}
\rho(r) = \frac{(3-\gamma)M}{4 \pi} \frac{a}{r^\gamma (r+a)^{4-\gamma}}.
\end{equation}
The parameter $\gamma$ in this profile affects primarily the slope of the density profile as $r \rightarrow 0$. If $\gamma = 1$, then the profile corresponds to a Hernquist profile, which features a central \emph{cusp} (i.e. a central divergence to infinity in the density profile). On the other hand, if $\gamma = 0$, then the profile gets flat as $r \rightarrow 0$, or, in other words, it features a central \emph{core}. This latter case is qualitatively similar to a $\beta$-model \citep{1976A&A....49..137C}, which is often used to model the ICM of observed galaxy clusters:

\begin{equation} \label{eq:betaprofile}
\rho(r) = \rho_0 \left[1 + \left(\frac{r}{r_c}\right)^2\right]^{-\frac{3\beta}{2}}.
\end{equation}

For stellar disks of disk galaxies, it is well known observationally that the surface brightness tends to fall exponentially with cylindrical radius \citep{1959HDP....53..311D,1970ApJ...160..811F}. This motivates the common use of the two following profiles in simulations of disk galaxies:
\begin{equation} \label{eq:rhodisksech}
\rho(R, z) = \frac{M}{4 \pi R_d^2 z_0} \exp\left(-\frac{R}{R_d}\right) \mathrm{sech}^2\left(\frac{z}{z_0}\right),
\end{equation}
as in \citet{2010MNRAS.406.2386M} and \citet{2015ApJ...814..131G}; or alternatively \citep[as in e.g.][]{2010MNRAS.409.1088B,2013MNRAS.436.1836R}:
\begin{equation}\label{eq:rhodiskexp}
\rho(R, z) = \frac{M}{4\pi R_d^2 z_0} \exp\left(-\frac{R}{R_d}\right) \exp \left(-\frac{z}{z_0}\right).
\end{equation}

\begin{figure} 
 \centering
 \includegraphics[width=0.8\textwidth]{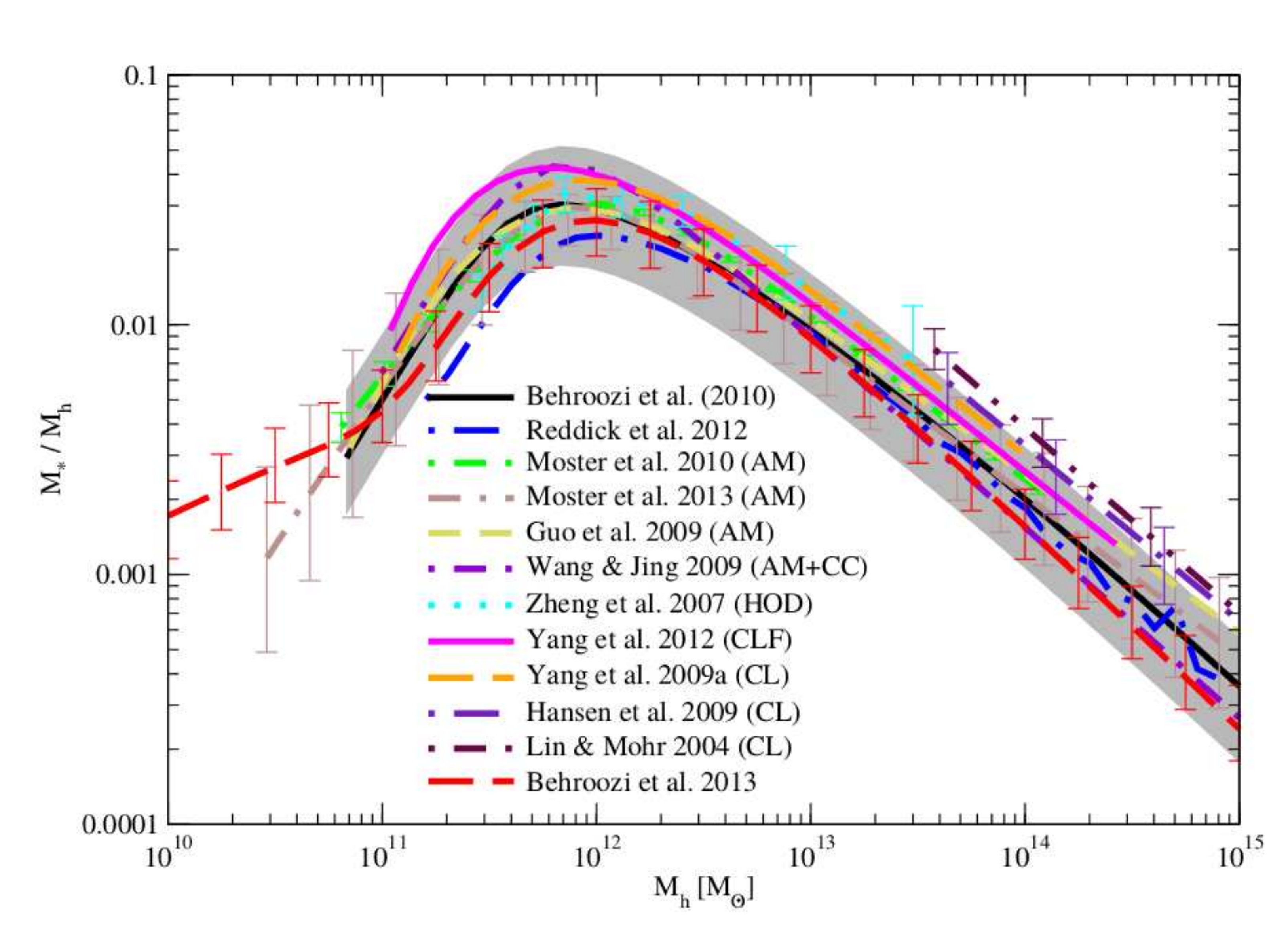}
 \caption{Abundance matching data from \citet{2013ApJ...770...57B}, which allows one to e.g. realistically assign stellar masses to dark matter halos extracted from dark matter only cosmological simulations.}
 \label{fig:abundancematching}
\end{figure}

Sometimes, in generating a model galaxy, it is necessary to realistically associate a certain stellar mass to a given dark matter halo, or a dark matter halo to a stellar disk with given mass. An example could be when one wants to use a dark matter halo extracted from a cosmological simulation for an idealized galaxy simulation. An useful result for this task comes from the technique of \emph{abundance matching}, which associates dark matter halos from cosmological simulations to observed galaxies \citep[see e.g.][]{2013ApJ...770...57B}. In essence, the technique starts by assuming that stellar mass grows monotonically with dark matter halo mass. One then computes dark matter halo mass bins for halos formed in cosmological simulations, and matches those bins to stellar mass bins from galaxy surveys, obtaining a $M_\mathrm{halo}$ -- $M_\star$ relation such as the one shown in Figure \ref{fig:abundancematching}. \grifac{Note that this figure shows a ratio of stellar mass to halo mass, and not stellar mass itself, and for this reason the curve does not grow monotonically with halo mass (even though the stellar mass does).}

\subsection{Initial conditions}

In generating initial conditions for an idealized simulation, positions must be assigned to the particles that will describe each system component. The density profiles usually have either spherical or cylindrical symmetry, meaning each profile is usually either written in the form $\rho(r)$ or $\rho(R, z)$. The process of assigning positions to the particles starts by interpreting the profiles as probability distribution functions; then, a sampling is made from from those distributions, usually by inverse transform sampling. To illustrate how it works, let's consider a radial profile $\rho(r)$ with total mass $M$. First a random number $c$ between $0$ and $1$ is sampled \grifa{with a flat probability distribution}, and then the radius $r'$ for which the normalized cumulative distribution function equals $c$ is found, i.e.
\begin{equation}
\frac{1}{M} \int_0^{r'} 4 \pi r^2 \rho(r) = c.
\end{equation}
This \grifac{method is known as \emph{inverse transform sampling}, and it} results in a sampling which is guaranteed to follow the original density profile. For the case of a cylindrical profile $\rho(R, z)$, a similar procedure is carried out, but two samplings must be made; \grifa{if the $z$ profile does not depend on $R$ (which is usually the case), then the samplings are made independently for each of the two marginalized distributions}:
\begin{equation}
\rho_R(R) \propto \int_{-\infty}^{\infty} \rho(R, z)\, dz
\end{equation}
and
\begin{equation}
\rho_z(z) \propto \int_{0}^{\infty} 4 \pi R^2 \rho(R, z)\, dR.
\end{equation}

In both cylindrical and spherical cases, angles must be additionally sampled in order for the 3D particle position to be completely specified, namely two angles $(\theta, \phi)$ in the spherical case and one angle $\phi$ in the cylindrical. In both cases, the angle $\phi$ must be simply sampled homogeneously between $0$ and $2\pi$. The angle $\theta$ in the spherical case, on the other hand, cannot be sampled the same way between $-\pi$ and $\pi$. Instead, a number $\mu$ between $-1$ and $1$ must be sampled first, and then be converted into $\theta$ by $\theta = \arccos{\mu}$, so as to assure that the point given by $(r, \theta, \phi)$ is sampled homogeneously over the surface of the sphere with radius $r$.  

Assigning velocities to the particles is a far more complex problem. The velocities must be such that the original density profiles of the constituent mass components of the simulated object will be preserved over time when it is simulated in isolation. For the special case of a system with spherical symmetry, a more straightforward approach exists. It is possible to show that collisionless spherical systems in equilibrium have particle energies which follow a distribution function called \emph{Eddington formula} \grifa{\citep{1916MNRAS..76..572E,1987gady.book.....B}}:
\begin{equation} \label{eq:eddington}
\text{DF}(\mathcal{E}) = \frac{1}{\sqrt{8} \pi^2} \left[ \int_0^{\mathcal{E}} \frac{d^2 \rho}{d \Psi^2} \frac{d \Psi}{\sqrt{\mathcal{E} - \Psi}} + \frac{1}{\sqrt{\mathcal{E}}} \left(\frac{d\rho}{d \Psi}\right)_{\Psi = 0}\right].
\end{equation}
Here $\Psi = -\Phi$ is the so called \emph{relative potential}, with $\Phi$ being the gravitational potential;
$\mathcal{E} = \Psi - v^2/2$ is the \emph{relative energy};
$\rho$ is the density of the component under consideration; and $v$ is the particle velocity. \grifa{In order for the system to have finite mass, $\rho(r)$ must fall faster than $r^{-1}$ as $r \rightarrow \infty$, or equivalently as $\Psi \rightarrow 0$. Meanwhile, $\Psi$ (which is proportional to the gravitational potential) falls as $r^{-1}$ for large radii. This implies that the second term in Equation \ref{eq:eddington} can always be neglected.}

This distribution function allows one to sample values of $\mathcal{E}$ consistent with the equilibrium of the system as a whole. This is done for each particle of each collisionless component. One way to sample velocities from this function would be to use the cumulative distribution method again, as we have described earlier for the particle positions, but this would involve calculating integrals of Equation \ref{eq:eddington}, which are computationally costly. Sampling by rejection \grifac{(also known as \emph{von Neumann sampling})} turns out to be more efficient in this case.

The idea is simple: values of $(\mathcal{E}, y)$ are sampled in an appropriate rectangle, and the pair is accepted if $y < DF(\mathcal{E})$ (i.e. if the pair is part of the distribution function area). The relative energy $\mathcal{E}$ of the particle is thus obtained, and as the gravitational potential in its position is known (since the mass distribution of the system is given), this also specifies its velocity $v$. Finally, a random direction of motion is assigned to this velocity (which is valid because of spherical symmetry), by sampling random angles $\theta$ and $\phi$ in the same way as for the particle positions.

In order to understand which rectangle is appropriate for the sampling, first note that the kinetic energy of the particle can be at most equal to its gravitational potential energy, so that it remains bound to the system. Thus, by definition of $\mathcal{E}$, this parameter must be a number in the interval [$\Psi$, $0$]. Moreover, the Eddington distribution function has a strictly negative derivative, such that the maximum value of $\text{DF}(\mathcal{E})$ in this interval is $\text{DF}(\Psi)$. Therefore, the appropriate interval for sampling of y is $[0, \text{DF}(\Psi)]$, noting that the values of the distribution function are, by definition, positive.

The methods which we have just described for initial conditions of spherical collisionless systems are implemented into our code \textsc{clustep}\footnote{\url{https://github.com/ruggiero/clustep}}, which we have used throughout this thesis to generate initial conditions for galaxy cluster halos, including dark matter and gaseous components (we will describe later how the gas is initialized). Now let's consider systems without spherical symmetry, such as a galaxy including a stellar disk. In this case, Equation \ref{eq:eddington} is not valid, and an alternative method must be used. The two major classes of methods which are used in the literature are iterative methods and methods based on making approximations of the velocity distributions using the Jeans equations. What the former methods do is to find the velocity distributions empirically, by integrating the system in time for short periods and updating the velocities each time until an unchanging phase space density is reached. An example of such method is described in \citet{2009MNRAS.392..904R}.

Methods based on the Jeans equations on the other hand are more commonly used, since they are somewhat simpler to implement and do not require several simulations to be run in the generation of initial conditions. The rationale is that those equations can be used to compute integrals of motion for the system, which can be associated with velocity dispersions $\sigma_i$ for each $v_i$ velocity component of each system component. The velocity components are then assumed to be distributed following a Gaussian distribution, allowing a sampling to be made. A commonly used method is described in \citet{2005MNRAS.361..776S} \citep[and also previously in][]{1999MNRAS.307..162S}, and in what follows, we will describe the version of that method which is implemented into our initial conditions code \textsc{galstep}\footnote{\url{https://github.com/ruggiero/galstep}}.

The model galaxy we will consider is made of a dark matter halo and stellar bulge, both with spherical symmetry, plus a stellar disk and a gaseous disk, both with cylindrical symmetry. Since the system as a whole has cylindrical symmetry, the natural coordinate system to be used in the initial conditions generation is cylindrical, with $(R, z, \phi)$ coordinates. In this system, the Jeans equations imply that the second-moments for the velocities $v_z$ and $v_R$ of the spherical components (halo and bulge) are given by
\begin{equation} \label{eq:sigmaintegral}
\left<v_z^2\right> = \left<v_R^2\right> = \frac{1}{\rho}\int_z^\infty
\rho(z',R) \frac{\partial \Phi}{\partial z'} \,{\rm d}z',
\end{equation}
where $\rho$ is the density of the component being considered. For the $\phi$ direction, the second-moment is given by
\begin{equation} \label{eq:sigmaphi}
\left<v_\phi^2\right> =
\left<v_R^2\right> + \frac{R}{\rho}\frac{\partial \left(\rho
\left<v_R^2\right>\right)}{\partial R} + v_c^2,
\end{equation}
where
\begin{equation}
v_c^2 \equiv R \frac{\partial\Phi}{\partial R}
\end{equation}
is the rotation speed. It is a basic statistical fact that the relation between the second-moment and the standard deviation of a distribution is:
\begin{equation} \label{eq:momentvssigma}
\sigma_i^2 = \left< v_i^2 \right> - \left<v_i\right>^2. 
\end{equation}
For both halo and bulge components, we choose for simplicity to not include any angular momentum, such that no net rotation is present (i.e. $\left<v_\phi\right> = 0$). With this assumption, Equation \ref{eq:momentvssigma} implies that $\sigma_\phi^2 = \left<v_\phi^2\right>$ for those components. Since naturally $\left<v_z\right> = \left<v_R\right> = 0$, it also holds that $\sigma_R^2 = \left<v_R^2\right>$ and $\sigma_z^2 = \left<v_z^2\right>$.

For the stellar disk, different equations are used because it is not spherical and because it rotates. The velocity dispersion in the $z$ direction is also calculated using the integral in Equation \ref{eq:sigmaintegral}, but this time the radial velocity dispersion is assumed to be proportional to the vertical one in a more general way, with $\sigma_R^2 = f_R \sigma_z^2$, where $f_R$ is a free parameter. This free parameter is useful because it allows one to adjust the stability of the stellar disk. In the $\phi$ direction, the \emph{epicyclic approximation} is used, in which
\begin{equation} \label{eq:epicycle}
\sigma_\phi^2 = \frac{\sigma_R^2}{\eta^2}, 
\end{equation}
where
\begin{equation}
\eta^2 =
\frac{4}{R} \frac{\partial \Phi}{\partial R} \left(\frac{3}{R}
\frac{\partial \Phi}{\partial R} + \frac{\partial^2 \Phi}{\partial
R^2} \right)^{-1}.  
\end{equation}
The mean rotation velocity is obtained using Equations \ref{eq:momentvssigma} and \ref{eq:epicycle}:
\begin{equation}
\left<v_\phi\right> = \left(\left<v_\phi^2\right>
- \frac{\sigma_R^2}{\eta^2}\right)^{1/2},
\end{equation}
with $\left<v_\phi^2\right>$ given by Equation \ref{eq:sigmaphi}

For the gaseous disk, no velocity dispersion is used, and its rotation velocity is set simply such that there is a radial balance between gravity on one hand, and thermal plus centrifugal support on the other:
\begin{equation}
v_{\phi,\rm gas}^2 = R \left(
\frac{\partial \Phi}{\partial R} + \frac{1}{\rho_{\rm
g}}\frac{\partial P}{\partial R} \right).  
\end{equation}

This method based on the Jeans equations has the disadvantage of neglecting higher order \grifa{moments} -- there is no a priori reason for the distribution of each velocity component to be Gaussian and also independent of other components. This can lead to transients at the beginning of the simulation, as the system relaxates into a true state of equilibrium\footnote{A very useful way to accelerate the relaxation is to first simulate the galaxy with all components frozen except the dark matter halo, for about one rotation period. This avoids transient rings in the stellar disk of the galaxy, which can appear if the halo turns out to not be properly in equilibrium.}. Indeed, \citet{2004ApJ...601...37K} has pointed out that the approximation of the velocity dispersion of the dark matter halo as being Maxwellian, i.e. Gaussian in each of its spatial components, is not valid\grifa{-- meaning higher order moments (e.g. kurtosis) are present in the velocity distribution for each of those components}. Iterative methods do not feature this problem, but they have the disadvantage of being much more computationally expensive.

Until now, we have only \grifa{discussed the} initial velocities. For a gaseous component, initial temperatures must also be set, in such a way as to ensure hydrostatic equilibrium. Let's first consider the case of a spherical gaseous halo, which could be the ICM of a model galaxy cluster. Its particle positions are set according to the chosen density profile, and the particle velocities are set to zero, since the system is kept from collapsing by local pressure, and not by velocity dispersion. Then the initial temperatures must be set; this is done by imposing hydrostatic equilibrium. If the gas is in hydrostatic equilibrium, then the sum of the local gravitational force and the \grifab{force due to the local pressure gradient} must be zero. Since there is spherical symmetry, this can be written as:
\begin{equation}
\frac{dP}{dr} = -\frac{G M(r) \rho(r)}{r^2},
\end{equation}
where $P$ is the gas pressure, $\rho$ is the gas density, and $M(r)$ is the total mass within the radius $r$. Integrating this equation from a given radius to infinity, and noting that $P(r) \rightarrow 0$ as $r \rightarrow \infty$, one obtains:
\begin{equation} \label{eq:equilibrium}
P(r) = \int_r^{\infty} \frac{G M(r') \rho(r')}{r'^2} \, dr'.
\end{equation}
Assuming an ideal gas,
\begin{equation}
P = \frac{\rho}{\mu m_\mathrm{H}} k T,
\end{equation}
\grifa{where}
\begin{equation}
\mu \equiv \frac{\bar{m}}{m_\mathrm{H}}
\end{equation}
\grifa{is the \emph{mean molecular weight}, defined as the average particle mass $\bar{m}$ in units of proton mass $m_\mathrm{H}$}\footnote{\grifa{The mean molecular weight depends on the ionization state of the atoms present in the gas, since the number of particles is a function of it. For instance, a cloud of neutral hydrogen has $\mu = 1$, but a cloud of ionized hydrogen has $\mu = 0.5$.}} \grifa{With this,} Equation \ref{eq:equilibrium} may be rewritten as
\begin{equation}
T(r) = \frac{\mu m_\mathrm{H}}{k\rho} \int_r^{\infty} \frac{G M(r') \rho(r')}{r'^2} \, dr',
\end{equation}
which is the equation that is used to calculate the temperature of each particle, given its radius $r$.

For a non-spherical case, a different procedure must be used. \citet{2005MNRAS.361..776S} uses an iterative method to set up the initial vertical density profile of their gaseous disk, which is assumed to be initially isothermal at a chosen temperature; in our simulations, we also initialize the gaseous disk as isothermal, but we set its density profile as equal to the one of the stellar disk, but with a smaller scale height (5 -- 10\% of that of the stellar disk). That is because we use radiative cooling, which makes the disk settle into a thin sheet in equilibrium within a very short timescale (a few tens of Myr), and also because we have only run simulations at resolutions of $\sim$100 -- 200 pc, which is not enough to resolve a detailed vertical structure within the gaseous disk.

\section{Sub-grid physics \grifa{and radiative cooling}} \label{sec:subgrid}

The typical resolution of a large scale astrophysical simulation nowadays, e.g. a cosmological simulation of galaxy formation or a galaxy cluster simulation in which the member galaxies are resolved, is in the range 100 pc -- 1 kpc. But there are processes taking place at scales much below this resolution which directly affect the evolution of the system as a whole. For instance, one such process is that of star formation. Star formation happens in cold clouds of molecular gas ($T \sim 10$ K), which fragment into stars due to their own self-gravity. The typical size of a star forming region is $\sim$1 pc, and the size of a star like the Sun is $\sim$10$^{-8}$ pc. Clearly those scales are much below the resolution of the \grifa{aforementioned} simulations, while star formation has a vital impact on the gas mass evolution of galaxies. This motivates the use of \emph{sub-grid} recipes. 

A sub-grid recipe is a recipe to take into account the result of a given small scale process while not actually resolving it \grifa{in space and in time}. For instance, in the case of star formation, a recipe inspired on the Schmidt law \citep{1959ApJ...129..243S} is often used, which converts the gas in a given volume element of the simulation into star particles at the following rate:
\begin{equation}
\dot{\rho_\star} = \epsilon \frac{\rho_{\mathrm{gas}}}{t_\mathrm{ff}},
\end{equation}
where 
\begin{equation}
t_\mathrm{ff} = \sqrt{\frac{3 \pi}{32 G \rho_{\mathrm{gas}}}}
\end{equation}
is the local free-fall time, and $\epsilon$ is the star formation efficiency. The recipe is usually only applied to cells which exceed a chosen density threshold, and star particles are formed by sampling from a Poisson distribution at each timestep. The free parameter $\epsilon$ must be calibrated according to the simulation resolution such that some desired observational constraint is met -- for instance, in the simulation of a Milky Way-like galaxy, one constraint could be that the global star formation of the simulated galaxy is similar to that of the real one. Typically used values of $\epsilon$ are in the range 1 -- 2\%.

Another important sub-grid recipe is that of feedback by Active Galactic Nuclei (AGN). The energy release \grifa{associated with the infall of mass into} supermassive black holes, which are believed to reside at the center of most galaxies, has been shown to be an important ingredient in models of galaxy formation -- without it, galaxies tend to end up with too large stellar masses in cosmological simulations of galaxy formation \citep{2017MNRAS.465.3291W}. The \grifac{Schwarzschild radius} of a typical supermassive black hole which resides at the center of an active galaxy, \grifa{and also of the accretion disk within which the bulk of the energy release takes place}, is $\sim$10$^{-3}$ pc, much below the resolution of such simulations ($\sim$100 pc -- 1 kpc). An example of recipe to take into account the feedback by AGN is to \grifac{assume spherical accretion, and} consider that at each timestep, an amount of gas equal to the amount given by the Bondi rate is removed from the surroundings of a black hole particle and added to its mass, and that a fraction $\epsilon$ of the rest energy of this accreted gas is converted into energy, which is deposited into the surrounding cells or particles of the simulation in the form of energy and/or momentum.

So far we have talked of stochastic recipes to take into account sub-grid phenomena. But there is also a case in which it is useful to do that by directly modifying the gas equation of state. On a scale intermediate between that of star formation and the simulation resolution, there are turbulent motions taking place in the dense gas which are not being resolved by the simulation. These motions have the effect of counterbalancing the effect of radiative cooling on small scales. This can be accounted for by assigning a polytropic equation of state to the gas that exceeds a density threshold $\rho_0$ (usually taken to be the same as the star formation threshold):
\begin{equation}
T = T_0 \left(\frac{\rho_\mathrm{gas}}{\rho_0}\right)^{\kappa-1}.
\end{equation}
In practice, this prevents an artificial runaway cooling of the gas (which would lead to an artificial surge in star formation), by introducing a temperature floor $T_0$, while also \grifac{enhancing the pressure exerted by the cold gas in a way supposed to mimic the aforementioned sub-grid motions}. Typical values of the two free parameters are $\kappa = 2$ \citep[e.g.][]{2015ApJ...812L..36B} and $T_0 = 10^4$ K  \citep[e.g.][]{2011MNRAS.414..195T}. 

\grifa{Other than sub-grid recipes, another important ingredient beyond ideal hydrodynamics in a fluid simulation in Astrophysics is radiative cooling. The rate of energy loss by radiative cooling depends on the density, temperature and also composition of the gas -- the presence of heavier elements (higher metallicity) greatly enhances the emission rate due to the efficient emission of energy through emission lines in the atomic orbitals of those elements. This can be accounted for by either assuming some constant metallicity throughout all the gas in the simulation, or by employing a chemical evolution recipe in which the metallicity is allowed to vary in space and time. In the latter case, metal deposit by supernovae events and stellar winds are usually employed. In \textsc{ramses}, both options are available; we have empirically found that introducing metals in a simulation makes it significantly slower, so in all our simulations we have chosen to use a constant metallicity of 1 Z$_\odot$ for simplicity.}

\section{Cosmological simulations} \label{sec:cosmologicalsimulations}

One of the most important classes of astrophysical simulations are cosmological simulations. In fact, the interest in these simulations has been such that they have motivated the development of many of the most widely used simulation codes today, such as \textsc{gadget} \citep{2005MNRAS.364.1105S}, \textsc{ramses} \citep{2002A&A...385..337T} and \textsc{arepo}\footnote{\grifab{Of those three examples, \textsc{arepo} is the only one which is currently not publicly available.}} \citep{2010MNRAS.401..791S}.

The goal of a cosmological simulation is to compute the evolution of a large volume of the universe, while self-consistently taking into account its nonlinear evolution on small scales. At high redshift \grifa{($z \gtrsim 100$)}, the universe is approximately homogeneous on all scales, and simplified approaches based on perturbation theories can be used to compute the time evolution of its density fluctuations; but as time passes, gravity makes matter clump on small scales. Once collapsed structures emerge, and orbits become relevant, the matter structure evolution cannot be reliably calculated with any perturbation theory currently available, so the use of cosmological simulations (which can be thought of as a brute force approach) becomes necessary.

All cosmological simulations are run into a simulation box with periodic boundary conditions, such as to emulate an universe which behaves as if it was infinite inside the limited memory of a computer. The initial conditions of the simulation consist of dark matter particles and optionally gas particles \grifa{-- in a grid simulation, no gas particles are present, so in this case those particles are deposited into the grid before the simulation is started}. All particles are initialized into a uniform grid with zero speed; then, for each particle, position and velocity displacements are made such as to ensure that the matter power spectrum and velocity distribution are consistent with those obtained from a perturbation theory (not necessarily a linear one). In the cosmological simulations we have run for the purpose of this thesis, we have used the code \textsc{music}\footnote{\url{https://bitbucket.org/ohahn/music}} \citep{2011MNRAS.415.2101H} for generating the initial conditions. We will return to \textsc{music} on Section \ref{sec:zoom}, to discuss how the initial conditions for a zoom simulation are generated.

\subsection{Halo finding}

\begin{figure} 
 \centering
 \includegraphics[width=\textwidth]{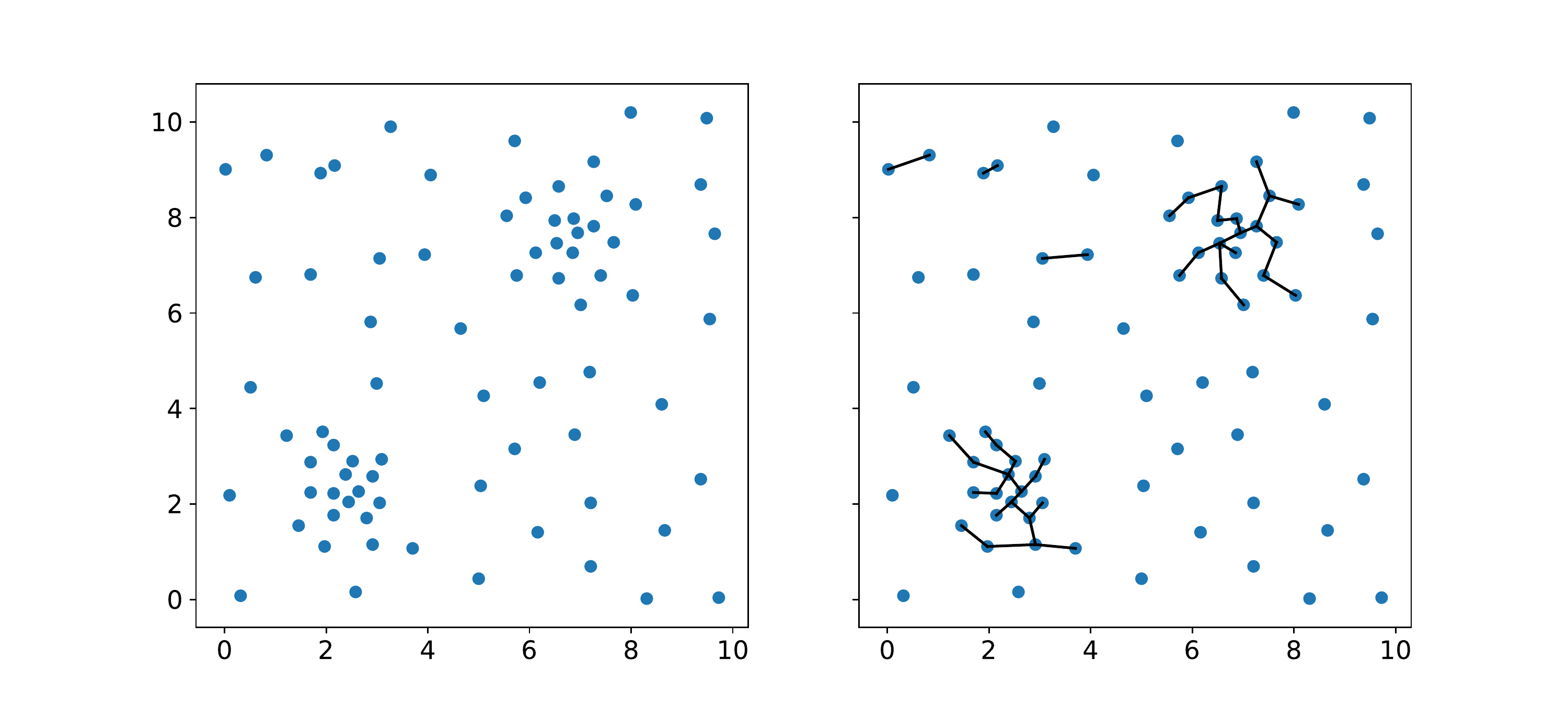}
 \caption{Illustration of the Friends of Friends (FoF) halo finding algorithm. Here the linking length is set to $l=1$, and the two clusters of particles in the image are detected, plus some spurious halos which should be discarded for having a low number of particles.}
 \label{fig:fof}
\end{figure}

In the analysis of the output of a cosmological simulation, one is often interested in characterizing the dark matter halos that are formed -- their internal structure, evolution over time, etc. The task of identifying halos from the particle distribution in a simulation is known as \emph{halo finding}, and several algorithms have been developed for that. Let's consider one of the simplest algorithms: the Friends Of Friends (FoF) algorithm. It starts with one single free parameter, which is called the \emph{linking length}, and which we will call $l$. For each particle in the simulation, one looks for all the particles which are at a distance smaller than $l$ from it, and considers that these two particles are connected. The final result is a graph like the one shown in Figure \ref{fig:fof}; the halos are then defined as isolated sets of connected particles. It can be noted that some halos made up of only two particles are found in this particular case; in a real simulation, it is commonplace to not consider halos containing less than a minimum amount of particles \grifab{(typically not less than $10^{3-4}$)} as statistically relevant, and to simply ignore them.

A more sophisticated halo finder, which we have employed in the cosmological simulations that we will describe later in this thesis, is the \textsc{rockstar} algorithm \citep{2013ApJ...762..109B}. Instead of considering only particle positions, \textsc{rockstar} finds halos by locating isolated sets of particles in 6D phase space (3 spacial coordinates plus 3 velocity coordinates). It works by first finding FoF halos within the simulation volume, so as to readily identify overdense regions. Then, for each of those regions, the particle positions are normalized according to the size of the group, and particle velocities according to the velocity dispersion. A linking length in 6D phase space is then adaptively found for the resulting coordinates, with the constraint that 70\% of particles end up linked together, generating as an output a list of subgroups. The process is recursively repeated for each subgroup, until no more subgroup is found. The final halos themselves are defined as being centered in the innermost subgroups found; each particle is categorized as belonging to the closest of those halos in phase-space, and after all particles have been categorized, particles unbound to their assigned halo are discarded before the halo properties are actually computed.

\subsection{Zoom methodology} \label{sec:zoom}

Astrophysicists are often interested in modeling the formation and evolution of particular systems, e.g. individual galaxies or galaxy clusters, in a cosmological context. An interesting methodology in this case is that of a \emph{zoom cosmological simulation}, defined as a cosmological simulation where more resolution elements are placed in a region of interest in the cosmological box than in the rest, so that most of the computation time is allocated to this region, and a greater spacial resolution can be achieved in it. This is usually done by placing particles with lower mass within the zoom region than outside it.

\begin{figure} 
 \centering
 \includegraphics[width=\textwidth]{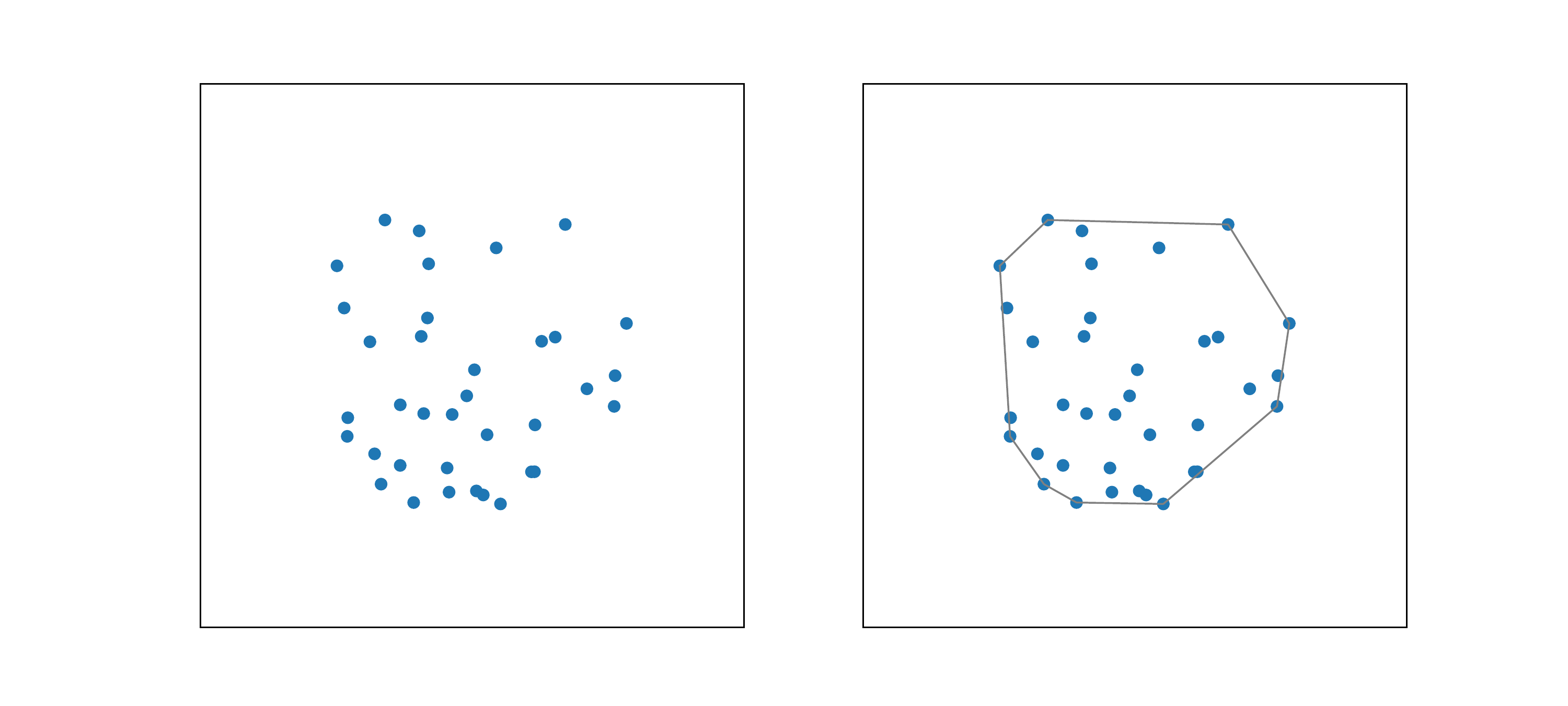}
 \caption{\grifa{An example of convex hull. On the left, a set of points, and on the right, the convex hull which encompasses those points. In 3D, the hull is built in a similar way, but convex polygons determine the segments of the hull surface instead of line segments.}}
 \label{fig:chull}
\end{figure}

A zoom simulation must be preceded by a unigrid cosmological simulation (meaning a simulation where all particles have the same mass), from which an appropriate object will be selected to be zoomed -- typically some dark matter halo. The particles within this object are identified, and their positions in the initial conditions of this preliminary simulation define the zoom region. The geometric shape itself of the region is usually defined as the \emph{convex hull} that encompasses all of these particles, that is, the smallest convex polyhedron which encompasses the particles. \grifa{An example of convex hull in 2D is shown in Figure \ref{fig:chull}.}

Once the convex hull has been defined, new initial conditions must be generated placing more resolution elements inside the hull than outside it, \grifac{so that the object which will be formed inside this hull will be resolved at higher resolution}. This means that higher frequency modes of the matter power spectrum must be added to the zoom region, \grifac{since its spacial resolution is higher}; care must be taken to preserve the lower frequency modes both in the zoom region and in the rest of the box, such that the zoom simulation is statistically equivalent to the uniform resolution one -- this is done by keeping the same random number seeds in the random number generator for the lower levels as those used in the unigrid simulation.

It can happen that, when the zoom simulation is executed, high mass particles from outside the zoom region end up falling into it. Since this can lead to numerical issues with refinement and to artificial gravitational effects, sometimes a procedure of \emph{decontamination} of the zoom region is employed, in which the volume of the convex hull is expanded to include \grifac{the contaminating particles identified at the end of the first zoom simulation}, and the simulation is run again with updated initial conditions. The procedure is then repeated until no more high mass particles leak into the zoom region.

\chapter{Interstellar medium evolution in the cluster environment} \label{c:interstellar}

This chapter includes results first presented in \hyperlink{Paper I}{Paper I}, so it is useful to define now what this paper consists of. In it, we have attempted to constrain the possible final states of Milky Way-like disk galaxy after it crosses a galaxy cluster. For that, we have run simulations in which an idealized galaxy, modeled after our own, falls radially into four different idealized, spherical galaxy clusters. The mass of each cluster was either $10^{14}$ M$_\odot$ or $10^{15}$ M$_\odot$, and it either had a cool-core or not. The properties of those clusters are shown in Table \ref{tab:clusters}. The galaxy started its infall from the $R_{200}$ of each cluster, with an initial speed equal to either $0.5\sigma$, $\sigma$ or $2 \sigma$, where $\sigma$ is the velocity dispersion of the cluster; and with a disk orientation relative to its velocity vector either face-on, edge-on or halfway (45$^\circ$). So on top of there being 4 different clusters, there were 9 different entry conditions for each cluster, the net result being that we have systematically probed 36 different scenarios. The simulations included sub-grid recipes for star formation and feedback, allowing us to model the gas mass and stellar mass evolution of the galaxy in those various scenarios. A sample of those simulations can be visualized in Figure \ref{fig:paper1:panels}.

\begin{table}
 \centering
 \captionsetup{position=above}
 \caption{The properties of the clusters used in Paper I.}
 \label{tab:clusters}
 \begin{tabular}{cccccc}
Mass (M$_\odot$) & cool-core & $R_{200}$ (kpc) & $\sigma$ (km/s) \\
\hline
10$^{14}$ & yes & 805.1 & 352.5 \\
10$^{14}$ & no & 796.5 & 344.8 \\
10$^{15}$ & yes & 1679.0 & 713.2 \\
10$^{15}$ & no & 1657.9 & 697.0 \\
 \end{tabular}
\end{table}

\section{Gas loss and consumption} \label{sec:gasloss}

The interstellar gas in disk galaxies that fall into clusters is subject to the ram pressure exerted by the ICM. If the ram pressure is greater than the gravitational restoring force per unit area of the disk, this gas can be removed from the galaxy, in a process called \emph{ram pressure stripping} \citep{1972ApJ...176....1G}. This condition can be expressed as:
\begin{equation} \label{eq:gunngott}
\rho_\mathrm{ICM} v^2 > 2 \pi G \,\Sigma_\mathrm{gas} \Sigma_\mathrm{stars},
\end{equation}
where the left side is the ram pressure and the right side can be shown to be the aforementioned restoring force in the case of a disk galaxy falling face-on into a cluster. \grifa{Here $\rho_\mathrm{ICM}$ is the ICM gas density, $v$ is the ICM speed relative to the galaxy, and $\Sigma_\mathrm{gas}$ and $\Sigma_\mathrm{stars}$ are the surface densities of the interstellar gas and of the stars within the disk of the galaxy, respectively.}

Surface density typically falls exponentially with cylindrical radius in a disk galaxy \citep[i.e. $\Sigma \propto e^{-R}$,][]{1959HDP....53..311D,1970ApJ...160..811F}. For this reason, in a ram pressure stripping scenario, the outer part of the ISM of these galaxies tends to be stripped first when they fall into clusters, giving rise to galaxies with truncated H$_{\mathrm{I}}$ disks. Indeed, such truncated galaxies have been observed e.g. in Virgo \citep{2004ApJ...613..866K,2008AJ....136.1623C} and Coma \citep{2015AJ....150...59K}.

\begin{figure} 
 \centering
 \includegraphics[width=1.0\textwidth]{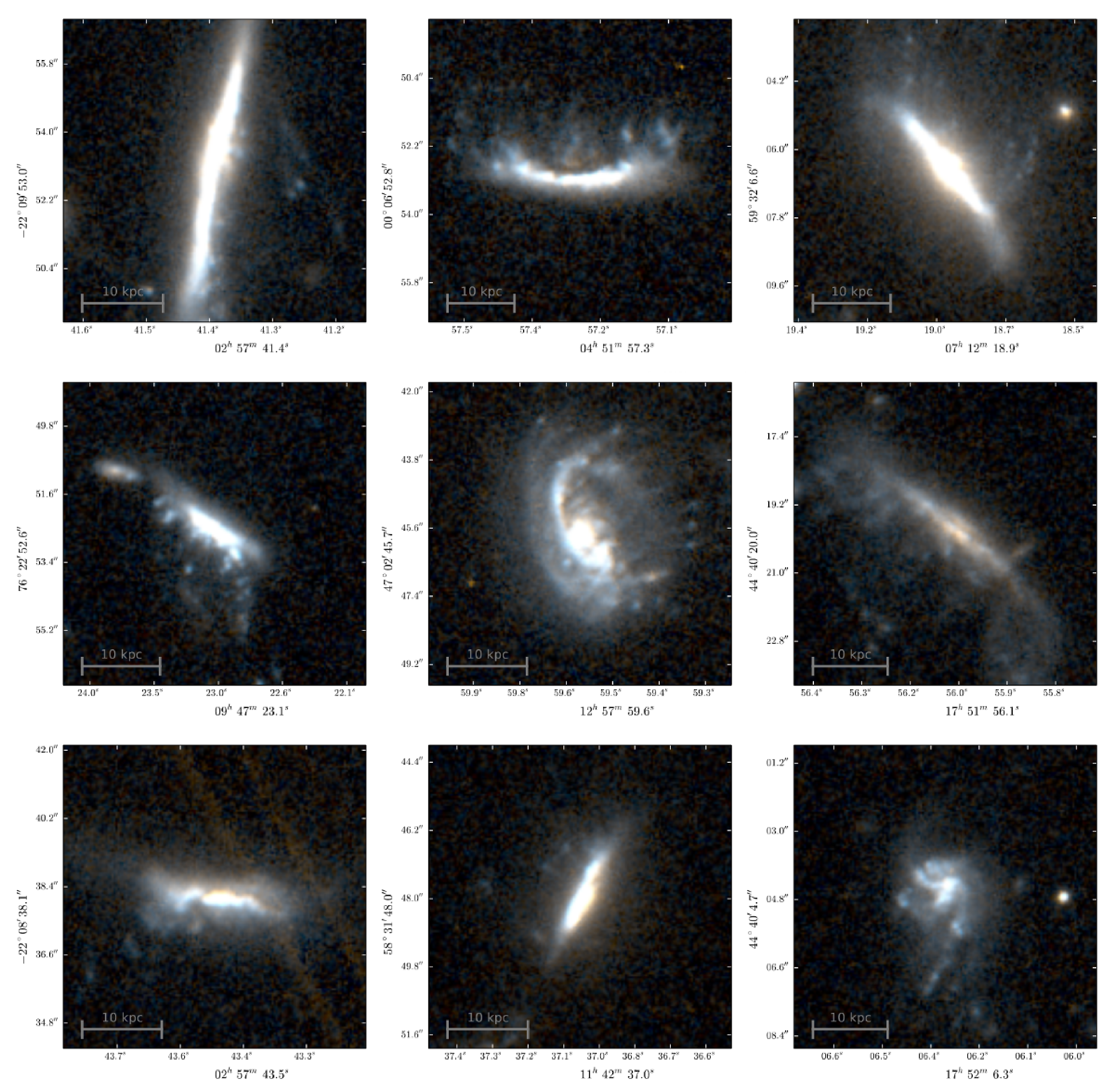}
 \caption{Examples of jellyfish galaxies. Reproduced from \citet{2016MNRAS.455.2994M}.}
 \label{fig:jellyfish}
\end{figure}

Sometimes galaxies undergoing ram pressure stripping are observed with so called \emph{jellyfish} morphologies, in which the galaxy features a tail with filaments and knots, corresponding to regions in the stripped gas where star formation is taking place. Such morphologies are found in galaxies undergoing a strong ram pressure stripping event, where a significant portion of its ISM mass is being ejected in a relatively short timescale -- otherwise the galaxy would feature an asymmetric and disturbed morphology without a prominent tail. Some examples of jellyfish galaxies, reproduced from \citet{2016MNRAS.455.2994M}, are shown in Figure \ref{fig:jellyfish}; more examples have been reported in \citet{ebeling14} and \citet{Poggianti16}.

The process of gas loss in a galaxy falling into a cluster is not due exclusively to ram pressure stripping. For instance, gas is also consumed by star formation, which can in some cases be enhanced due to the ISM compression provided by the ram pressure. We will talk more about this in Section \ref{sec:starformation}. The rate at which star formation consumes ISM gas can also be indirectly affected by the ram pressure in a process called \emph{starvation} \citep{1980ApJ...237..692L,2000ApJ...540..113B}, in which the gaseous halo of a galaxy is removed \grifa{by ram pressure stripping. The gaseous halo acts as a source of gas for the disk of the galaxy -- as the halo cools by radiative emission, gas is accreted into the disk, replenishing it with new gas and counterbalancing the gas loss by star formation. Once the halo has been lost, this gas replenishment ceases to take place, leading} to a higher rate of ISM gas depletion by star formation. Simulations have shown \citep{2016A&A...591A..51S} that the gaseous halo of an infalling galaxy is lost in $\sim$100 Myr, which is a relatively short timespan -- for comparison, the galaxy will take 1 -- 2 Gyr to cross a typical cluster. This means that, in practice, as soon as the galaxy enters \grifa{the inner region of} a cluster \grifa{(which we take as its $R_{200c}$)}, starvation is already taking place.

\begin{figure} 
 \centering
 \includegraphics[width=\textwidth]{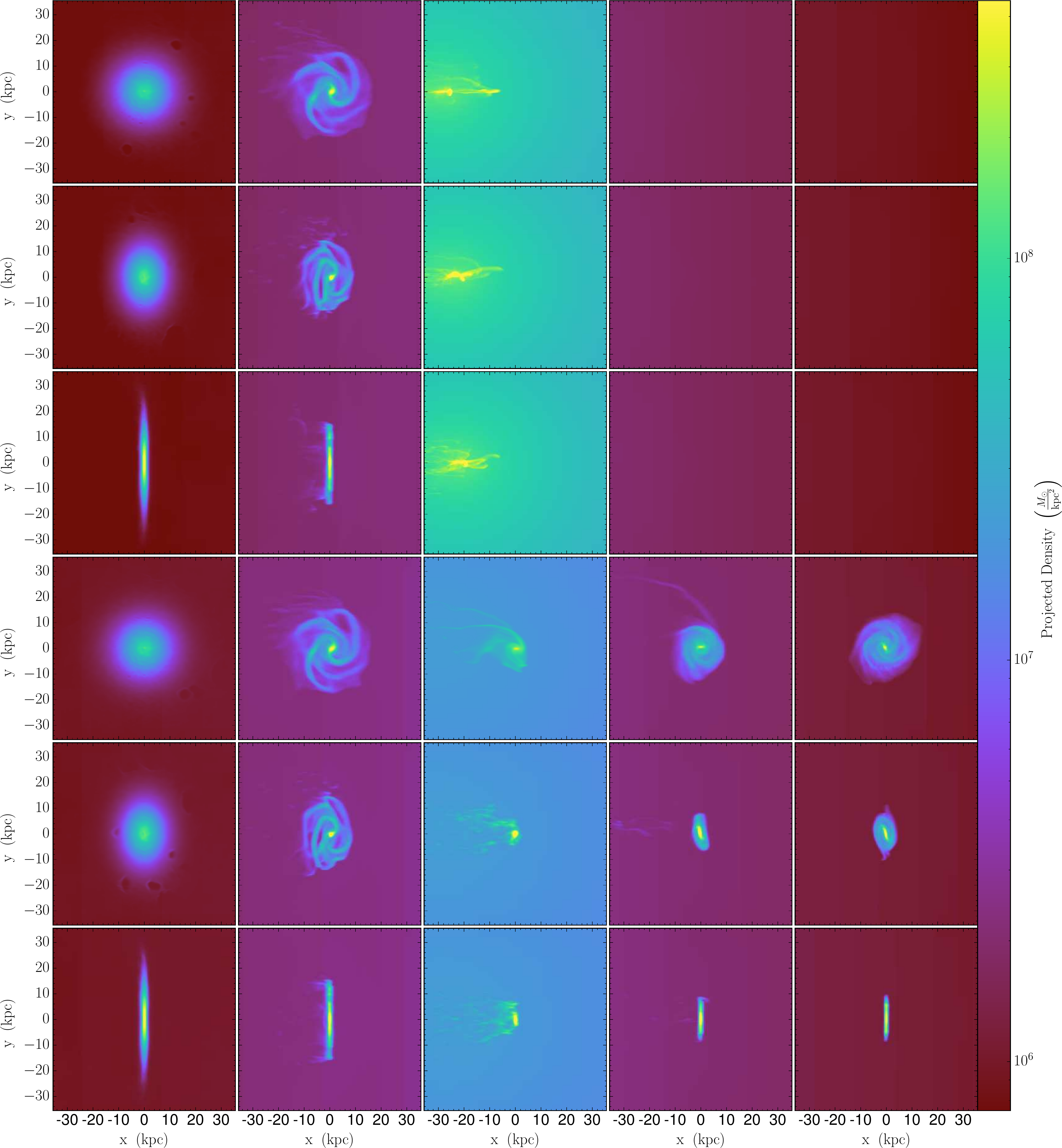}
 \caption{Panel showing some of the simulations of Paper I. Each row shows a zoom into the galaxy as it crosses a cluster moving from left to right. In the first 3 rows, the galaxy is crossing a cluster with a cool-core starting from 3 different disk orientations, and in the last 3 rows it is the same but for a cluster without a cool-core. In all cases, the cluster mass is  $10^{14}$ M$_\odot$ and the initial speed of the galaxy is equal to the velocity dispersion of the corresponding cluster. \grifac{In the first three rows, the gas is completely lost because of the high density in the central region of the cool-core cluster. We find that this result is valid no matter the entry conditions of the galaxy and the cluster mass.}}
 \label{fig:paper1:panels}
\end{figure}

\begin{figure} 
 \centering
 \includegraphics[width=0.6\textwidth]{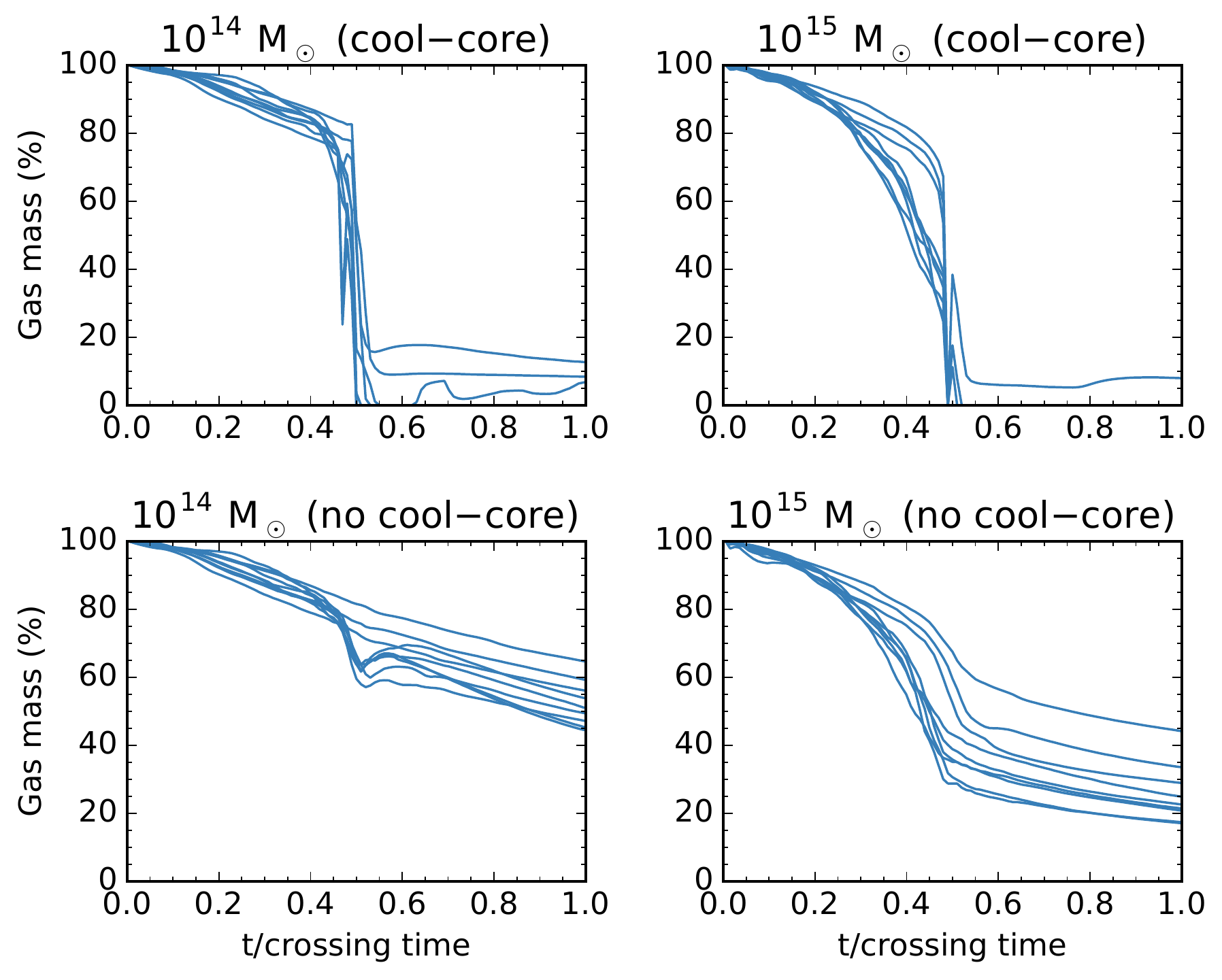}
 \caption{Gas mass still in the disk of the galaxy as it crosses the clusters. The different lines in each subplot represent different entry speeds and disk orientations. Reproduced from Paper I.}
 \label{fig:paper1:gasmass}
\end{figure}

\begin{figure} 
 \centering
 \includegraphics[width=0.6\textwidth]{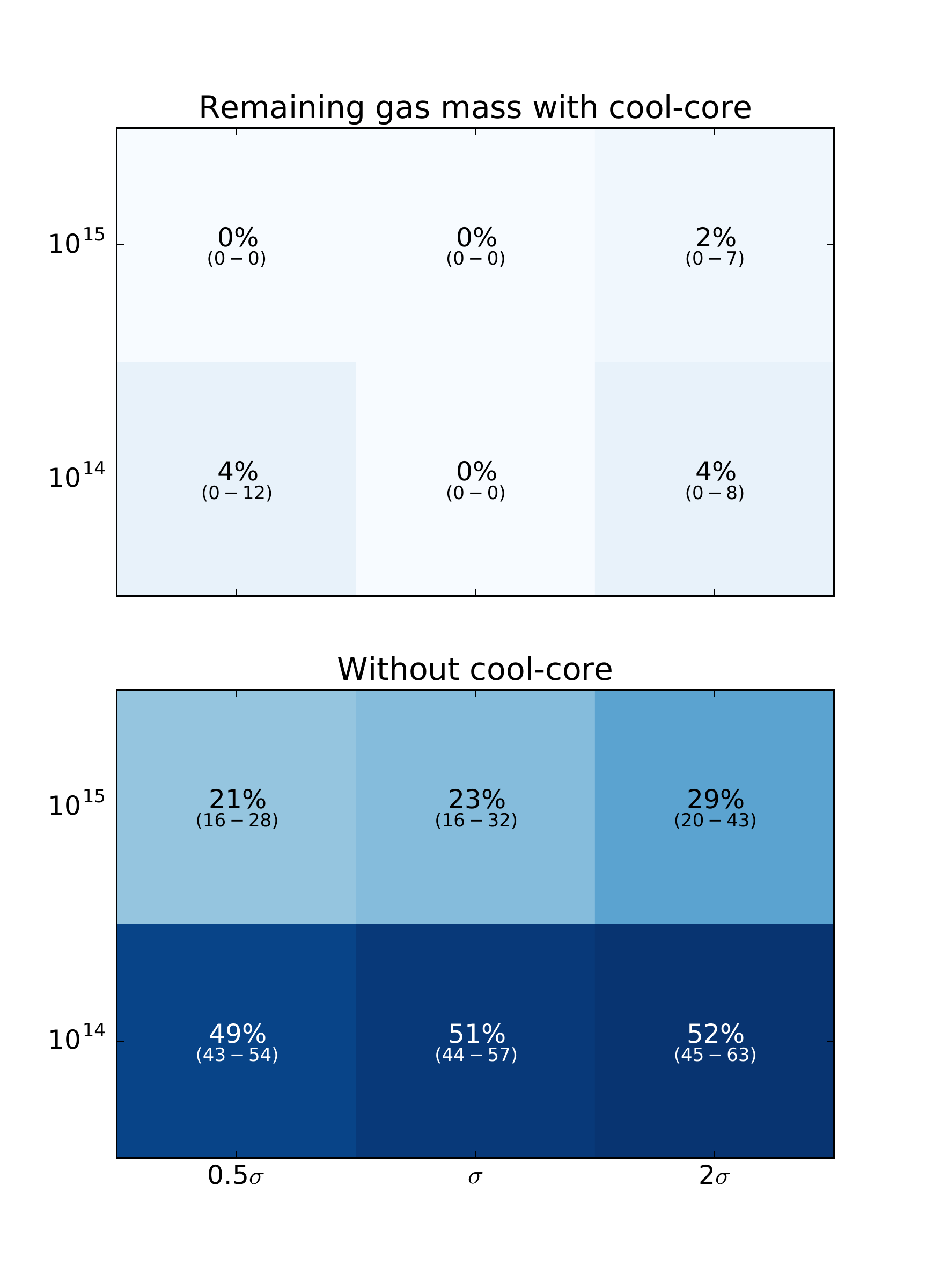}
 \caption{Gas mass still attached to the disk of the galaxy after it has crossed the four different clusters, in terms of its original gas mass, as a function of cluster mass in M$_\odot$ (left) and entry speed (bottom). The numbers in parenthesis are the minimum and maximum values among the 3 disk orientation angles at entry. Reproduced from Paper I.}
 \label{fig:paper1:gas}
\end{figure}

Now we turn to considering the gas mass evolution of the Milky Way-like galaxy falling into clusters in the simulations of \hyperlink{Paper I}{Paper I}. \grifa{Snapshots for some of the simulations are shown in Figure \ref{fig:paper1:panels}, for the sake of illustration. Each row shows a different simulation, and the columns show the galaxy at 0, 0.25, 0.5, 0.75 and 1.0 of its crossing time in that case. As expected, in cases where the galaxy is not completely stripped, its disk ends up truncated. The gas mass evolution of the galaxy in each of the four cluster models considered} is shown in Figure \ref{fig:paper1:gasmass}, with different lines in each subplot representing different entry speeds and disk orientations. The first major pattern that can be noted is that when the galaxy crosses a cluster with a cool-core, it is completely stripped as soon as it reaches the center of the cluster, regardless of its entry conditions \grifa{(as exemplified in the first three rows of Figure \ref{fig:paper1:panels})}. This is a direct consequence of the high density in a cool-core, and of the fact that the ram pressure is proportional to the medium density (see Equation \ref{eq:gunngott}). The second major pattern is that, after crossing a rich cluster with $10^{15}$ M$_\odot$, the galaxy loses gas at a faster rate than when it crosses a less rich cluster with $10^{14}$ M$_\odot$. This is expected because\grifab{, by construction, the galaxy enters the larger cluster at larger speeds than it enters the smaller cluster (the initial speeds are proportional to the cluster velocity dispersion), while the ram pressure strongly depends on the speed, more so than on the medium density (Equation \ref{eq:gunngott}); and also because} although the larger cluster contains a factor of 10 more gas mass, its size is a factor of unity larger (both have a radius within the 1 Mpc order of magnitude), so that the mean density grows with cluster mass. An alternative way to view the results in Figure \ref{fig:paper1:gasmass} is shown in Figure \ref{fig:paper1:gas}, where the ISM gas still inside the galaxy after one crossing is shown for each of the clusters and for each entry speed. 

\grifac{A detail of the gas mass evolution shown in Figure \ref{fig:paper1:gasmass} is that, after the central passage, in some cases the galaxy regains part of the gas mass which had been lost. This happens because in those cases, large clouds of dense material are stripped from the disk at the central passage but remain gravitationally bound to the halo; once the galaxy moves away from the cluster center and the ram pressure decreases, those clouds manage to be accreted back into the disk.}

\begin{figure} 
 \centering
 \includegraphics[width=0.6\textwidth]{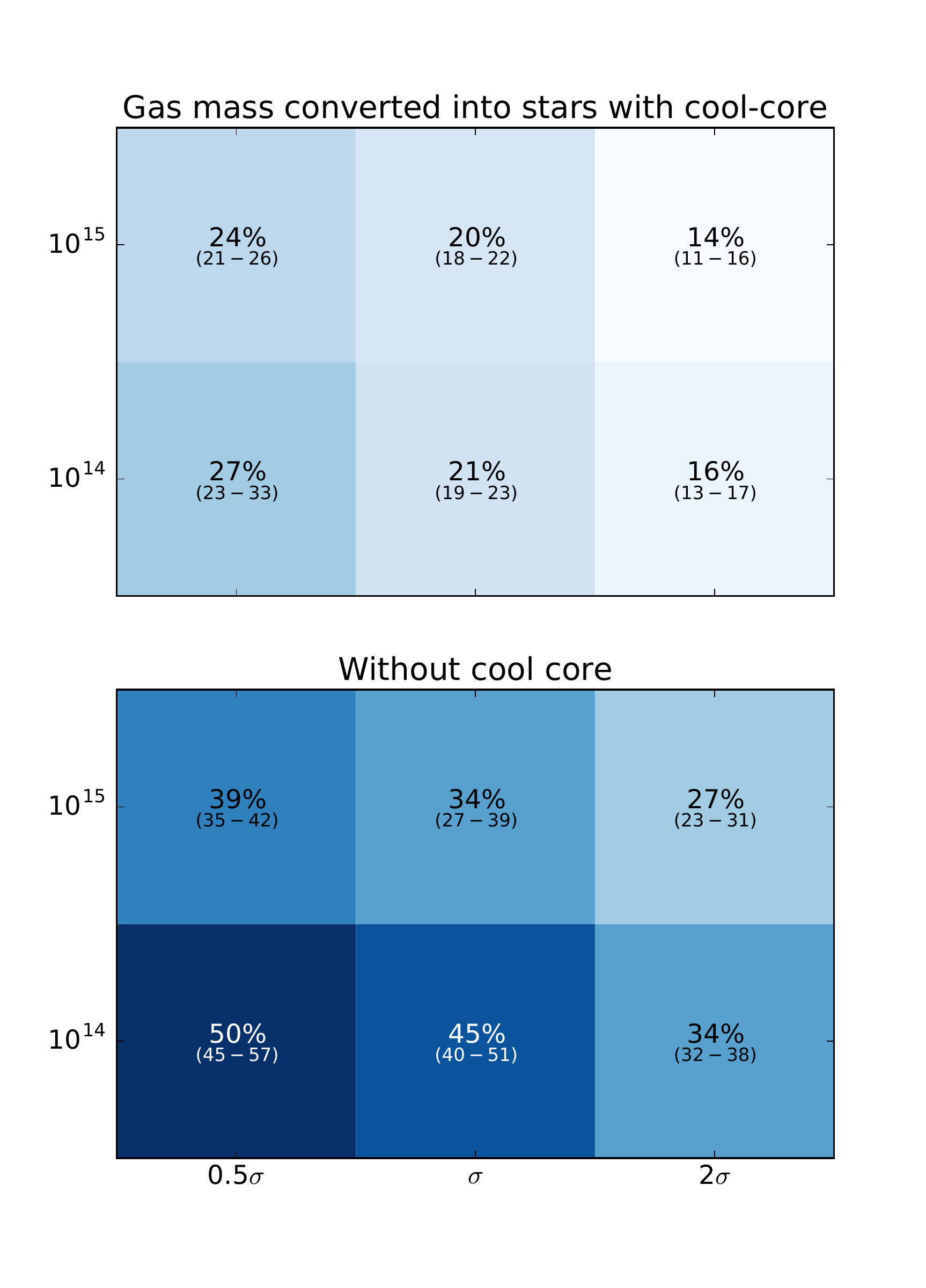}
 \caption{Same as Figure \ref{fig:paper1:gas}, but for the gas mass that has been converted into stars after the galaxy has crossed the four different clusters. Reproduced from Paper I.}
 \label{fig:paper1:stars}
\end{figure}

Note that the gas loss data we have shown also considers the gas mass lost to star formation. The net result of star formation over one crossing can be seen in Figure \ref{fig:paper1:stars}, which shows the fraction of the original gas mass in the disk of the galaxy which has been converted into stars. Overall, the greater the ram pressure encountered by the galaxy, the smaller the net amount of gas mass converted into stars after one crossing: less stars are formed after crossing a cool-core cluster than a cluster without a cool-core; after crossing a more massive cluster than a less massive one; and after entering the cluster with a greater speed rather than a smaller one, all three being manifestations of greater ram pressure intensities being faced by the galaxy, which translate into a greater rate of gas loss, and consequently into a smaller conversion of ISM gas into stars.  

\section{Star formation rate alterations} \label{sec:starformation}

Let's now consider more explicitly the changes in star formation rate and their consequences. Alterations in star formation rate are naturally expected in a ram pressure stripping scenario because if less gas is present in a galaxy, then less molecular clouds will be available for star formation, hence the global star formation rate would tend to decrease. But it turns out that the gas that remains bound to the galaxy is also subject to the ram pressure -- as it is not stripped, the effect of the ram pressure is to compress this gas. As star formation rate correlates with gas density \citep{1959ApJ...129..243S}, this compression (when present) translates into an increase in \emph{specific star formation rate} (i.e. star formation rate per unit area) in the non-stripped region of the disk, and possibly into a global increase in star formation rate.

\begin{figure} 
 \centering
 \includegraphics[width=0.6\textwidth]{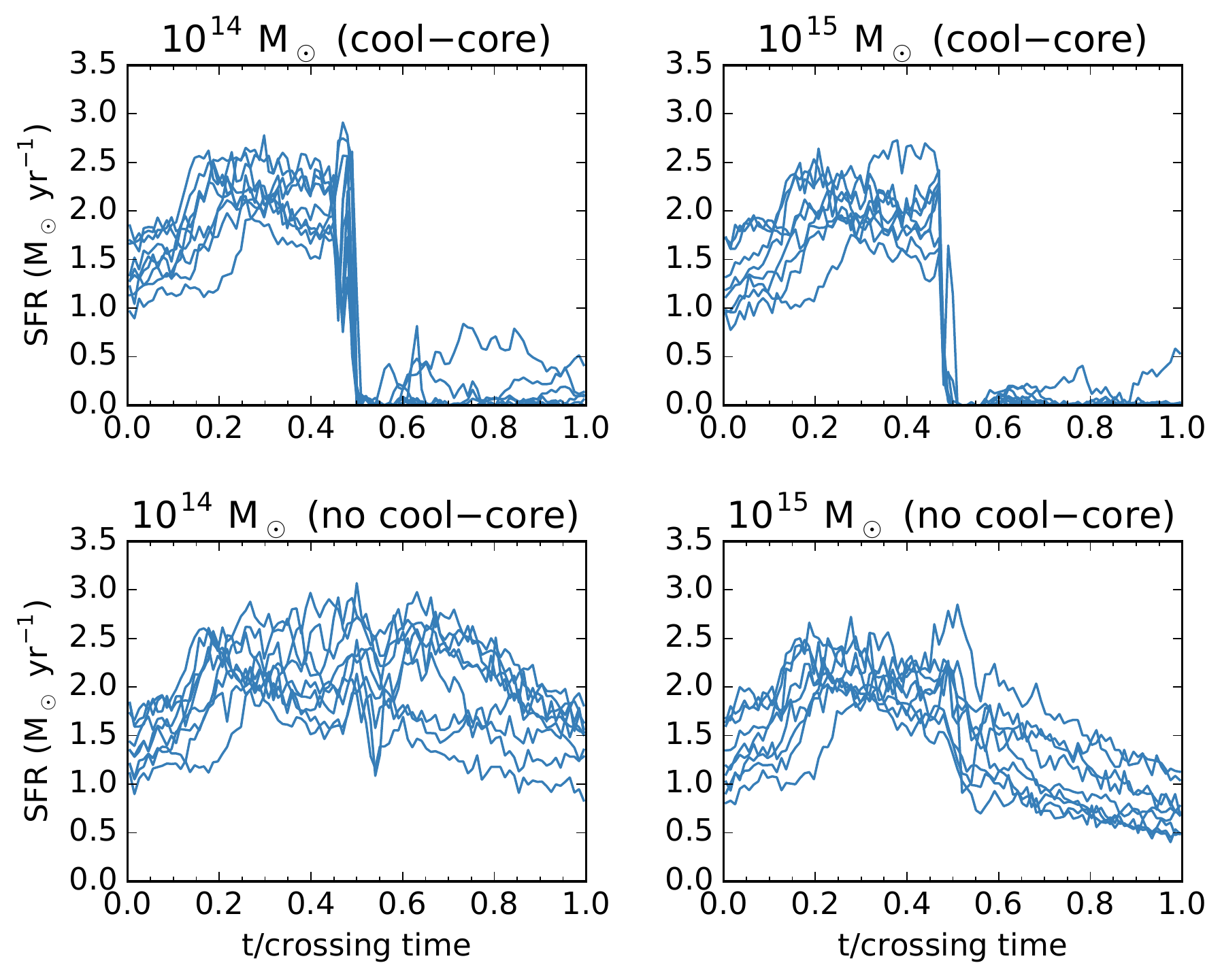}
 \caption{Star formation rate alterations as the galaxy falls into the four clusters. It can be noted that, initially, the effect of the ram pressure is always to increase the SFR of the galaxy, which is due to a compression of its ISM. Reproduced from Paper I.}
 \label{fig:paper1:sfr}
\end{figure}

This process is more straightforwardly imagined in the case of a galaxy falling face-on into a cluster, i.e. with the plane of its disk perpendicular to its direction of motion. One could wonder if in cases in which the disk is inclined an increase in specific star formation rate is still expected, since the hydrodynamics in this case is far less trivial. We can answer this question using Figure \ref{fig:paper1:sfr}, which shows the star formation rate of our simulated galaxy as a function of time as it crosses the clusters. Each subplot contains different lines for each entry speed and disk orientation, and it can be noted that in all cases there is an increment in SFR, meaning that there is a global effect of gas compression even in non-trivial cases in which the galaxy is inclined. This result is in agreement with previous models available in the literature \citep[][]{1999ApJ...516..619F,2008A&A...481..337K}. \grifa{Our results add new infall scenarios to the ones previously available, which for instance have not considered the role of inclination angle on the star formation rate change at entry; also our numerical scheme (AMR simulation) is different from the ones previously used, which were an analytical model in \citet{1999ApJ...516..619F} and a SPH simulation in \citet{2008A&A...481..337K}, adding robustness to the finding that a star formation rate increase takes place at entry}.

\subsection{Effect on luminosity} \label{sec:luminosity}

When star formation takes place in a molecular cloud, stars with different masses are formed \grifa{almost} simultaneously. The mass distribution of those newly formed stars is known as the \emph{Initial Mass Function}, or \emph{IMF}. Different IMFs have been proposed in the literature \grifa{\citep[e.g.][]{1955ApJ...121..161S,2001MNRAS.322..231K,2003PASP..115..763C}}, but one unifying characteristic of all of them is that the more massive a star is, the less likely it is to be formed. From models of stellar evolution, it is also well known that massive stars are bluer, more luminous and live less than less massive ones.

The implication of this in a ram pressure stripping scenario is that, if the ram pressure leads to an overall increase in the star formation rate of a galaxy, this will be associated with an increase in luminosity and also a change in color towards blue. Indeed, in some more extreme cases, galaxies have been observed to temporarily become the most luminous in the entire cluster through this process, even more luminous than the cluster's BCG \citep{2014ApJ...781L..40E}.

We have quantified this effect in the second paper published during this thesis (\hyperlink{Paper II}{Paper II}), using an idealized simulation of realistic disk galaxies falling into a Virgo-like cluster, which will be described in more detail in Section \ref{sec:paper2}; the properties of those galaxies are shown in Table \ref{tab:galaxies}. Before getting into the results regarding luminosity changes, let's first describe how to calculate the luminosity of a simulated galaxy. The galaxy contains star particles, which represent ensembles of stars (their masses are typically in the range $10^{3-5}$ M$_\odot$ nowadays); the luminosity of those star particles depends on their ages, masses and metallicities. Star particles which were formed in the simulation time by a star formation recipe have an intrinsic age, but stars which were already present in the initial conditions of the simulation do not. For this reason, some hypothesis must be made about the past star formation history of an idealized simulated galaxy before its luminosity can be calculated. Once this has been defined, ages are randomly sampled for the initial stellar particles based on the chosen star formation history, and the luminosity of each individual particle is calculated using a stellar evolution model such as \citet{2017ApJ...835...77M}\footnote{A web interface is available for generating magnitude tables using this model, at \url{http://stev.oapd.inaf.it/cgi-bin/cmd}}. Finally, the luminosity of the galaxy as a whole is simply calculated as the sum of the luminosities of its constituent star particles.

\begin{figure} 
 \centering
 \includegraphics[width=0.6\textwidth]{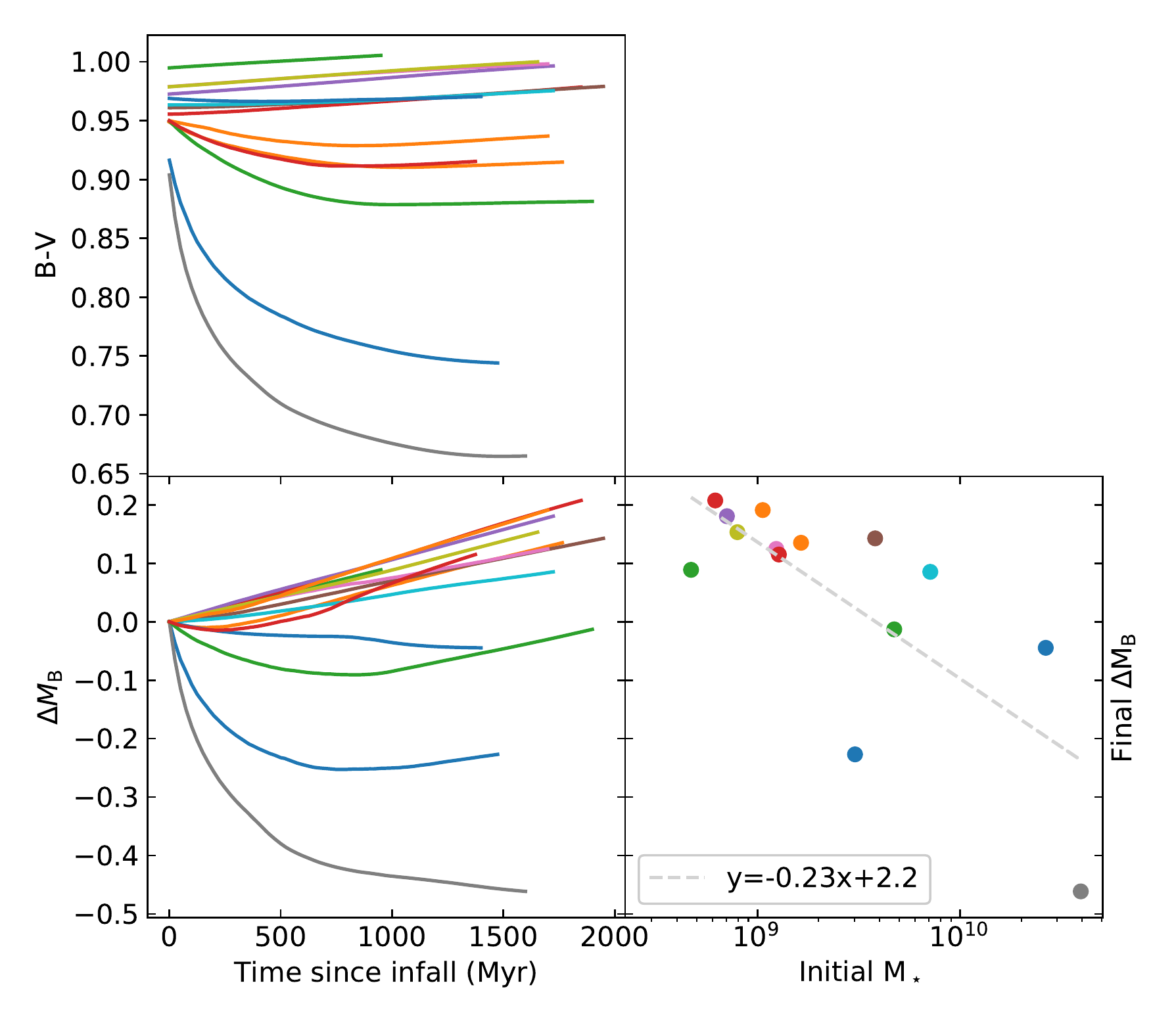}
 \caption{Luminosity and color evolution of simulated galaxies falling into a Virgo-like cluster, over one crossing time. \grifa{Each color represents a different galaxy. Here $\Delta M_\mathrm{B}$ is the difference between the B band luminosity of the galaxy at a given time and its B band luminosity in the initial conditions. The right plot shows this quantity after one crossing as a function of stellar mass, along with a linear fit to the resulting points, which shows a trend for the luminosity gain being larger for larger stellar masses.} Reproduced from Paper II.}
 \label{fig:paper2:luminosities}
\end{figure}

An analysis of the luminosity evolution of the simulated galaxies presented in \hyperlink{Paper II}{Paper II} is shown in Figure \ref{fig:paper2:luminosities}, \grifa{in which the B band luminosity and the B$-$V color of those\ galaxies are shown over time.} The less massive galaxies ($\lesssim 10^9$ M$_\odot$) lose nearly all their gas just after entering the cluster, and after that proceed to get monotonically less luminous and more red over time. \grifab{This complete removal of gas is consistent with semi-analytic models of galaxy formation and evolution, which show that $\sim$70\% of satellite galaxies within the virial radius of a cluster have been completely stripped \citep{2010MNRAS.408.2008T}.} More massive galaxies, on the other hand, tend to gain luminosity and get bluer after entering the cluster, but this depends on their orbit -- a massive galaxy falling into a radial orbit will lose gas faster and gain less luminosity. In the B band, the largest gain in luminosity we found was of 0.5 magnitude. For comparison, \citet{2017A&A...605A.127P} has calculated a gain of up to 0.9 magnitude in the B band based on the SFR data presented for the infalling galaxies in \hyperlink{Paper I}{Paper I}.

\begin{table}
 \centering
  \captionsetup{position=above}
  \caption{The infalling galaxies used in the simulations presented in Paper II. The first four columns show data extracted from a cosmological simulation, and the last two were computed having as an input the $M_{200c}$. We will describe the procedure used for that in more detail in Section \ref{sec:paper2}. Here $v_0$, $t_0$ and $b_0$ are the velocity, simulation time and impact parameter at entry respectively.}
 \label{tab:galaxies}
 \resizebox{0.6\columnwidth}{!}{
 \begin{tabular}{llllll}
  \hline
  $M_{200c}$ (M$_\odot$) & $v_0$ (km/s) & $t_0$ (Myr) & $b_0$ (kpc) &  $M_\star$ (M$_\odot$) & $R_d$ (kpc) \\
  \hline
1.8 $\times 10^{12}$ & 1026 & 0 & 507 & 3.9 $\times 10^{10}$ & 3.48\\
1.1 $\times 10^{12}$ & 980 & 2330 & 248 & 2.6 $\times 10^{10}$ & 2.92\\
3.8 $\times 10^{11}$ & 917 & 1920 & 424 & 7.1 $\times 10^{9}$ & 1.65\\
3.1 $\times 10^{11}$ & 524 & 490 & 351 & 4.7 $\times 10^{9}$ & 1.38\\
2.9 $\times 10^{11}$ & 928 & 1760 & 720 & 3.8 $\times 10^{9}$ & 1.25\\
2.6 $\times 10^{11}$ & 1102 & 90 & 483 & 3.0 $\times 10^{9}$ & 1.13\\
2.0 $\times 10^{11}$ & 991 & 430 & 764 & 1.6 $\times 10^{9}$ & 0.86\\
1.7 $\times 10^{11}$ & 979 & 490 & 114 & 1.3 $\times 10^{9}$ & 0.77\\
1.7 $\times 10^{11}$ & 1032 & 2990 & 721 & 1.2 $\times 10^{9}$ & 0.76\\
1.6 $\times 10^{11}$ & 935 & 490 & 600 & 1.1 $\times 10^{9}$ & 0.71\\
1.4 $\times 10^{11}$ & 1066 & 2990 & 639 & 7.9 $\times 10^{8}$ & 0.63\\
1.3 $\times 10^{11}$ & 948 & 2540 & 678 & 7.1 $\times 10^{8}$ & 0.60\\
1.2 $\times 10^{11}$ & 993 & 1450 & 738 & 6.2 $\times 10^{8}$ & 0.56\\
1.1 $\times 10^{11}$ & 997 & 4070 & 464 & 4.7 $\times 10^{8}$ & 0.50\\
\hline
 \end{tabular}}
\end{table}

\section{Contamination with ICM gas} \label{sec:contamination}

\begin{figure} 
 \centering
 \includegraphics[width=0.7\textwidth]{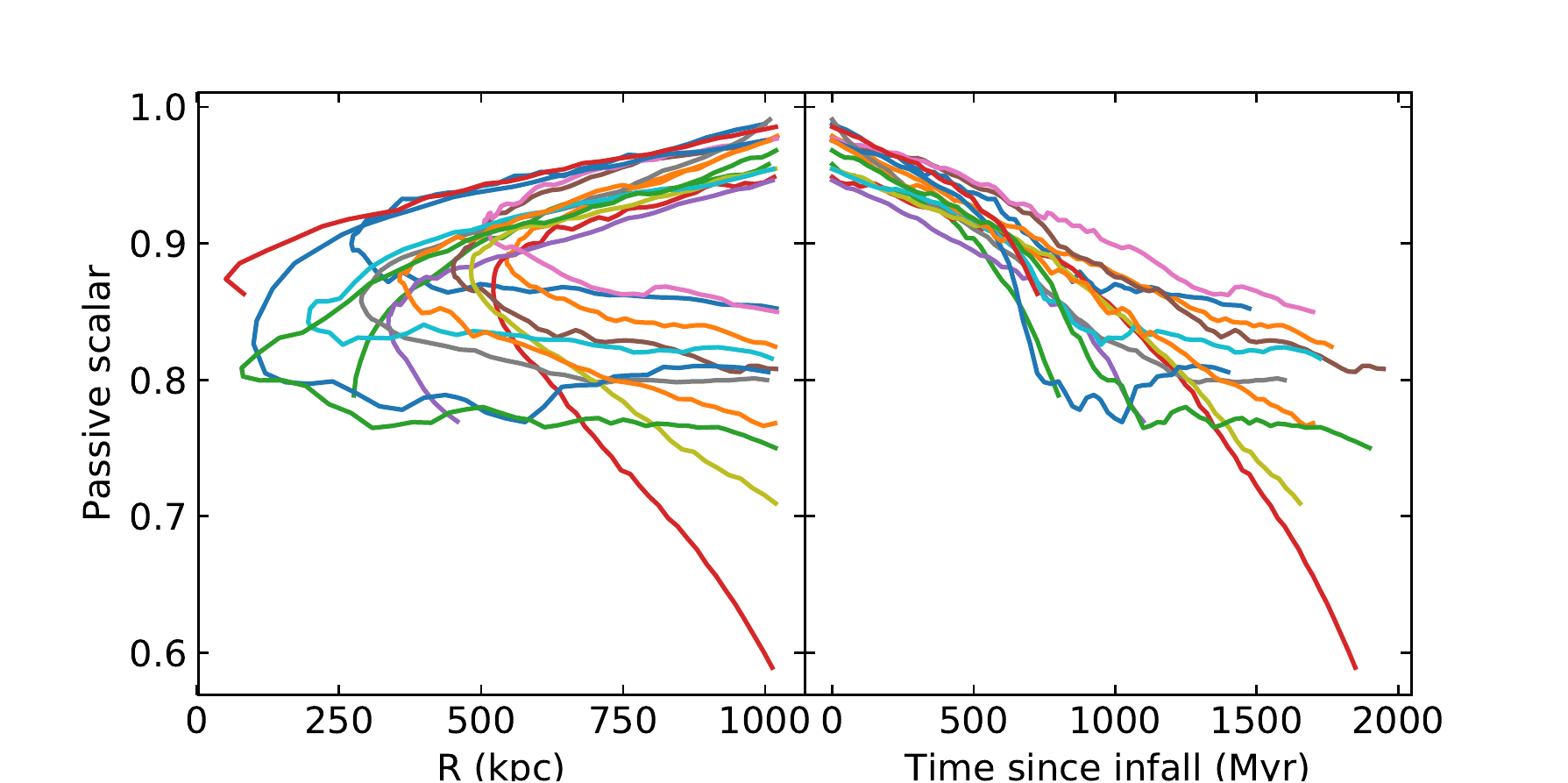}
 \caption{Contamination with ICM gas as simulated infalling galaxies cross a galaxy cluster. \grifa{Each color represents a different galaxy.} Typically $\sim 20$\% of the gas still in the disk of a spiral galaxy after one crossing comes from the ICM. Reproduced from Paper II.}
 \label{fig:paper2:contamination}
\end{figure}

\begin{figure} 
 \centering
 \includegraphics[width=0.7\textwidth]{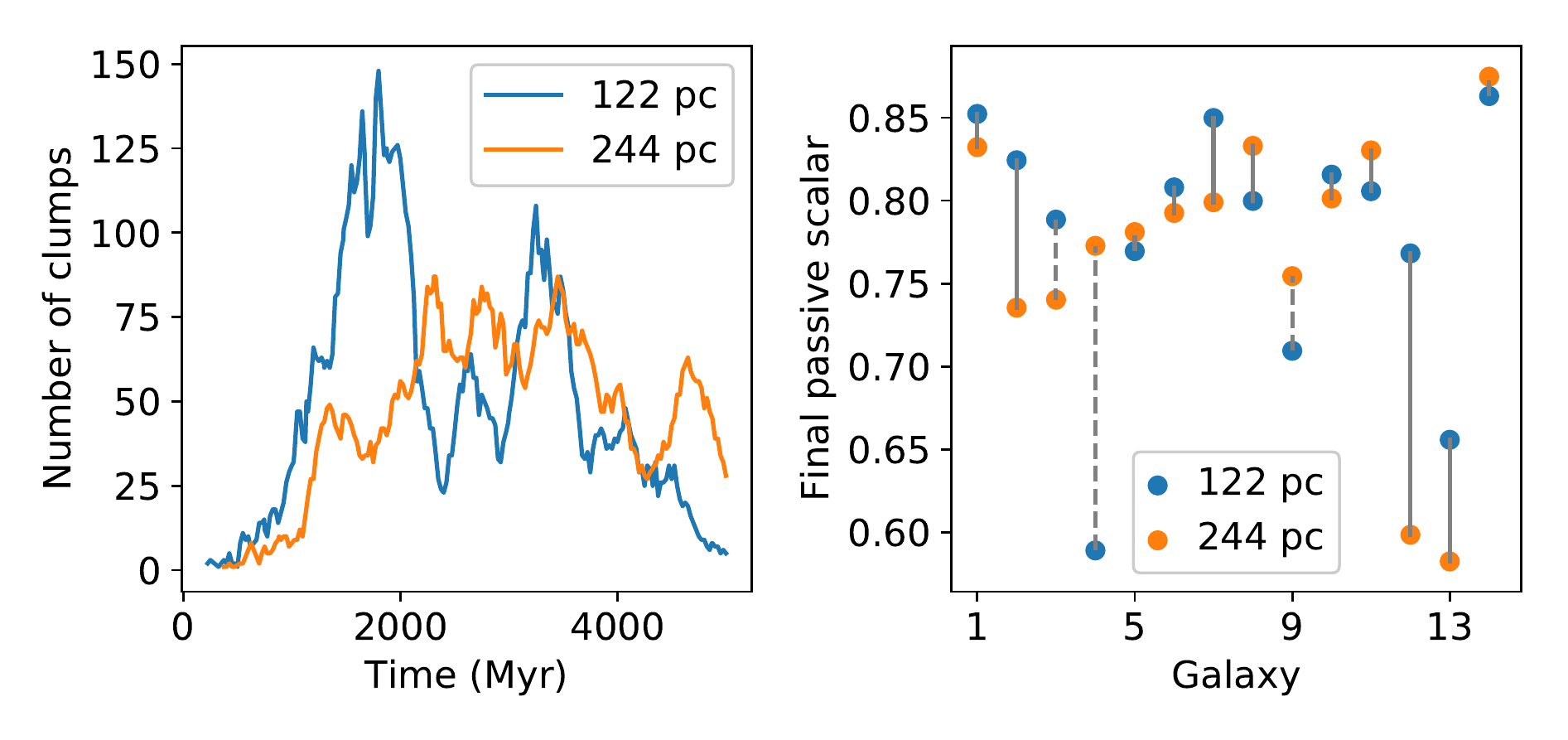}
 \caption{Convergence test showing the robustness of the results shown in Figure \ref{fig:paper2:contamination}. The result that 20\% of the gas in the disk of a galaxy comes from the ICM after it crosses a galaxy cluster once is sustained against a change in resolution. Reproduced from Paper II.}
 \label{fig:paper2:convergence}
\end{figure}

The gas that is lost by galaxies via ram pressure stripping is later mixed with the ICM, resulting in some increase in its metallicity -- the simulations of \citet[][]{2006A&A...452..795D} show that infalling galaxies are responsible for $\sim$10\% of the metal enrichment within the ICM of a cluster. But it turns out that the opposite can also happen: as the galaxy moves within the cluster, ICM gas can cool and become part of its ISM, so that after the galaxy has crossed the cluster, some of its remaining ISM gas was not there originally. This has been shown for the first time in the simulations presented in \hyperlink{Paper II}{Paper II}, which we have mentioned in the previous \grifa{section}. Those same infalling galaxies for which we tracked the color and luminosity evolution had their ISM gas composition tracked over time through a special \emph{passive scalar} technique. In the simulation, a scalar was advected along the flow without influencing its dynamics; this scalar is initially set to 1 in the the gas cells belonging to the ISM of the galaxies and 0 in the cells belonging to the ICM of the cluster. At any point of the simulation, the value of the scalar in a cell then allows one to know how much of that gas came from the ICM -- for instance, a value of 0.7 means that 30\% of that gas came from the ICM. This has allowed us to quantify the contamination with ICM gas within the disks of the galaxies over time, as shown in Figure \ref{fig:paper2:contamination}. What we found was that, after one crossing of the cluster, typically 20\% of the ISM gas still in any given galaxy comes from the ICM. We have also verified that this result is robust against changes in the spacial resolution of the simulation, as shown in Figure \ref{fig:paper2:convergence}.

\grifac{As in our analysis we only take into account the dense and cold gas within the disks of the galaxies, the presence of ICM gas within that gas implies that the ICM must have cooled at the ISM/ICM interface to then be advected into the disk. The likely cause of this phenomenon is that, at this interface, the ICM partially mixes with the ISM gas, forming an intermediate density layer around the gaseous disk; as the rate of energy loss by radiative emission quickly increases with the medium density, the formation of this interface leads to a subsequent runaway cooling of the intermediate density gas, which then ends up depositing into the gaseous disk.}

\section{Caveats of our models} \label{sec:ismcaveats}

\grifac{The results presented in this section deserve mentioning of some caveats. The numerical treatment and sub-grid recipes we have employed in our simulations are by no means the only possible choices, and changes in either of those could lead to different results.}

\grifac{Regarding the numerical treatment, an important point is that our resolution in the gaseous disks of the galaxies is in the range $\sim$100 -- 200 pc. This resolution is insufficient for resolving a multi-phase interstellar medium, with cold molecular clouds embedded into a diffuse and warmer medium, as in a real galaxy. The importance of this is that, as shown in the simulations presented in \citet{2009ApJ...694..789T}, the molecular clouds are not as easily stripped as the diffuse gas. Thus, every result we have presented which depends on the rate of gas loss could feature some degree of bias due to this numerical limitation.}

\grifac{Other than resolution concerns, different sub-grid recipes for e.g. stellar feedback could also lead to different results. The supernova feedback recipe we have employed is a simple one, including only thermal feedback by type II supernovae. The topic of how to include stellar feedback in a simulation is not a solved problem, and many alternative implementations are available, as discussed in \citet{2017MNRAS.466...11R}. An important point is that the thermal energy released by supernovae tends to be quickly lost when radiative cooling is included in the simulation; some implementations aim to circumvent this effect by suppressing the rate of energy loss in the cells around a supernova, so as to allow this energy to have a greater effect on the gas. The use of such stronger recipes for feedback in a ram pressure simulation could lead to different results, as that greater amount of energy has an impact on the gas dynamics. This caveat also applies to additional feedback sources which we have not included in our simulations, such as feedback by Active Galactic Nuclei (AGN).}

\chapter{Free-floating clumps of molecular gas} \label{c:clumps}

In this chapter, we will describe the population of free-floating clumps of molecular gas which have emerged in the galaxy cluster simulation including infalling galaxies presented in \hyperlink{Paper II}{Paper II}. The setup of this simulation will be described in detail in Section \ref{sec:paper2}. Those clumps are objects which are formed at the tails of the galaxies after they undergo ram pressure stripping, and which proceed to live independently within the cluster's ICM for up to 300 Myr, often without an obvious association to any galaxy, making them possibly a new class of objects. \grifac{Similar objects have been observed in both galaxy groups and clusters, where isolated H$_\mathrm{I}$ clouds without any apparent connection to a galaxy, sometimes hosting young stellar populations, have been observed \citep[e.g.][]{2008AJ....135..319D,2009A&A...507..723T,2010ApJ...725.2333K,2017ApJ...843..134S}.} Such objects have not received a comprehensive treatment in the literature so far, except within the local group, where analogous objects have been named \emph{Ultra Compact High Velocity Clouds} \citep[UCHVCs, see e.g.][]{2013ApJ...768...77A,2015ApJ...806...95S}. Our simulations indicate that clumps of molecular gas should be ubiquitous in galaxy clusters and groups.

\section{Physical considerations} \label{sec:clumpdefinition}

In Astrophysics, one recurrent scenario is that of a multiphase gaseous medium, in which cold and dense gas is embedded in a hot and diffuse surrounding medium. An example is a gas-rich galaxy falling into a cluster, where the cold phase is represented by the galaxy's ISM and the hot phase by the ICM. In this particular case, the cold gas is being internally affected by sources of gravity other than itself, namely the dark matter halo and stars of the galaxy; but cases also exist where the opposite takes place, and the internal structure of the object is only affected by its own self-gravity and by the local hydrodynamic forces. 

\begin{figure}%
    \centering
    \subfloat[]{{\includegraphics[width=0.45\textwidth]{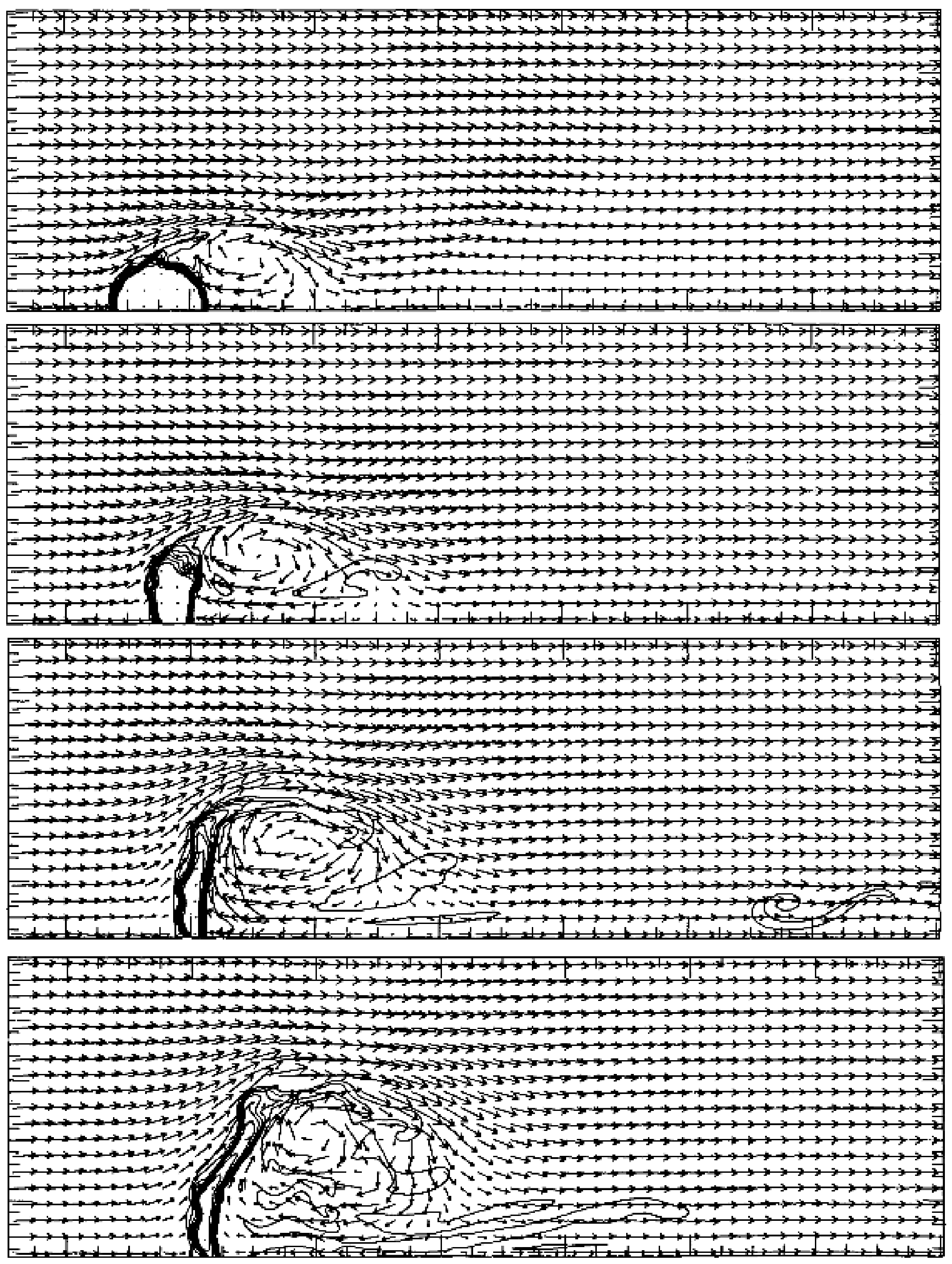} \label{fig:unstableclump} }}%
    \qquad
    \subfloat[]{{\includegraphics[width=0.45\textwidth]{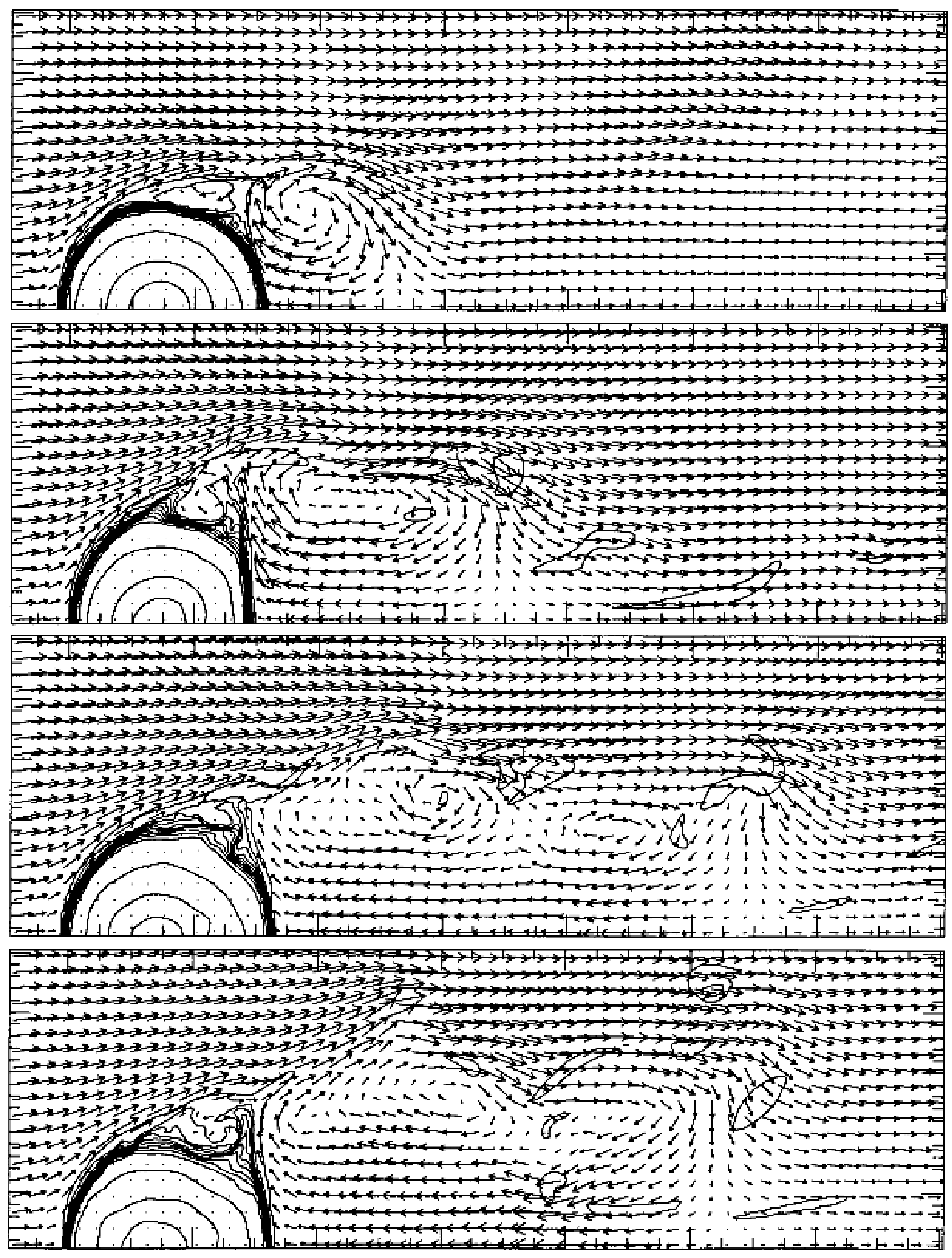} \label{fig:stableclump} }}%
    \caption{Simulations showing a dense clump moving within a diffuse medium. a) The mass of the clump is below the gravitational collapse threshold. It is quickly dissolved by Kelvin-Helmholtz instabilities. b) Its mass is above that threshold, making it stable against KH instabilities, and consequently allowing it to survive for much longer. Reproduced from \citet{1993ApJ...407..588M}.}%
    \label{fig:clump2d}%
\end{figure}

In order to analyze the dynamics of the latter kind of objects, let's consider as a starting point a pressure-bound, spherical and homogeneous gaseous clump moving in a diffuse gaseous medium. At the interface with the medium, there are two layers of gas with different densities moving at different speeds in contact with each other, which is the exact scenario where Kelvin-Helmholtz instabilities take place. It can be shown that, in such scenario, the timescale for the clump to be dissolved by the KH instability is \citep{1959flme.book.....L}:
\begin{equation} \label{eq:KH}
T_{\mathrm{KH}} \sim \frac{\rho_{\mathrm{ICM}}+\rho_\mathrm{clump}}{\sqrt{\rho_{\mathrm{ICM}}\rho_\mathrm{clump}}} \frac{\lambda}{v},
\end{equation}
where $v$ is to be regarded as the relative velocity between the clump and its surrounding medium, and $\lambda$ is the wavelength of the instability -- an upper bound for the timescale can be obtained by setting $\lambda$ to be the clump's size (e.g. its radius). The disruption of the clump would ultimately lead to its mixing with the surrounding medium. But if the clump's self-gravity is strong enough, it can have a stabilizing effect that countereffects that of the KH instability. It can be shown \citep[][see also \citealt{1956MNRAS.116..351B}]{1993ApJ...407..588M} that there is a critical mass for this to be relevant, namely
\begin{equation}
M_{G} = 1.18 G^{-3/2} \left(\frac{k_B T_\mathrm{clump}}{\mu m_H}\right)^{3/2} \frac{\rho_\mathrm{ICM}^{3/2}}{\rho_\mathrm{clump}^2}.
\end{equation}
This is simply the mass above which the clump is expected to not be stable against gravitational collapse. \grifa{This threshold takes place when, at the surface of the clump, the force associated with the pressure gradient between the interior of the clump and the diffuse medium plus the gravitational force by the clump itself are equal.} As an illustration, consider the 2D simulations shown in Figure \ref{fig:clump2d}, taken from \citet{1993ApJ...407..588M}. In Figure \ref{fig:unstableclump}, the evolution of a clump with mass below this critical mass is shown, where it can be noted that it is quickly destroyed and mixed with the surrounding medium. On the other hand, Figure \ref{fig:stableclump} illustrates the evolution of a clump with mass comparable to the critical mass -- in this case, the clump does not mix very effectively, surviving for much longer.

It turns out that at the tails of ram pressure stripped galaxies, concentrations of cold gas exist, which can be detected either directly through radio observations or through the presence of young stellar populations \citep{2015A&A...582A...6V,2017ApJ...839..114J}. In the simulations we will discuss in the next Section, we will describe how this cold gas proceeds to live within the ICM of a cluster in the form of clumps much like the one shown in Figure \ref{fig:stableclump}, which can survive for long enough to distance themselves hundreds of kpc from their progenitor galaxies, appearing for this reason as isolated objects. 

\section{Case study of a Virgo-like simulated cluster} \label{sec:paper2}

The results we present in this section are based on \hyperlink{Paper II}{Paper II}. We have already discussed results from this paper on Sections \ref{sec:luminosity} and \ref{sec:contamination}, where we focused on its results regarding the ISM evolution of galaxies in clusters. Now it is useful to define in more detail the setup of the simulations used in this paper. The starting point was a dark matter only, zoom cosmological simulation of a galaxy cluster with $M_{200c}$ similar to the Virgo cluster. From this simulation, the infalling halos which crossed the virial radius of the cluster for the first time within the last 5 Gyr of its evolution were extracted. Then, an idealized re-simulation of this same cluster was run, with the cluster halo replaced by a spherical dark matter potential fit to the mass distribution in the cosmological simulation plus an ICM modeled by a $\beta$-profile fit for Virgo X-ray data taken from \citet{2011MNRAS.414.2101U}. 
The infalling halos of the cosmological simulation were replaced by \grifa{a gas rich model of} disk galaxies with observationally reasonable properties: their stellar masses were chosen using abundance matching data from \grifa{\citet[][see also Section \ref{sec:idealizedsimulations}]{2013ApJ...770...57B}}, and their disk scale lengths from the observational relation between stellar mass and disk scale length shown in \citet{2010MNRAS.406.1595F}, which was obtained using a sample of galaxies from the SDSS survey. The properties of the resulting model galaxies are shown in Table \ref{tab:galaxies}. Those model galaxies were set up to enter the virial radius of the idealized cluster at the same positions and times as in the cosmological run. This has allowed us to model the hydrodynamic evolution of the galaxies which fall into the cluster as a function of cosmic time, which as we will see shortly was essential for describing the population of clumps of molecular gas which emerged in the simulation.

The \grifa{best} resolution in the idealized simulation was 122 pc, which was reached by all the cold gas in the simulation by employing a refinement criterion based on the Jeans length. With this high resolution in the cold gas, we have found that the gas that is stripped from the infalling galaxies does not mix immediately with the surrounding medium. In fact, clumps of molecular gas similar in nature to the ones described in Section \ref{sec:clumpdefinition} are left behind lurking in the ICM. The first step in characterizing those clumps was identifying them in the first place, which was done using an algorithm called \emph{PHEW} \grifa{(Parallel HiErarchical Watershed)}, described in \citet{2015ComAC...2....5B}. This algorithm is a clump finder, which was designed to find localized density peaks in a simulation, originally with the goal of finding halos in cosmological simulations; it is implemented in the current version of \textsc{ramses}, and can be used to find clumps either in the dark matter distribution of a simulation or in its gas distribution.

\begin{figure} 
 \centering
 \includegraphics[width=0.8\textwidth]{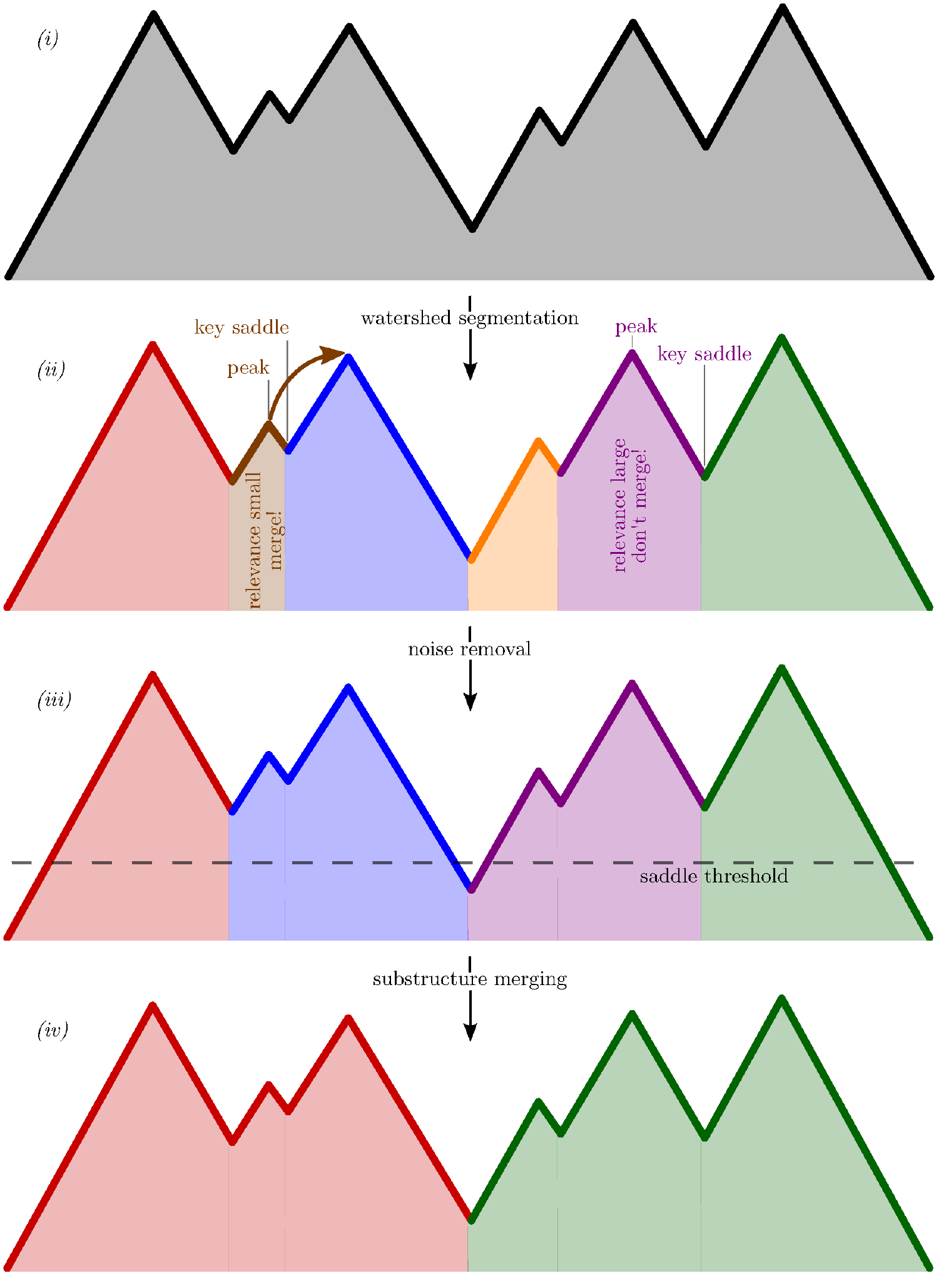}
 \caption{Illustration of how the \textsc{PHEW} clump finding algorithm works. First the density peaks are found, then noisy peaks are merged, and finally substructures are merged into halos. Reproduced from \citet{2015ComAC...2....5B}.}
 \label{fig:phew}
\end{figure}

The algorithm contains three free parameters. They are the density threshold above which a density peak is to be considered as a clump; the relevance threshold, which is the minimum density ratio between a density peak and its nearest density local minimum in order for this peak to be considered a separate clump, and not a statistical fluctuation; and the saddle threshold, which is used to group neighboring clumps together into a halo, where the individual clumps are substructures. An illustration of those parameters at work is shown in Figure \ref{fig:phew}.

In our case, we have used PHEW to find clumps in the gas matter distribution. The relevance threshold used was 3, a low number because, as we will describe ahead, spurious clumps will be later discarded; and the saddle threshold used was $1 \times 10^{-26}$ g cm$^{-3}$, a value which we have empirically found to cause clumps which visually overlap with each other to be merged. The resulting clumps (which are of the ``halo'' kind, since a saddle threshold was used) were further selected according to a set of criteria. But before getting into those, we should define two things. First, a \emph{self-shielding} threshold is a density above which hydrogen is expected to be in a cold and mostly molecular state. Following \citet{2018MNRAS.tmp.1018C}, we have chosen a value of $n_\mathrm{H} = 3 \times 10^{-3}$ cm$^{-3}$ (or $\rho = 5 \times 10^{-27}$ g cm$^{-3}$) for this threshold. Secondly, we have defined what we call a \emph{bounding sphere} in the following manner. For each clump, a sphere is created centered on its position, with radius equal to the smallest cell size in the simulation. Then the sphere's radius is gradually increased until exactly 50\% of the cells it encapsulates are self-shielded -- this way, the boundary of clumps which are not exactly spherical will be contained within the sphere. Then we discard clumps:

\begin{itemize}
\item For which there are no self-shielded cells, since we are interested in clumps of molecular gas and not spurious overdensities in the ICM plasma, or clumps with a high degree of mixing with the ICM.
\item For which the bounding sphere overlaps with the bounding sphere of another clump, since we were focused on isolated clumps.
\item Which are located within $10 R_d$ of any galaxy, where $R_d$ is the scale length of the disk of each galaxy, such as to remove the highly irregular and entangled gas which has just been stripped from a galaxy.
\end{itemize}

The remaining clumps are the ones which we will be talk about in the following subsections. A video of the simulation, showing the clumps moving within the galaxy cluster, can be watched at \url{http://www.astro.iag.usp.br/~ruggiero/paperdata/2/}.

\subsection{Clump properties and dynamics} \label{sec:paper2clumps}

\begin{figure} 
 \centering
 \includegraphics[width=0.8\textwidth]{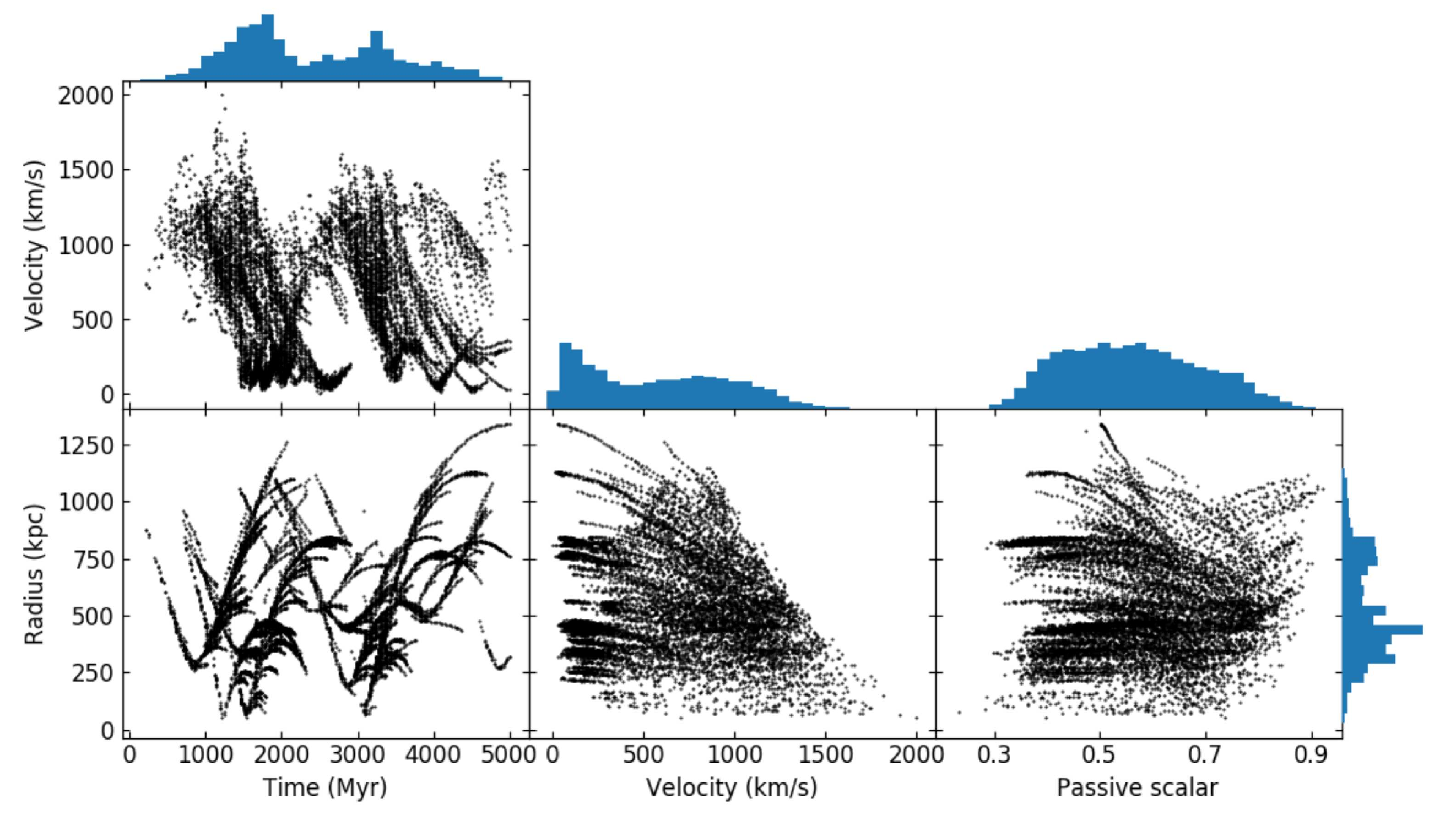}
 \caption{Kinematic properties of the clumps formed in the galaxy cluster simulation presented in Paper II. \grifab{Here the velocity, cluster-centric radius and passive scalar distributions are shown for the clumps detected at all times of the simulation. This is the reason why parabolas appear in the left panels and lines in the middle and right panels -- clumps appear more than once at different stages of their orbits in this plot.}}
 \label{fig:paper2:clumpskinematics}
\end{figure}

Let's now describe in more detail the clumps of molecular gas which we have identified in our simulation. In Figure \ref{fig:paper2:clumpskinematics}, we show the kinematic properties of the clumps at all times of the simulation. The time histogram shows two major peaks, which correspond to the central passage of the two most massive galaxies in the simulation (both of which have $M_{200c} \sim 10^{12}$ M$_\odot$). Those peaks fade away in a timescale of a few hundred Myr, which is the clump lifetime (most live between 100 and 300 Myr). The velocity histogram shows two regimes: below 500 km/s, there is a concentration of clumps moving at low speeds, which are clumps that were once at high speed but decelerated and remained for a significant part of their lifetime at low speed before mixing with the ICM. At speeds above that, there is a population of clumps which were ejected recently from a galaxy and which are still moving at a speed close to the original speed of the galaxy -- the galaxies move \grifa{supersonically} at $\sim$1000 km/s. \grifa{This might seem like a high speed for a Virgo-like cluster, but note that infalling galaxies are not a virialized population.}

\begin{figure} 
 \centering
 \includegraphics[width=0.65\textwidth]{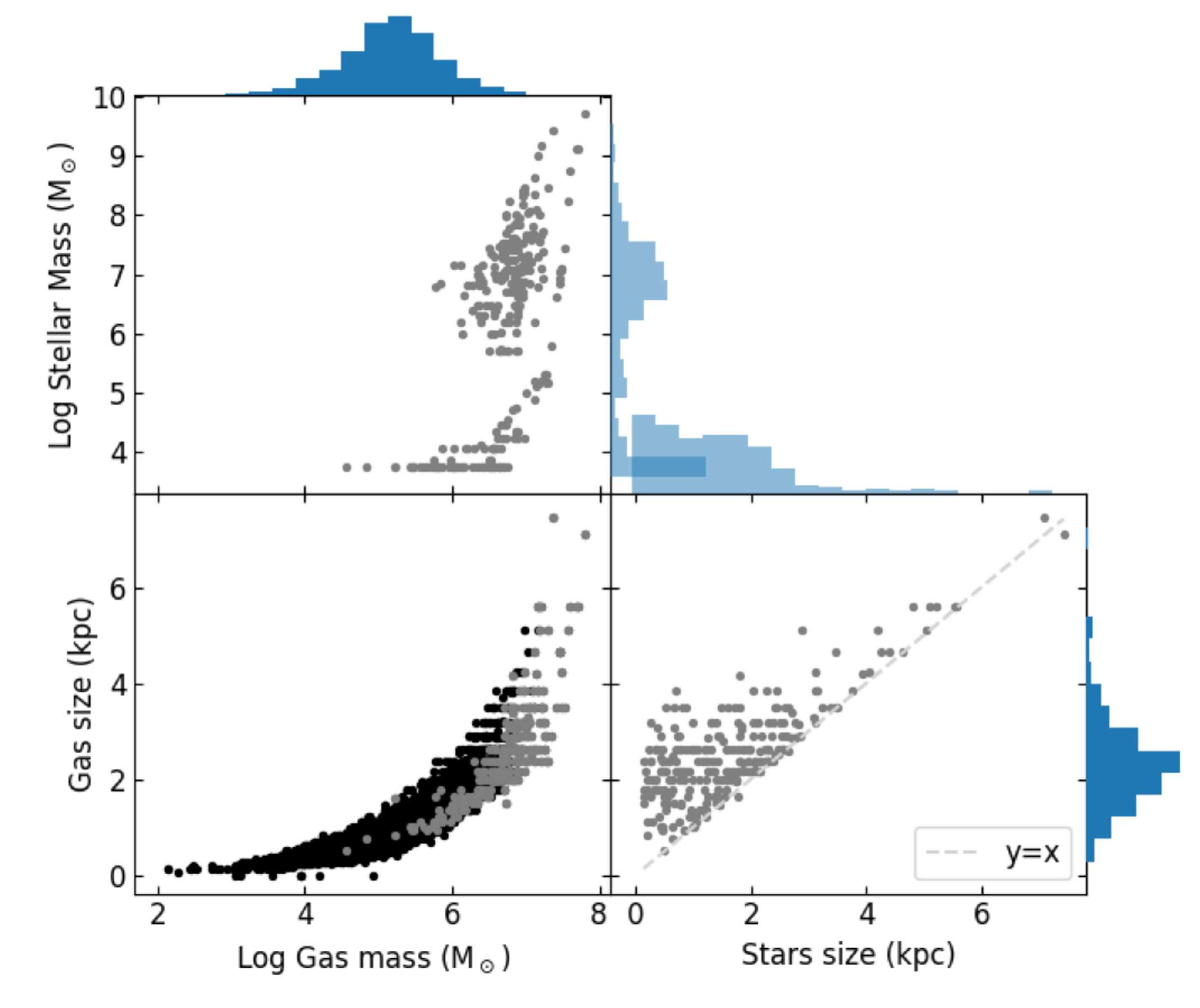}
 \caption{Structural properties of the clumps formed in the galaxy cluster simulation presented in Paper II. \grifab{Grey points represent clumps which contain star particles, and black points represent the remaining clumps. The right panel shows that the extent of the stars within any given clump is in almost all cases smaller than the extent of the gas in that same clump, showing that the stars are formed at the center of the clump and remain dynamically cold.}}
 \label{fig:paper2:clumpsproperties}
\end{figure}

The structural properties which we have found for the clumps are shown in Figure \ref{fig:paper2:clumpsproperties}. \grifa{The gray points in that figure represent clumps which contain star particles, \grifac{which must have formed in situ because in our simulation no stars were stripped from the galaxies}, and the black points are clumps without any star particle. The right panel shows the extent of the star particles as a function of the clump size derived from the clump gas content, with the extent of the stars defined in the same way as the clump size: the largest distance between a star within a clump and the clump center. It can be noted that the points are all above the line $y = x$ in that plot, implying the stars are centrally concentrated relative to the gas, and suggesting that those stars are formed at the densest region of the clump and remain dynamically cold after that.}

In general lines, the structural properties of the clumps are the following:

\begin{itemize}
\item Gas mass: most common value is $1.6 \times 10^{5}$ M$_\odot$, and most clumps range from $2.7 \times 10^{3}$ M$_\odot$ to $4.6 \times 10^{6}$ M$_\odot$.
\item Size: most commonly $2.2$ kpc; most clumps range from $0.3$ kpc to $4.1$ kpc. We have defined the clump size as the largest distance from its center to any of its self-shielded cells. The largest clumps are usually simply elongated.
\item Stellar mass: among the clumps which have formed stars, the most common stellar mass is $10^{7}$ M$_\odot$, and the most common gas mass is $7 \times 10^{6}$ M$_\odot$.
\end{itemize}

Those properties are consistent with observations of similar objects previously reported in the literature:

\begin{itemize}
\item In the Virgo cluster, two clouds of neutral gas have been observed and reported in \citet{2010ApJ...725.2333K}. The H$_\mathrm{I}$ masses of those clouds are $10^{7.28}$ and $10^{7.85}$ M$_\odot$ -- both numbers are within our clump mass distribution, but at the upper end, which could possibly mean an observational bias. Another cloud has been reported in \citet{2017ApJ...843..134S}, with a stellar mass of $5.4\pm1.3 \times 10^4$ M$_\odot$ (also within our predicted range), and a stellar age of $\sim$ 7 -- 50 Myr, relatively small compared to our clump lifetime, but consistent with it.
\item Also in the Virgo cluster, clumps of molecular gas have been reported within the tail of NGC 4388 \citep{2015A&A...582A...6V}, with masses between $7 \times 10^5$ and $2\times 10^6$ M$_\odot$, which are also within our mass distribution.
\item \citet{2009A&A...507..723T} describe compact clouds of star forming gas in three different galaxy groups, with stellar ages below 100 Myr (consistent with our clump lifetime).
\item Similar objects were found in the Hickson Compact Group, with ages below 200 Myr \citep{2008AJ....135..319D}. In this case, the stellar masses of the compact objects are in the range $10^{4.3-6.5}$ M$_\odot$, consistent with our findings, although the H$_\mathrm{I}$ mass reported for those objects ($10^{9.2-10.4}$ M$_\odot$) is larger than any clump mass we obtain.
\end{itemize}

Regarding our results on the stellar mass within clumps, one caveat should be noted, which is that the minimum star particle mass in our simulation was $5.6 \times 10^3$ M$_\odot$. Naturally, clumps with mass below that will not form any stars in the simulation, but that does not mean that in reality they would not have any stars. With that pointed out, we still find that a disproportionately small amount of clumps (3\%) form stars, implying that the density within those objects is such that star formation in them is not efficient. This is consistent with observations of star forming clumps in the tails of ram pressure stripped galaxies, which have gas depletion timescales (defined as $\tau \equiv M_\mathrm{gas}$/SFR) larger than the age of the universe \citep{2015A&A...582A...6V,2017ApJ...839..114J}.

\subsection{Contribution to the intracluster light}

As we have 	discussed in Section \ref{sec:paper2clumps}, star formation does take place inside the clumps of molecular gas left behind by galaxies that are accreted into a cluster. Naturally, many of those stars will live far longer than the clump lifetime (which is $\lesssim 300$ Myr), and will be left as members of the intracluster light after all the cold gas in the clump has been mixed with the ICM. 

\begin{figure} 
 \centering
 \includegraphics[width=0.7\textwidth]{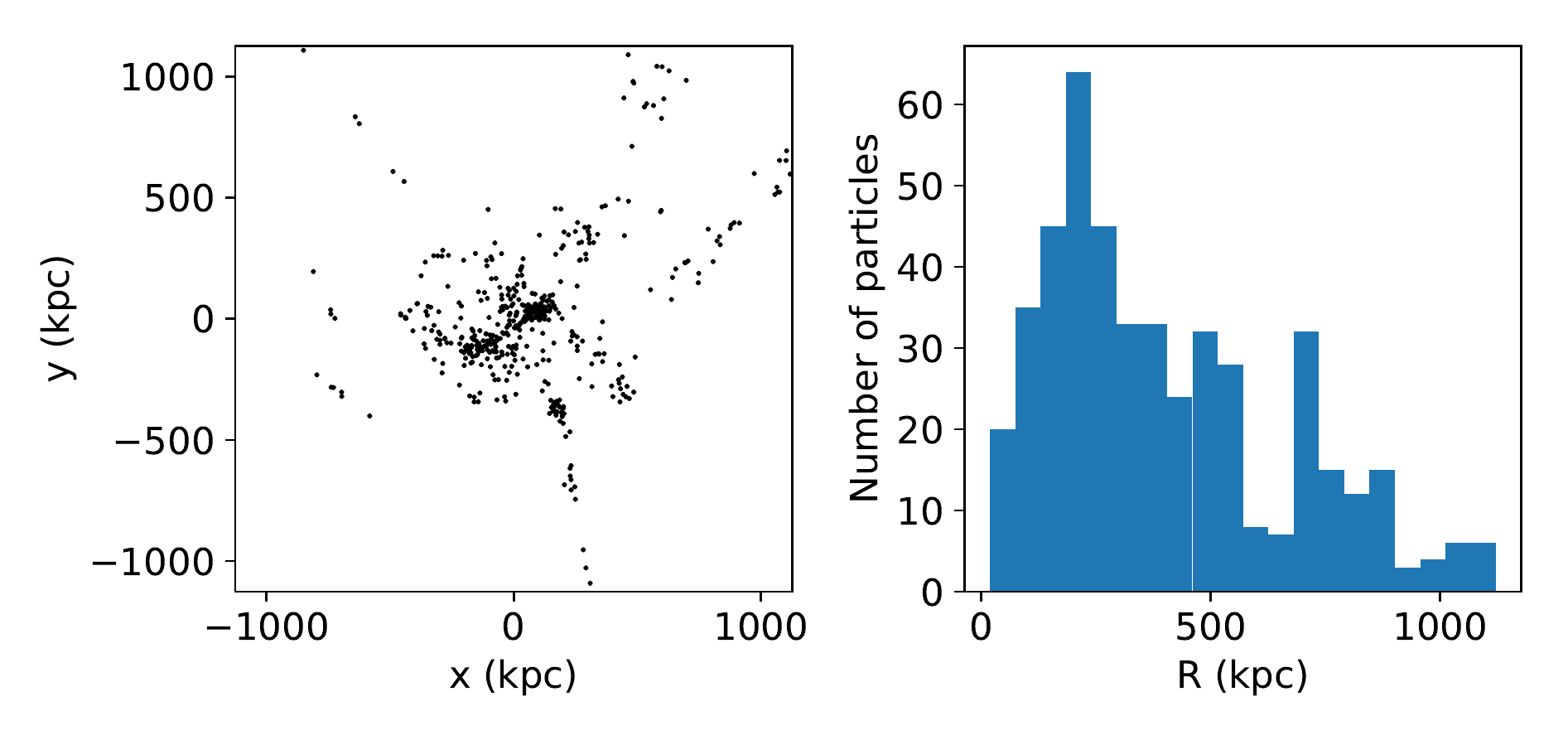}
 \caption{Cloud of free moving stars, which were formed inside clumps, seen at $z = 0$ occupying the galaxy cluster volume. Reproduced from Paper II.}
 \label{fig:paper2:diffusestars}
\end{figure}

In Figure \ref{fig:paper2:diffusestars}, we show the spacial distribution of such stars at $z = 0$. It can be noted that they are centrally distributed, forming a cloud of radius of roughly 500 kpc, or half of the virial radius of the cluster. This is directly related to the higher concentration of clumps in the inner region of the cluster ($r < 500$ kpc) shown in  Figure \ref{fig:paper2:clumpskinematics}, which emerge within the tails of galaxies after their central passage. The combined $M_\mathrm{V}$ of all those stars is -10.8, a very small value \grifa{which would convert into a surface brightness of just 62 mag/arcsec$^2$ even if all of the stars were within $r < 500$ kpc}, showing that the stars formed in ram pressure clumps are not a significant component of the intracluster light of a cluster.

\chapter{Formation of jellyfish galaxies in galaxy cluster \grifab{merging} systems: case study of the A901/2 system} \label{c:a901}

This chapter is based on the results presented in \hyperlink{Paper III}{Paper III}. In this paper, we have modeled the A901/2 multi-cluster system with a galaxy cluster \grifab{collision} simulation, which was used to investigate the ICM conditions at the locations of a sample of candidate jellyfish galaxies \grifa{(see Section \ref{sec:gasloss} for a definition)} within the system, with the goal of explaining why the jellyfish are found at the locations where they are found. In what follows, we will first describe the system in general terms, and then we will proceed to describe the mechanism for the generation of the jellyfish which we have inferred from the simulation model. As will become clear later, the mechanism we propose is based on general enough arguments to be applied to any galaxy cluster merger in its early stages. 

\section{About A901/2}

\begin{figure} 
 \centering
 \includegraphics[width=\textwidth]{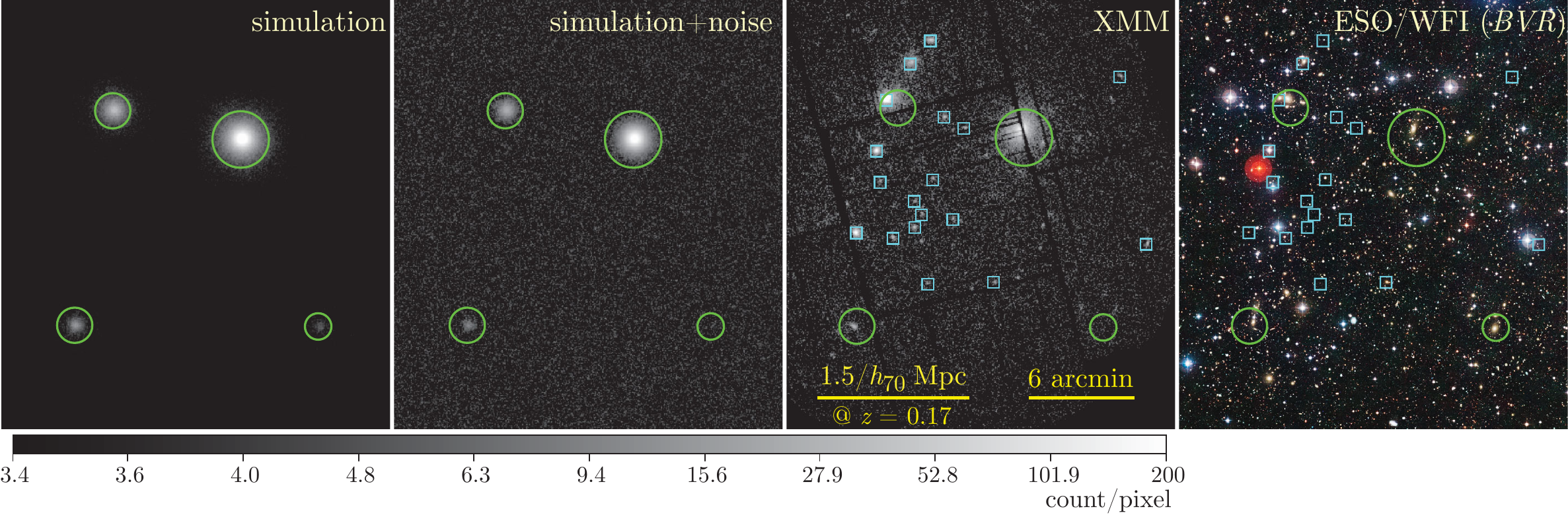}
 \caption{Comparison between the X-ray emission of our simulation model and the X-ray emission of the A901/2 system from a XMM-Newton observation. In the last panel, an optical image of the system area, taken with the ESO/WFI instrument, is shown. \grifa{The substructures are, in order (from left to right and from the top to bottom): A901a, A901b, A902 and the SW Group. Reproduced from Paper III.}}
 \label{fig:a901mock}
\end{figure}

The A901/2 system is a multi-cluster system in which two galaxy clusters, A901a and A901b, and two smaller groups, A902 and the SW Group, are seen in close proximity to each other in the sky. Those four substructures have similar redshifts \citep[e.g.][]{Weinzirl2017}, indicating that they are actually a \grifab{merging} system; since no shocks or strongly asymmetrical features can be observed in the X-ray emission of the system \citep{Gilmour2007}, the merger seems to be in its early stages. \grifa{Images of the system in X-rays and in the optical, \grifab{taken with the XMM-Newton telescope and the ESO/WFI instrument respectively}, can be seen in the last two panels of Figure \ref{fig:a901mock} (the remainder of the figure will be discussed later).}

The galaxy population inside the system is very rich and diverse, including galaxies across a range of evolutionary stages \citep{gray04,gray09}. Inside each of the four subclusters, a visual classification of member galaxies, based on \grifa{HST/ACS images in the F606W band} and presented in \citet{roman-oliveira18}, has been able to identify a sample of $\sim$100 galaxies with asymmetrical features indicative of jellyfish morphologies. This is a very large number of jellyfish galaxies for a single system, which indicates that the interaction between the four substructures may be acting as a trigger for the formation of such morphologies. Those galaxies have been classified in the so-called \emph{JClass} system \grifa{\citep[introduced in][]{2016AJ....151...78P}}, in which an integer number from 1 to 5 is used to quantify their degree of asymmetry, with higher values corresponding to larger degrees of asymmetry, and in principle to a greater likelihood of a given galaxy actually being a jellyfish.

The presence of such a large sample of jellyfish candidates makes the system particularly suitable for studying the mechanism for the generation of jellyfish morphologies in galaxy cluster \grifab{collisions}. It has been suggested in \citet{2016MNRAS.455.2994M} that the larger relative velocities found in \grifab{merging systems} might make them the main locations where jellyfish galaxies are generated in the Universe, as opposed to isolated galaxy clusters in which galaxies fall. Indeed, \citet{owers12} has reported the observation of 4 individual jellyfishes in the very inner region of a galaxy cluster \grifab{collision}, suggesting a connection between the two phenomena.

\section{Hydrodynamic model and the jellyfishes}

In order to probe the exact mechanism behind the generation of jellyfish galaxies in the A901/2 system, we have in \hyperlink{Paper III}{Paper III} generated a galaxy cluster \grifab{collision} model for the system as a whole, in which 4 spherical galaxy clusters in equilibrium, with virial masses similar to the observed clusters, collide in such way that the relative positions between the four subclusters are reproduced to a good approximation. \grifac{Contrary to the simulations presented in the previous sections of this thesis, this simulation was run with the code \textsc{gadget-2} \citep{2005MNRAS.364.1105S} instead of \textsc{ramses}, which employs the Smoothed Particle Hydrodynamics \citep[SPH,][]{1977MNRAS.181..375G,1977AJ.....82.1013L} method instead of the Adaptive Mesh Refinement (AMR) method}. \grifa{The four substructures are at similar redshifts \citep[][]{Weinzirl2017}, so we assume in our model for simplicity that they are all in the same plane.} The gas fractions were chosen such that the X-ray emission of the system as a whole is also consistent with observational expectations; the properties of the 4 subclusters are shown in Table \ref{tab:a901table}, and a mock X-ray image of the best fit snapshot, where it is compared to the X-ray emission of the A901/2 system, is shown in Figure \ref{fig:a901mock}. Each model subcluster is made of a dark matter and a gaseous halo component, both spherical.

\begin{table}
\caption{Properties of the four simulated subclusters used in Paper III to model the A901/2 system.}
\label{tab:a901table}
\begin{center}
\begin{tabular}{l c c c c c}
\hline
         &   $M_{200}$     & $r_{200}$ &  $f_{\rm gas}$\\ 
         &  (${\rm M}_{\odot}$)  & (kpc) &     \\
\hline
subcluster A  (A901a)   & $1.3 \times 10^{14}$ & 1034 & 0.08 \\
subcluster B (A901b)    & $1.3 \times 10^{14}$ & 1036 & 0.15 \\
subcluster C (A902)     & $0.4 \times 10^{14}$ & 688  & 0.08 \\
subcluster D (SW Group) & $0.6 \times 10^{14}$ & 788  & 0.06 \\  
\hline
\end{tabular}
\end{center}
\end{table}

In order to probe the connection between the gas conditions in the simulation model and the locations of the jellyfish candidates in the real system, the fundamental quantity to be analyzed is the ram pressure, since it is what is responsible for the gas stripping of cluster galaxies \citep{1972ApJ...176....1G}:
\begin{equation} \label{eq:rampressure}
P_{\mathrm{ram}} = \rho_{\mathrm{ICM}} v_{\mathrm{ICM}}^2,
\end{equation}
where $v_{\mathrm{ICM}}$ is the ICM velocity relative to a particular galaxy.

\begin{figure} 
 \centering
 \includegraphics[width=\textwidth]{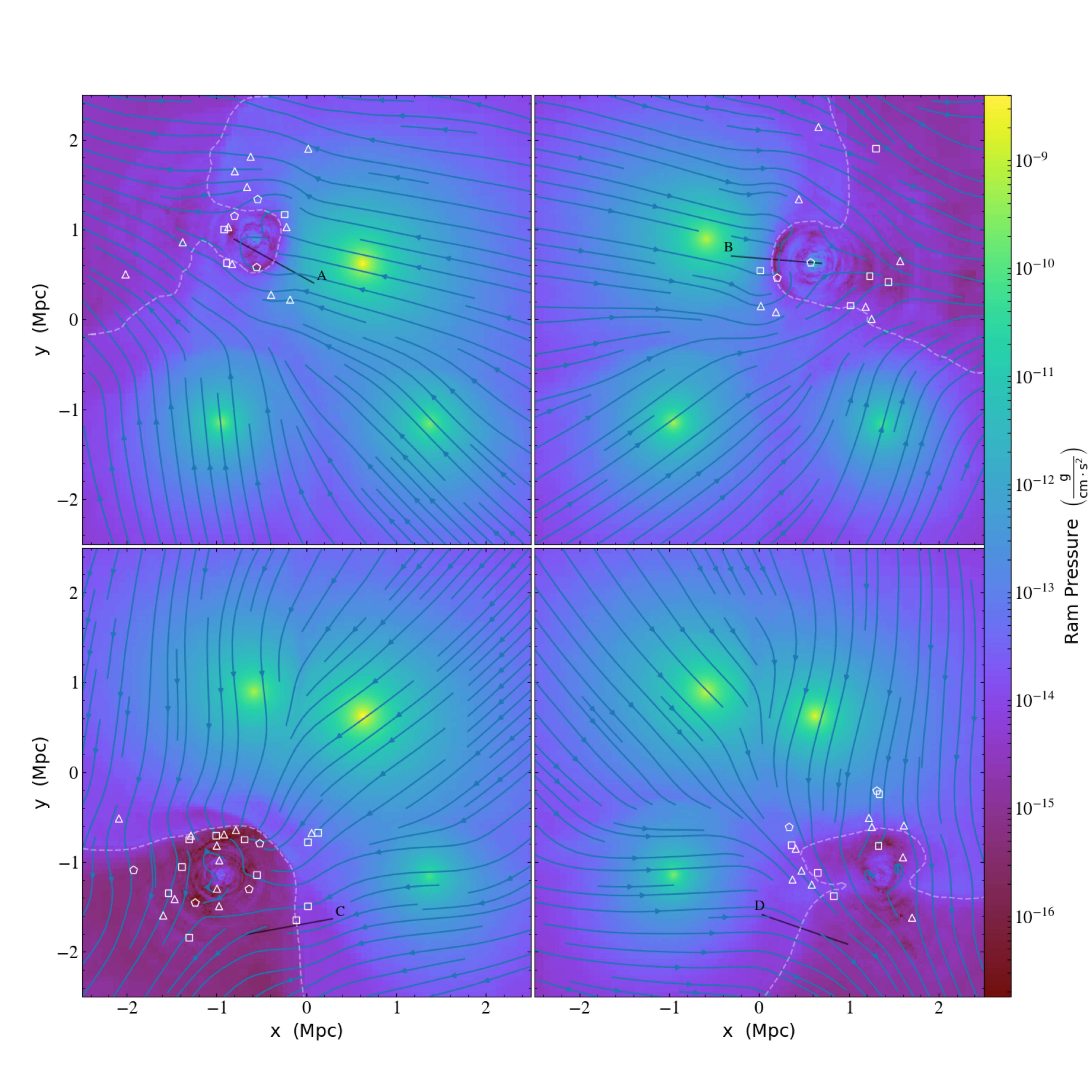}
 \caption{Ram pressure in the reference frame of each of the four subclusters in the A901/2 system. The dashed lines are the locations where we have detected a discontinuity in the ram pressure, and the annotations are the locations of the jellyfish candidates in the system. Triangles, squares and pentagons are used for JClass 3, 4 and 5 galaxies respectively. Reproduced from Paper III.}
 \label{fig:rampressures}
\end{figure}

\begin{figure} 
 \centering
 \includegraphics[width=0.6\textwidth]{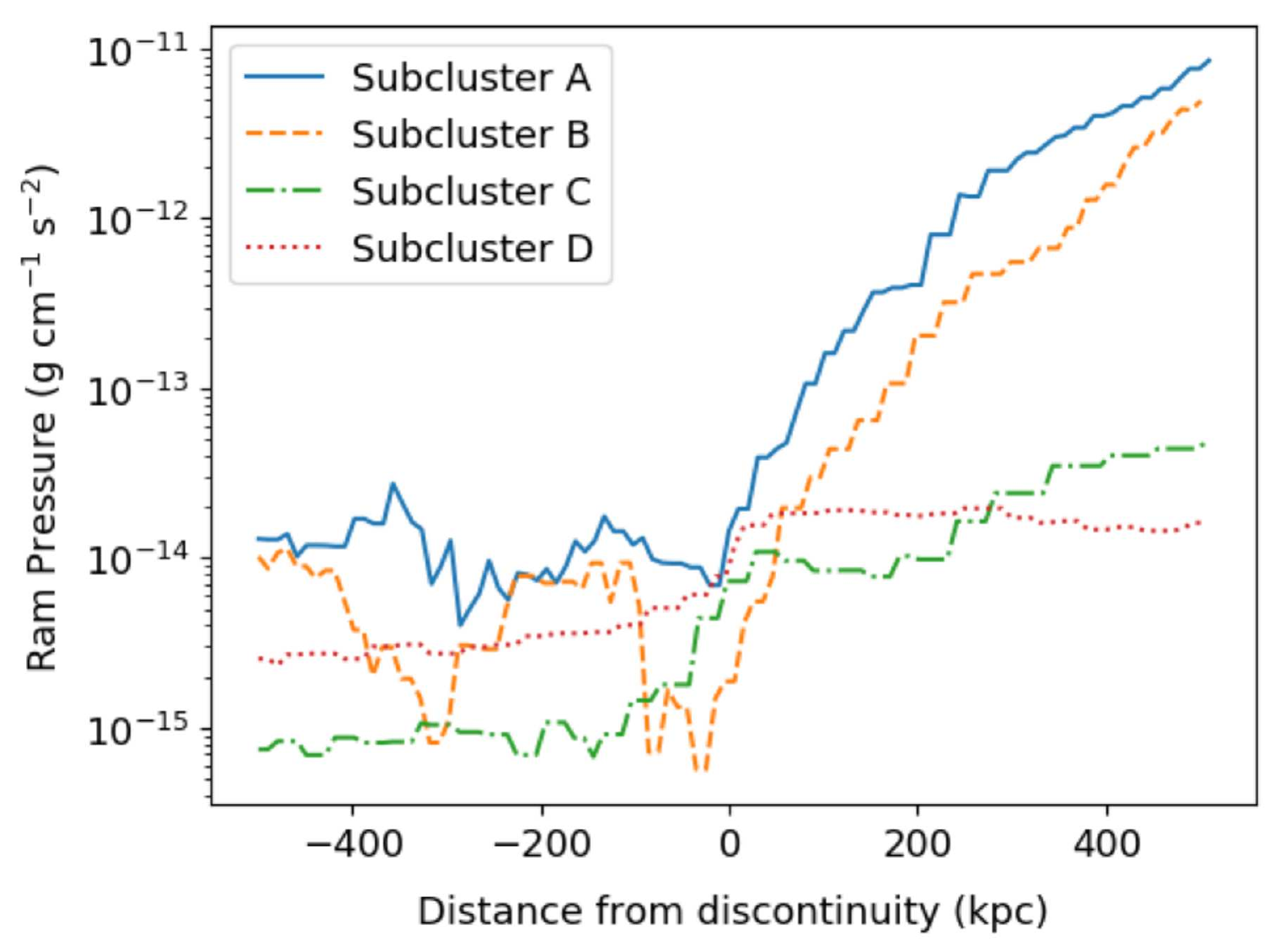}
 \caption{Ram pressure profiles along the four black lines shown in Figure \ref{fig:rampressures}. Reproduced from Paper III.}
 \label{fig:discontinuity}
\end{figure}

Given the local density, the ram pressure experienced by a member galaxy depends on its speed relative to the surrounding gas; as a first approximation for the reference frame of all galaxies within a given subcluster, we have considered the reference frame defined by the average velocity of the dark matter particles within the central region of that subcluster. Defining this velocity as $\vec{v}_{\mathrm{cl}}$, the ram pressure at any given location in the system, using Equation \ref{eq:rampressure}, is then
\begin{equation}
P_{\mathrm{ram}} = \rho_{\mathrm{ICM}} \left(\vec{v}_{\mathrm{ICM}} - \vec{v}_{\mathrm{cl}}\right)^2.
\end{equation}
In Figure \ref{fig:rampressures}, we show the resulting ram pressure maps in the reference frames of each of the four subclusters. We find that in each subcluster, a boundary exists where a sharp increase takes place in the ram pressure. These boundaries correspond to regions where gas originally from that subcluster and gas from the remainder of the system, which are moving at $>$1000 km/s relative to each other, meet. The locations of those boundaries, indicated by the dashed lines, were obtained by smoothing out ram pressure isocontours at the appropriate locations. After the boundaries are crossed, an increment of a factor of 10 -- 1000 in the ram pressure takes place, depending on the subcluster. A way to better visualize this is to consider ram pressure profiles along the boundaries; this is shown in Figure \ref{fig:discontinuity} for the 4 illustrative black lines indicated in Figure \ref{fig:rampressures}.

Given that such large increments in the ram pressure take place within a few hundred kpc, and given that the member galaxies are expected to be moving at $\sim$1000 km/s, a galaxy which crosses one of those boundaries is expected to encounter greatly enhanced ram pressure within a short timescale, of $\sim$100 Myr. Under these conditions, it would be reasonable to expect a gas-rich galaxy to undergo an intense ram pressure stripping event, and \grifa{temporarily} become a jellyfish. \grifab{Given that the jellyfish morphology depends on ongoing gas loss to manifest, this timescale should also be that of the morphology itself.}

\begin{figure} 
 \centering
 \includegraphics[width=0.6\textwidth]{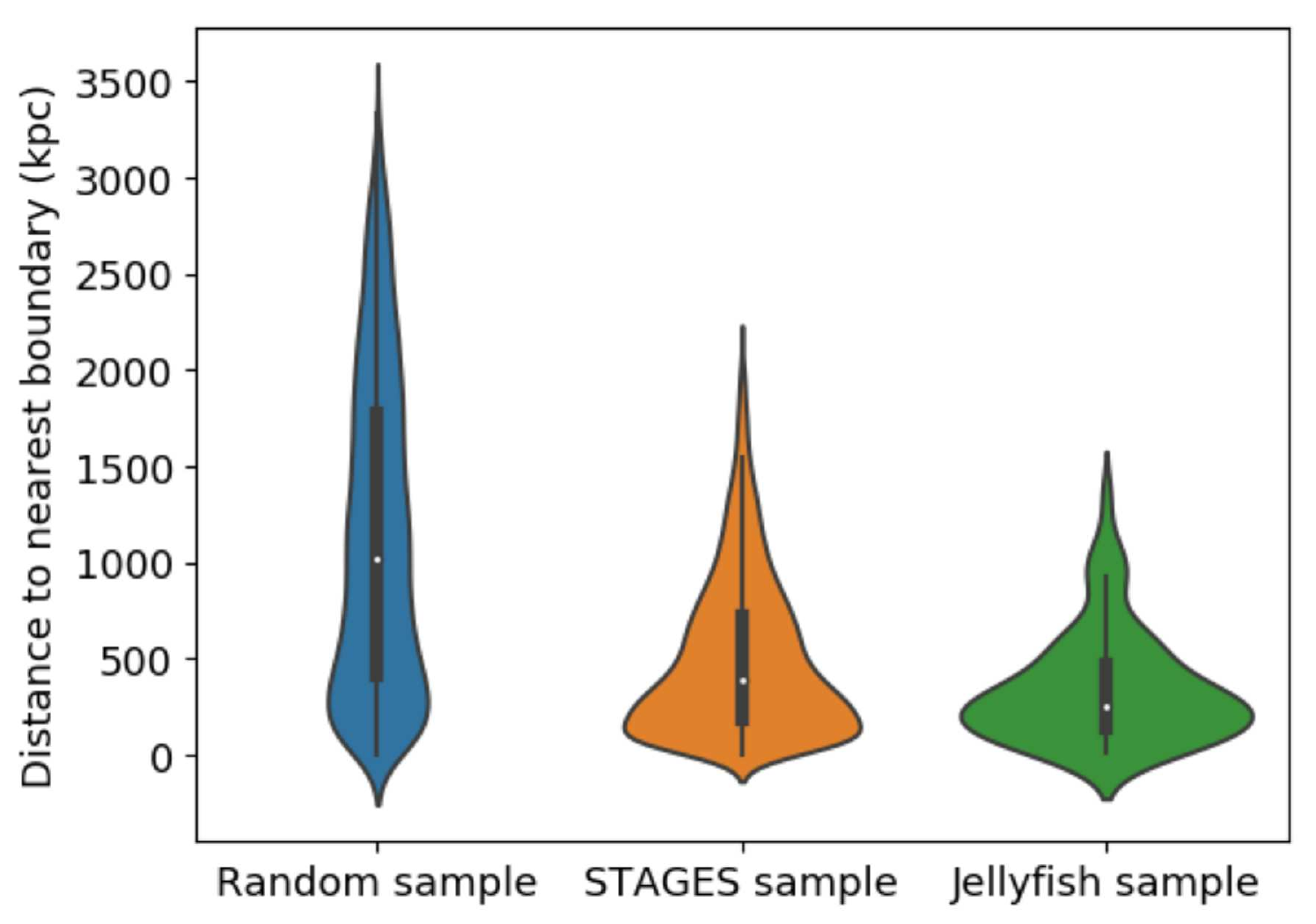}
 \caption{Comparison of the distance to nearest boundary distribution for the jellyfish sample, the entire STAGES sample of member galaxies in the A901/2 system, and a random set of points occupying the same area as the jellyfish. The jellyfish are systematically closer to the boundaries than the other two distributions. Reproduced from Paper III.}
 \label{fig:randomdistances}
\end{figure}

A way to probe whether this effect is likely to be taking place or not is to consider the positions of the jellyfish in A901/2 relative to the boundaries we report. Those positions are annotated in Figure \ref{fig:rampressures} for the candidates categorized as JClass 3, 4 or 5, which are the most asymmetrical and hence the most likely to be jellyfish. In total, there are 73 such galaxies. Visually, many of them  seem to be near the boundaries we report. We have quantified this effect by calculating the distance from each galaxy to its nearest boundary, and comparing the distribution of such distances to the distributions obtained considering instead either the entire STAGES sample of member galaxies in the A901/2 system \citep{Gray2009}, or a random cloud of points occupying the same area as the jellyfish. The result is shown in Figure \ref{fig:randomdistances}; we find that the jellyfish galaxies are systematically closer to the boundaries than both the other two samples. Kolmogorov-Smirnov tests applied to compare the jellyfish distribution and the corresponding STAGES and random distributions yield p-values of 0.13 per cent and $10^{-6}$ percent respectively, evidencing that the jellyfish distribution is different from the other two. The median, lower quartile and upper quartile of the jellyfish distribution are also all smaller than in those distributions.

A caveat of this result is that not all jellyfish are close to one of the boundaries -- for instance, several galaxies in the A902 subcluster are far from the respective boundary. What we propose is that, in galaxy cluster \grifab{collisions}, jellyfish galaxies may also be formed in mechanisms without a direct association to the \grifab{collision}, e.g. through passively falling into a subcluster. But the fact that a statistical difference exists between the distance to the nearest boundary distribution for the jellyfish and for both the entire STAGES sample and the random sample suggests that a significant fraction of the jellyfish are indeed being generated by crossing the boundaries. We note here that the presence of such boundaries depends solely on the subclusters approaching each other at high speed; hence, everything we have discussed here should also apply to galaxy cluster \grifab{collisions} in general, at least the ones in their early stages, like A901/2.

\begin{figure} 
 \centering
 \includegraphics[width=0.6\textwidth]{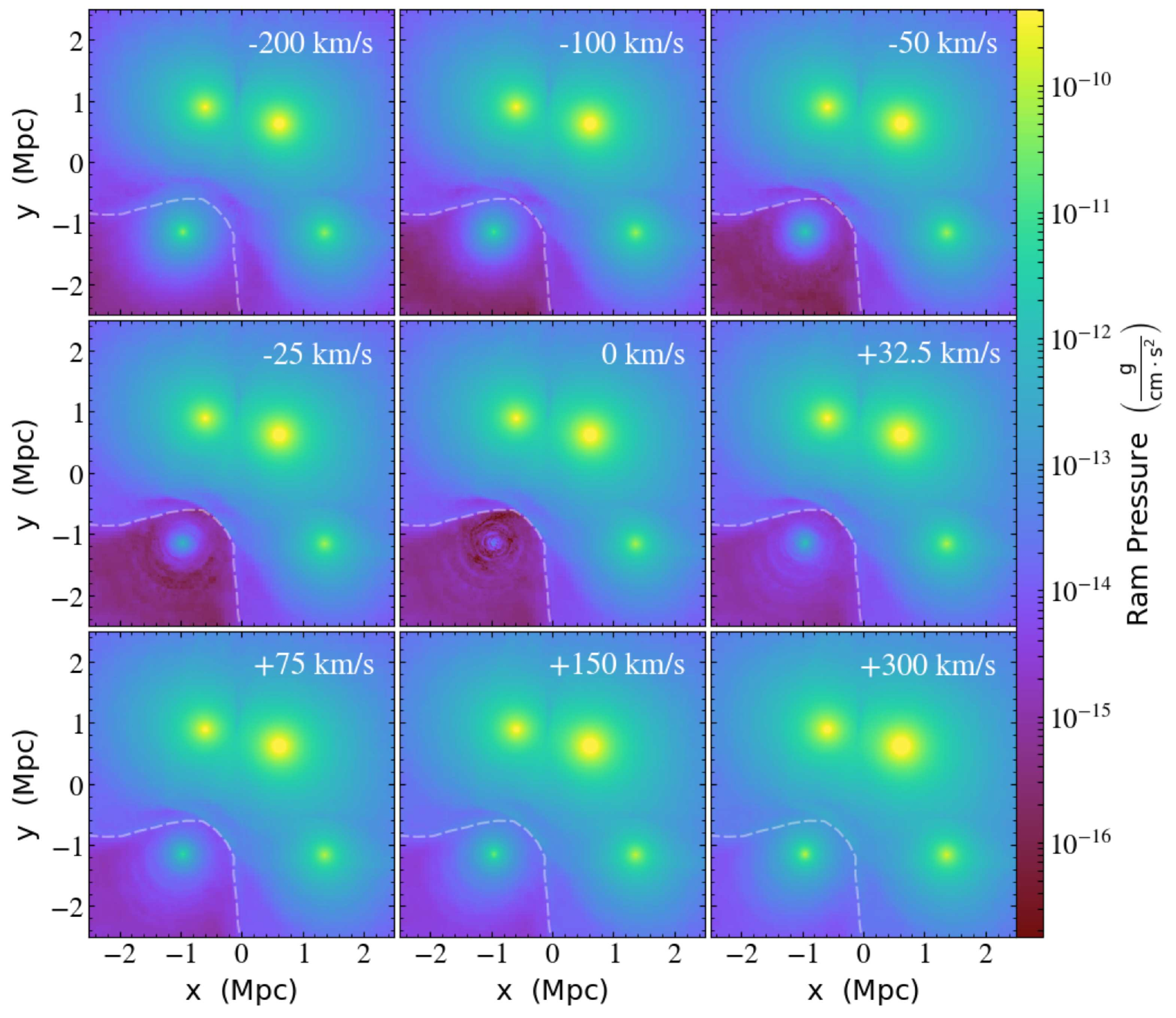}
 \caption{Effect of speeds relative to the parent subcluster on the ram pressure maps. The reference frame of each subplot is defined by the average A902 subcluster velocity added to the indicated velocity in the same direction. A discontinuity in the ram pressure exists for relative velocities within 100 km/s of the average cluster velocity. Reproduced from Paper III.}
 \label{fig:relativevelocity}
\end{figure}

This whole discussion was based on analyzing the ram pressure in the reference frame of the subclusters; but the galaxies in general have peculiar velocities relative to their parent subclusters. We have probed the effect of those velocities by generating ram pressure maps for A902 similar to the ones in Figure \ref{fig:rampressures}, but with different velocities added to the average cluster velocity. This subcluster was chosen for illustration because it is the one with the simplest ram pressure boundary; the comparison is shown in Figure \ref{fig:relativevelocity}. We find that a galaxy moving within 100 km/s of the average cluster velocity is expected to still find a ram pressure discontinuity at the locations of our reported boundaries. For peculiar velocities larger than that, the boundaries fade away, and the ram pressure becomes correlated with the local density. From this, we infer that our scenario applies specifically to galaxies which are moving at relatively low speed relative to their parent subcluster, perhaps at their apocentric passage -- \grifa{\emph{backsplash} galaxies \citep[see e.g.][]{2006MNRAS.366..645P,2011MNRAS.411.2637P}, which are galaxies that have crossed the cluster for the first time, could be particularly suitable candidates.}

\chapter{Summary and outlook} \label{c:summary}

In this thesis, we have explored the phenomenon of galaxy evolution in galaxy clusters using hydrodynamic simulations including spherical galaxy clusters within which realistic model galaxies move and evolve. With these simulations, we were able to constrain the final state of a Milky Way-like galaxy after crossing a galaxy cluster, including how much of its gas is converted into stars and how much of it is lost after one crossing; establish that the ISM of galaxies crossing a cluster becomes contaminated with ICM gas, with typically 20\% of the gas still in the ISM having come from the ICM after one crossing; characterize the luminosity evolution of galaxies falling into a cluster, finding that the most massive galaxies can gain up to 0.5 magnitude in the B band after entering the cluster due to a compression of their gaseous disks, while least massive galaxies are quickly stripped and proceed to only get redder and less luminous; and find that infalling galaxies leave behind a population of clumps of molecular gas lurking inside a cluster in isolation for up to 300 Myr. We have also proposed a mechanism for the generation of jellyfish galaxies in galaxy cluster \grifab{collisions}, based on a multi-cluster \grifab{collision} simulation tailored to mimic the A901/2 system, which was used to infer that jellyfish galaxies found in that system are systematically located near a boundary inside each subcluster where diffuse gas moving along with the subcluster and diffuse gas from the remainder of the system meet; this could provide a mechanism for the generation of jellyfish galaxies in galaxy cluster \grifab{collisions} in general.

Regarding possible extensions of the work presented here, we can name a few:

\begin{itemize}
\item \grifac{The theoretical literature on ram pressure stripping of galaxies in clusters could be overall improved if the detailed effect of supernova and AGN feedback on the gas evolution of galaxies undergoing ram pressure stripping were characterized, by comparing a simulation including each of those effects to a control simulation without feedback.}
\item It is not clear at this point what is the expected gas mass fraction $f_{\mathrm{gas}}$ \grifab{within the disks of} disk galaxies in clusters as a function of cluster mass, infall time and redshift, an information that could be obtained from radio surveys and compared to cosmological simulations of galaxy formation or to semi-analytical models.
\item How common should jellyfish galaxies be as a function of those same parameters? Is the frequency predicted by $\Lambda$CDM in accordance with observations? Only very recently such cosmological treatment of ram pressure stripping has started to gain attention in the literature \citep{2018arXiv181000005Y}.
\item Our mechanism for the generation of jellyfish galaxies in galaxy cluster \grifab{collisions} could be further tested in other systems, using an analysis similar to the one carried out in \hyperlink{Paper III}{Paper III}. A particularly interesting extension of this work would be to carry out the same analysis for a galaxy cluster \grifab{collision} in a later stage of its evolution, in which shocks are present, in order to probe if perhaps a different mechanism for the generation of jellyfish takes place in that case.
\item \grifac{We believe that the spectroscopic properties of galaxies which have undergone or are undergoing ram pressure stripping could be better characterized. Although the photometric properties of such galaxies have been theoretically explored in the literature \citep[e.g.][]{2016A&A...591A..51S}, a more detailed study could be carried out to check whether their stellar populations and spectroscopic properties feature some kind of distinct signature as a consequence of the ram pressure.}
\end{itemize}

       						%
						%
						%
\bibliography{tex/bibliografia}	
						%
						%
\end{document}